\documentclass[conference]{IEEEtran}
\IEEEoverridecommandlockouts
\usepackage{cite}
\usepackage{amsmath,amssymb,amsfonts}
\usepackage{graphicx}
\usepackage{textcomp}
\usepackage{xcolor}
\def\BibTeX{{\rm B\kern-.05em{\sc i\kern-.025em b}\kern-.08em
    T\kern-.1667em\lower.7ex\hbox{E}\kern-.125emX}}

\usepackage{algpseudocode,algorithm}
\usepackage{latexsym}
\usepackage{url}
\usepackage{subcaption}

\def\|{\verb|}
\newcommand{\wdq}{0.24}
\newcommand{\wdd}{0.6}

\graphicspath{{./Figs/}}

\newtheorem{Def}{Definition}
\newtheorem{Theorem}{Theorem}
\newtheorem{Lemma}{Lemma}
\newtheorem{Corollary}{Corollary}
	{\trivlist \item[\hskip \labelsep{\bf Proof:}]}%
	{\hfill $\Box$ \endtrivlist}
\newenvironment{ProofOf}[1]%
	{\trivlist \item[\hskip \labelsep{\bf Proof of #1}]}%
	{\hfill $\Box$ \endtrivlist}

\begin{document}

\title{{\em Castell:} Scalable Joint Probability Estimation of
  Multi-dimensional Data Randomized with Local Differential Privacy}

\author{
  \IEEEauthorblockN{Hiroaki Kikuchi}
  \IEEEauthorblockA{\textit{School of Interdisciplinary},
      \textit{Mathematical Sciences}, \textit{Meiji University/URV}\\
    Nakano, Tokyo, \url{kikn@meiji.ac.jp}}
}

\maketitle

\begin{abstract}
  Performing randomized response (RR) over multi-dimensional data is subject 
  to the curse of dimensionality. As the number of attributes increases, 
  the exponential growth in the number of attribute-value combinations 
  greatly impacts the computational cost and the accuracy of the
  RR estimates.
  In this paper, 
  we propose a new multi-dimensional RR scheme
  that randomizes all attributes independently,
  and then aggregates these randomization matrices 
  into a single aggregated matrix.
  The multi-dimensional joint probability distributions
  are then estimated. 
  The inverse matrix of the aggregated randomization matrix
  can be computed efficiently 
  at a lightweight computation cost (i.e., 
  linear with respect to dimensionality)
  and with manageable storage requirements.
  To overcome the limitation of accuracy, 
  we propose two extensions to the baseline protocol, called  {\em hybrid}
  and {\em truncated}  schemes. 
  Finally, 
  we have conducted experiments using synthetic 
  and major open-source datasets for 
  various numbers of attributes, domain sizes, and numbers of respondents. 
  The results using UCI Adult dataset give
  average distances between the estimated and the real
  (2 through 6-way) joint probability are $0.0099$ for {\em truncated}
  and $0.0155$ for {\em hybrid} schemes, whereas they
  are $0.03$ and $0.04$ for LoPub \cite{LoPub}), which is
  the state-of-the-art multi-dimensional LDP scheme.
\end{abstract}


\section{Introduction}

With today's widespread application of Internet of things (IoT) devices, 
our daily activities are continuously being scanned and monitored.
This generates a huge amount of personal data,
most of which contain values of for many personal attributes.
These high-dimensional big data are useful for
improving human life. 
For example, Shen et al.~\cite{Shen2017} proposed a method 
for aggregating high-dimensional data to improve the response
to the demand for smart grids. 
Saint-Maurice et al.~\cite{jama2020} found that
a greater number of steps per day was associated
with a significantly lower risk of all-cause mortality in US adults. 
However, the downside of such an accumulation of personal big data
is that the data are very often highly privacy sensitive. 


Local anonymization 
has been recognized as a good approach to privacy-preserving data collection
since at least 1965, when randomized response (RR) was first 
proposed~\cite{RR2}. 
Under RR, the respondents individually anonymize their responses locally
before the responses are sent to the data controller, 
who can thereafter accurately estimate the frequencies 
of the true responses from the collected RRs.
Much more recently, 
local differential privacy (LDP) has added a differential privacy (DP)
guarantee to RRs. 
For example, 
Erlingsson et al. at Google~\cite{RAPPOR} proposed an LDP algorithm
called the randomized aggregatable privacy-preserving
ordinal response (RAPPOR), which is used by 
Google Chrome to collect user data 
in a privacy-guaranteed manner. 


Unfortunately, neither the RR nor LDP algorithms
can estimate the joint probability distribution of
high-dimensional data because of
{\em the curse of dimensionality}, which entails several issues:
\begin{itemize}
\item {\em Exponential domain growth.} 
  The number of values (categories) of the Cartesian product
  of multiple domains grows exponentially.
  The analysis of the aggregated randomization matrix entails a high
  computational and communication costs.

\item {\em Loss of dependency.} 
  A simple way to circumvent the previous problem with
  the Cartesian product is to independently randomize
  each attribute in the response. However, doing so means
  losing any non-negligible dependencies among the attributes. 
  Independently randomized values can be distributed almost
  uniformly and strongly associated pairs of data may be
  hidden over the aggregated domains. 

\item  {\em Domain sparsity.}
  Domain sparsity is an additional undesirable consequence of the exponential growth of
  attribute combinations.
  Here, the combination of values increases exponentially,
  while the number of respondents remains constant.
  The number of respondents answering any specific
  combination therefore becomes very small.
  As the number of attributes increases,
  the distribution of the RR becomes sparse,
  which implies a loss of accuracy when estimating
  the frequencies of the original responses. 
\end{itemize}


There have been many studies on high-dimensional data with DP or LDP guarantee
including 
\cite{Ding-SIGMOD2011},
\cite{Qardaji2013}, \cite{Qardaji2014},
\cite{Zhang-PrivBayes17},
\cite{Chen-KDD2015},
\cite{Day-AsiaCCS2015},
\cite{Zhang-CCS2018}, and
\cite{Xu-IFS2017}.
Most of these studies aim to inject DP noise
and focus on the optimality of subsets of attributes to minimize
estimation error.
But, the dimensionality issues were not fully examined and 
estimation accuracy loss with dimensionality was not evaluated. 
Some recent works \cite{LoPub}, \cite{LoCop}, and \cite{WAE}
use RR for high-dimensional data in an LDP guarantees. 
Ren et al. \cite{LoPub} studied an LDP scheme called LoPub, 
estimating multi-dimensional joint probability distributions.
Wang et al.~\cite{LoCop} proposed an improvement scheme, called LoCop, 
which leveraged a multivariate Gaussian copula to
estimate cross-attribute dependencies.
Jiang et al.~\cite{WAE} introduced DP-FED-WAE, which
combined a generative Wasserstein autoencoder (WAE)~\cite{WAE2018}
with federated learning. 
However, the estimation accuracies were low 
and the dimensionality issue remained. 


To address the dimensionality issues, 
Domingo-Ferrer and Soria-Comas~\cite{MDRR} considered
two extreme RR schemes, called {\sf RR-Joint} and {\sf RR-Independent}.
{\sf RR-joint} performs RR on the full domain, 
with respondents perturbing their data according to 
a predefined probability over the full domain and
the server estimating a joint probability 
via 
the inverse probability matrix. 
The estimate is accurate but does not scale well in terms of dimensionality. 
Obtaining the inverse of the exponentially grown matrix is infeasible 
because of the complexity
of the computation and communication costs.
Conversely, 
{\sf RR-Independent} performs a separate RR for each attribute and then 
estimates the marginal probabilities, 
whose product gives the estimated joint probability. 
This is scalable with dimension size in the sense that
the computation cost is linear with respect to dimensionality, and
the estimation accuracy is independent of dimensionality. 
However, its overall estimation accuracy is low, 
particularly when multiple attributes happen to be correlated strongly. 
To summarize,
Table~\ref{tbl.rrs} shows the pros and cons of the two RR schemes. 

\begin{table}[tb]\centering
  \caption{The pros and cons of various RR schemes}\label{tbl.rrs}
  \begin{tabular}{r|cc|cc} \hline
       & \multicolumn{2}{|c}{accuracy} 	& 
    \multicolumn{2}{|c}{efficiency} 	\\
    RR schemes & low dim. & high dim. & comp.  & comm.  \\ \hline
    {\sf RR-Joint}~\cite{MDRR} 		& $\surd$  & $\times$ & $\times$ & $\times$ \\
    {\sf RR-Independent}~\cite{MDRR}	& $\times$ & $\surd$ & $\surd$ & $\surd$  \\
    {\sf RR-Ind-Joint} (\S\ref{sec.ij})	& $\surd$  & $\times$ & $\surd$ & $\surd$  \\ 
    {\em hybrid}(\S\ref{sec.hybrid})	& $\surd$  & $\surd$ & $\surd$ & $\surd$  \\ 
    {\em truncated}(\S\ref{sec.truncated})& $\surd$  & $\surd$ & $\surd$ & $\surd$ \\\hline
  \end{tabular}
\end{table}


To overcome the drawbacks of these two RR schemes, 
this paper proposes a new multi-dimensional RR scheme \textsf{RR-Ind-Joint},
whereby respondents randomize (in RR) all attributes {\em independently},
the server then aggregates these individual randomization matrices 
into a single aggregated matrix,
and estimates {\em jointly} multi-dimensional joint probability distributions. 
As shown in Table~\ref{tbl.rrs}, the baseline {\sf RR-Ind-Joint}
runs efficiently at a lightweight computational cost for high-dimensional data
and estimates the joint probability as accurately as {\sf RR-Joint}.

However, because of the domain sparsity, 
the estimation loss increases exponentially with dimensionality.
To address this limitation on accuracy, we propose two extensions 
to the baseline protocol, called  {\em hybrid} (in Section~\ref{sec.hybrid})
and {\em truncated} (in Section~\ref{sec.truncated}) schemes. 
The former combines the baseline scheme with {\sf RR-Independent}
to cover a broad range of dimensions
and 
the latter truncates the joint probability of each of the $w$ attributes
at an upper limit estimated by the joint probability of the
lower ($w-1$)-dimensional data. 
Both extensions are efficient ways of improving the estimation accuracy and
compensating for the estimation loss at high dimension, as shown in Table~\ref{tbl.rrs}. 

Our schemes have the following advantages:
\begin{enumerate}
\item There exists a unique nonsingular accumulated randomization matrix
  for any attribute-independent randomized multi-dimension data. 
  We show a simple way to construct the accumulated randomization matrix
  using the Kronecker product and a necessary condition for
  having an inverse matrix (Theorem~\ref{th.otimes}). 

\item Our joint probability estimation is accurate. 
  All randomizations added independently are aggregated exactly
  and are removed completely by producing the inverse matrix. 
  Our experiments with the Adult dataset~\cite{Adult}
  showed that
  the average variant distances between the estimated and the real
  (2-way through 6-way) joint probabilities are $0.0099$ for the {\em truncated}
  scheme
  and $0.0155$ for the {\em hybrid} scheme.
  These are $0.03$ and $0.04$, respectively, of that of LoPub\cite{LoPub}, state-of-the-art
  multi-dimensional LDP scheme.
  The errors were also smaller than those for other schemes,
  including LoCop~\cite{LoCop} and WAE~\cite{WAE}. 
  
\item The inverse of the aggregated randomization matrix is scalable. 
  For the inverse matrix computation, 
  we present an efficient algorithm, called {\em castell}%
  \footnote{
    A {\em castell} is a traditional human tower built during
    festivals in Catalonia. 
    The tower has a multi-tiered structure, 
    whereby members of a team 
    first link together to form a base layer, 
    above which more layers are added one at a time
    until they reach the top.
    Disassembly is performed as the inverse of assembly. 
    Our algorithm takes input of many respondents,
    repeats the randomization for attributes, and
    discards the reverse order like as for the disassembly of 
    the {\em castell}.}%
  , that
  requires  ${\cal O}(wd^{2.807})$ computational cost
  (linear with respect to dimensionality)
  and $d^w$ memory for the matrix (the same size as for the
  $w$-way contingency table), where $w$ is the dimensionality of the
  data and $d$ is the domain size per attribute. 
  Our experiments demonstrated the rapidity of the algorithm
  (0.007 seconds for the 6-way joint probability using the Adult dataset).
  The processing time increased $2\times 10^{-5}$ per dimension.

\item The estimation accuracy is guaranteed.
  We derive an {\em the upper bound of estimation error} for both
  {\sf RR-Ind-Joint} and {\sf RR-Independent}
  (see Theorem~\ref{th.otimes} and \ref{th.rrind}, respectively). 
  This error estimation analysis enables the design of an
  optimal hybrid RR scheme that can 
  switch between two estimation schemes 
  to select the one best suited to the given set of
  parameter values for 
  dimensionality, privacy budget, and  number of respondents. 

\item Privacy is guaranteed. 
  We prove that the attribute-independent randomization
  of multi-dimensional data satisfies the LDP (see Theorem~\ref{th.ldp}). 
\end{enumerate}

Our contributions to this work are as follows:
\begin{itemize}
\item We propose a new LDP scheme \textsf{RR-Ind-Joint}
  for multi-dimensional data and
  an algorithm that estimates the joint probability distribution from
  the observed frequencies of the randomized data. 
  \textsf{RR-Ind-Joint} comprises
  an attribute-independent randomization ({\sf RR-Ind})
  and an inverse matrix computation algorithm {\em castell} that
  executes with low computational and communication costs . 

\item We calculate an upper bound for the estimation error
  of {\sf RR-Ind-Joint} and {\sf RR-Independent}
  in terms of dimensionality, the domain size,
  the privacy budget, and the number of respondents.
  Having established a formula for the estimation error,
  we can adopt a hybrid scheme
  involving {\sf RR-Ind-Joint} and {\sf RR-Independent}
  that contains an optimal estimation algorithm for any dimension. 

\item We conducted experiments to evaluate the performance 
  of the proposed schemes
  using both synthetic and open-source datasets.
  Our results show that the proposed scheme can deal
  with a wide range of datasets
  and can estimate the joint probabilities as the estimated accuracy.
\end{itemize}

The rest of the paper is organized as follows.
In Section~\ref{sec.def}, we present some fundamental definitions
and review some existing work related to multi-dimensional
anonymization.
Section~\ref{sec.proposed} outlines our scheme and
describes an algorithm for estimation.
We also discuss privacy and 
the primary factors causing estimation errors.
Section~\ref{sec.eval} presents our experimental results using
synthetic and open-source data,
which verify that our model's estimation errors are as claimed analytically. 
In Section~\ref{sec.conc}, we conclude our study based
on the proven theorems and the experimental results. 

\section{Fundamental Definitions}\label{sec.def}
\subsection{Randomized Response}

An RR is a local anonymization mechanism whereby 
each data subject/respondent masks their
own data/responses before forwarding them to a data controller. 
Each response item is randomly replaced 
by a new item with probabilities determined by a randomization matrix.  

\begin{Def}
  Let $X$ be a set of $d$ elements, 
  labeled $1, \ldots, d$ without loss of generality. 
  A $d \times d$ matrix of probabilities 
  \[
  P = \left( \begin{array}{rrr}
  p_{11} & \cdots & p_{1d} \\
  \vdots & \ddots & \vdots \\
  p_{d1} & \cdots & p_{dd}
  \end{array}\right),
  \]
  is a {\em randomization matrix} of $X$ if and only if
  $p_{i1} +\cdots+ p_{id} = 1$ for $i = 1,\ldots,d$ and
  $p_{uv}$ is the conditional probability of a 
  randomized element being $v$, given that the true element
  is $u$, (i.e., 
  \(
  p_{uv}  = Pr(Y=v|X=u)$ for all $u,v \in \{1,\ldots,d\}).
  \)
\end{Def}

An RR is a randomized mechanism
whereby input $X$ of $d$ possible values $a_1,\ldots, a_d$ is
randomized to the response $Y$ according to $P$. 
By $Y = \textsf{RR}_P(X)$,
we denote the algorithm defined in Algorithm~\ref{alg.rr}. 
The goal of RR is to estimate the frequency of $a$ in $X$.

More specifically, if we let
$\pi_1, \ldots, \pi_d$ be the proportions of respondents
whose true values fall in each of the $d$ values in $X$
and let
$\lambda_a$ be the empirical probabilities of the
observed values, we can write 
\(
(\lambda_1,\ldots,
\lambda_d)^T = P^T (\pi_1,\ldots, \pi_d)^T. 
\)
According to Warner~\cite{RR2}, an unbiased estimator $\pi$
can be computed as
\(
\hat\pi = (P^T)^{-1}\hat\lambda,
\)
where $\hat\lambda = (\hat\lambda_1, \ldots, \hat\lambda_d)^T$ is
the vector of observed empirical probabilities for $Y$. 

\begin{algorithm}[tb]\centering
  \caption{Randomization \textsf{RR}(X)}\label{alg.rr}
  \begin{algorithmic}[1]
    \State $x_i \gets$ input of party $i$ for attribute $X$.
    \State $P \gets$ a randomization matrix for attribute $X$.
    \ForAll{respondents $i = 1,\ldots, n$} 
    \State \(y_i \gets \left\{
    \begin{array}{ll} 
      x_i & \mbox{\rm w.p. $= p_{uu}$, $x_i$ is $u$-th element} \\
      v & \mbox{\rm w.p. $= p_{uv} = q$}
    \end{array}\right. \)
    \EndFor
   \State \Return the randomized response $y_1,\ldots, y_n$. 
  \end{algorithmic}
\end{algorithm}

\subsection{Multi-Dimensional RRs}\label{sec.mdrr}

Here, we review in more detail the methods proposed by Domingo-Ferrer and
Soria-Comas~\cite{MDRR}.

\subsubsection{RR-Joint}

This is the natural way to apply RRs to multiple attributes. 
Given attributes $(A_1,\ldots, A_m)$, 
we consider the Cartesian product $A_1\times \ldots \times A_m$
as a single attribute and perform RR on it. 
The distribution of the true
data is estimated as
\begin{equation}\label{eq.joint}
  \hat\Pi^{(X_1,\ldots,X_m)}_{\sf RR-Joint} = 
  (P^T)^{-1} \hat\lambda^{X_1,\ldots,X_m}. 
\end{equation}
However, {\sf RR-Joint} is severely affected by the curse of dimensionality,
because the number of value combinations of 
$A_1\times \ldots \times A_m$ grows exponentially with the number
$m$ of attributes. 
In computational terms, we must deal with matrices and 
vectors whose size is exponential in $m$, 
which is intractable except for small values of $m$. 
However, even if we had enough computational power, 
we face a more fundamental limitation, whereby
Domingo-Ferrer and Soria-Comas~\cite{MDRR} show that, 
for a fixed number of respondents, 
the error of the estimated frequencies also grows exponentially
with $m$.

\subsubsection{RR-Independent}

This is a basic approach in which RR is applied separately 
to each attribute, and the joint distribution
is estimated by assuming that the attributes are independent of each other.
Each party applies RR independently for each of 
the $m$ attributes $X^1,\ldots, X^m$
in a dataset as
$Y = (Y^1, \ldots, Y^m)$, where
$Y^j = \textsf{RR}_{P^j}(X^j)$.
After estimating the marginal probabilities for the $j$-th attribute
as
\(
\hat\pi^j = {({P^j}^{T})}^{-1} \hat\lambda^j, 
\)
the joint probability distribution for $X^1,\ldots, X^m$
is estimated by the product of the marginal distributions as
\begin{equation}\label{eq.ind}
\Pi^{(X_1,\ldots, X_m)}_{\sf RR-Ind}(a_1,\ldots,a_m)  =
  \hat\pi^1(a_1) \cdots \hat\pi^m(a_m).
\end{equation}

Algorithm~\ref{alg-rr-ind} shows the steps for this method.
The issue with {\sf RR-Independent} is that it only 
yields an accurate estimate when the independence assumption among 
attributes is (approximately) true.

\begin{algorithm}[tb]\centering
  \caption{Estimation \textsf{RR-Independent}$(Y)$}\label{alg-rr-ind}
  \begin{algorithmic}[1]
    \State $\hat{\lambda}^j \gets$ observed empirical probability for attribute $A^j$. 
    \ForAll{$j = 1,\ldots, m$}
    \State $\hat{\pi}^j \gets ((P^j)^T)^{-1} \hat{\lambda}^j$
    \EndFor
    \State 
    \(
    \hat\Pi_{\sf RR-ind}^{(1,\ldots,m)} \gets \hat{\pi}^1(x^1) \dots \hat{\pi}^m (x^m)
    \)
    for $A^1 \times \cdots \times A^m$.
    \State \Return $\hat\Pi_{\sf RR-ind}^{(1,\ldots,m)}$
  \end{algorithmic}
\end{algorithm}

\subsubsection{RR-Clusters}

To overcome the issues with {\sf RR-Joint} and {\sf RR-Independent},
\cite{MDRR} proposed {\sf RR-Clusters}.
This method splits attributes into clusters according
to their mutual dependence.
That is, attributes within a cluster are highly dependent, whereas the dependence among attributes in different clusters is low. 
The method then proceeds by performing {\sf RR-Joint} within each of the clusters,
and assumes independence across clusters to
estimate the joint distribution.

As a measure of independence, \cite{MDRR} used
Cramer's V statistics~\cite{cramerV}, which gives
a value between 0 and 1, with 0 indicating complete independence
between two attributes. Cramer's $V_{ij}$ is defined as
\[
V_{ij} = \sqrt{\frac{\chi^2_{ij}/n}{\min(d_i - 1, d_j - 1)}},
\]
where $d_i$ is the number of values in attribute $A^i$
and $\chi^2_{ij}$ is the chi-squared independence statistic
defined as
\[
\chi^2_{ij} = \sum^{d_i}_{a=1} \sum^{d_j}_{b=1} \frac{(o^{ij}_{ab} - e^{ij}_{ab})^2}{e^{ij}_{ab}},
\]
for which $o^{ij}_{ab}$ and $e^{ij}_{ab}$ are
the observed and the expected frequencies of the combination $a$ and $b$, respectively.

\section{Proposed Method}\label{sec.proposed}

\subsection{Problem Statement}

Our goal is to perturb multi-dimensional data 
in order to obtain an LDP privacy guarantee,
while being able to use 
the perturbed data
to estimate the joint probability distributions for the true data. 

Consider that $n$ respondents, each with a record of $m$ attributes
(their respective true answers).
Each attribute has a domain $\Omega_i$ of possible values.
The full domain for the $m$ attributes is $\Omega = \Omega_1 \times \cdots \times
\Omega_m$.
Each respondent uses RR to perturb
their private answer $x_i^1, \ldots, x_i^m$  into
$y_i^1, \ldots, y_i^m$ and submits the latter to a central server.
Given this perturbed data $Y^1,\ldots, Y^m$, where $Y^i = (y^i_1,\ldots, y^i_n)$,
and a randomization
mechanism (dependent on privacy budget $\epsilon$ of LDP),
the central server aims to estimate the $w$-way joint probability
distribution $\hat\Pi^S$
of a subset $S$ of $w \leq m$ attributes 
without having access to the respondents' true data $X^1,\ldots, X^m$. 

We wish to obtain a solution with the following properties.
\begin{enumerate}
\item {\em Accuracy}. The estimated probability should be close to
  the true one. Namely,
  $\hat\Pi^S \approx \Pi^S$ for any $S$. 

\item {\em Scalability}. The scheme scales dimension $w$ from computational
  and communicational (storage) perspectives. Since the full domain size $|\Omega|$
  grows exponentially to $w$, we should manage to the domain expansion. 

\item {\em Generality}. The scheme can be applied to a general multi-domain
  data without requiring any limitation. 
\end{enumerate}

\subsection{Idea}

\begin{figure}[tb]\centering
  \includegraphics[width = \linewidth]{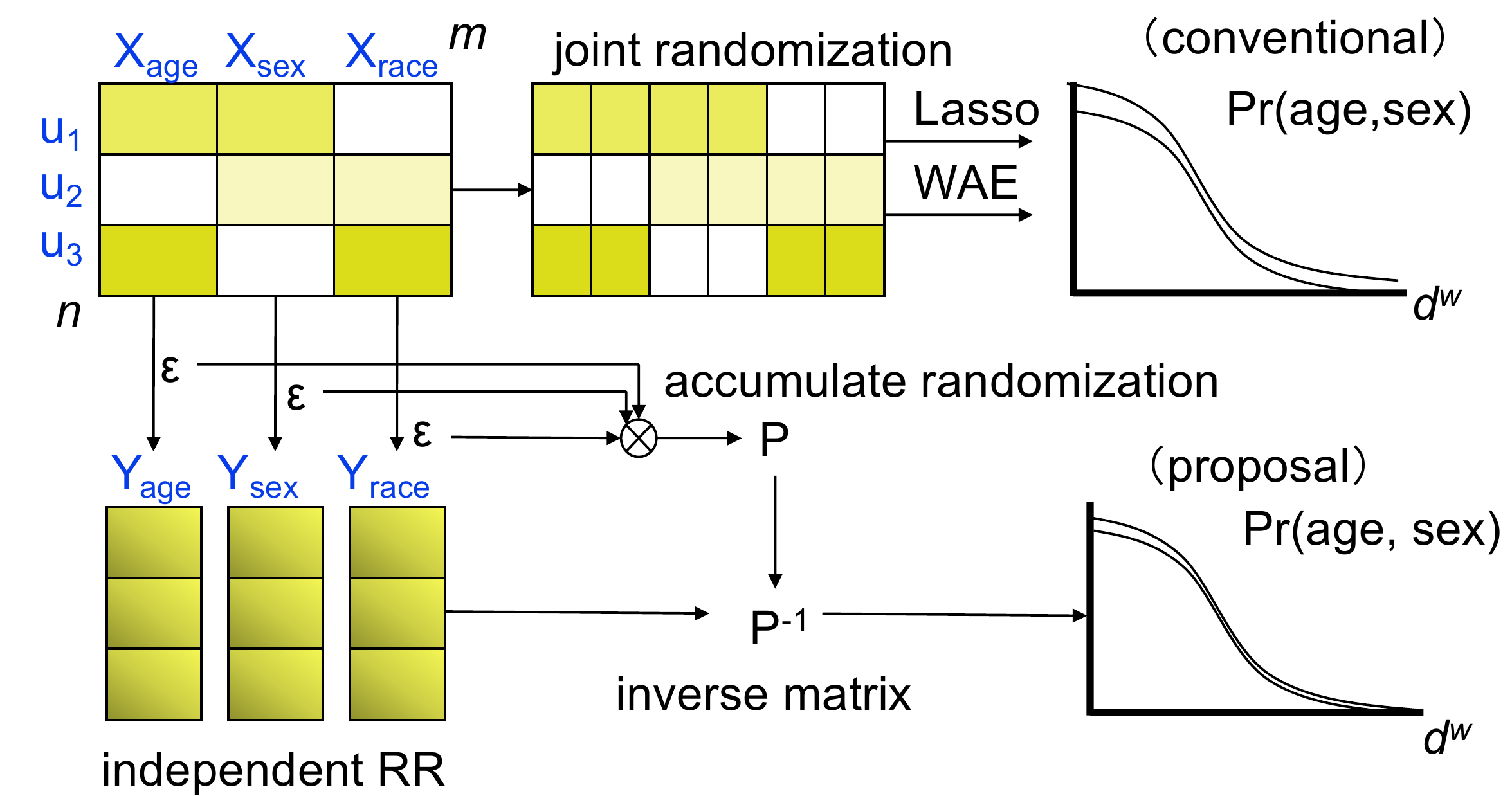}
  \caption{Overview of the proposed scheme}\label{fig-overview}
\end{figure}

We illustrate the overview of our proposed scheme in Fig.~\ref{fig-overview},
where $n = 3$ respondents have $(m = 3)$-dimensional records comprising
values for three attributes, Age, Sex and Race. Some conventional studies
randomize the matrix jointly and $w$-way joint probabilities are
estimated via the Lasso regulation~\cite{LoPub}, or the WAE~\cite{WAE}.
In contrast, our approach randomizes each attribute independently, according
to the privacy budget $\epsilon$.
The inverse of the aggregated randomization matrix allows us to
revise the randomized processes added to the original high-dimensional data
and estimate the $w$-way joint probability distribution. 

To estimate the joint probability, we must overcome the following three difficulties.
(1) The independently randomized attributes lose their dependencies. 
(2) The aggregated matrix grows exponentially with the dimensionality of the data.
With $d$ elements per attribute, the aggregation of $w$ matrices
leads to $d^{2w}$ dimensionality. 
It is therefore hard to compute the inverse matrix due to
the computation and the communication complexities. 
(3) Because of the domain sparsity, 
the estimation loss increases as the dimensionality increases.

First, we construct
the aggregated randomization matrix 
using the Kronecker product and the $w$ randomized matrixes. 
The aggregated randomization matrix recovers
the hidden associations among attributes. 
We also give a necessary condition for the aggregated matrix
to be nonsingular. 

Second, we divide the problem of inverting the aggregated matrix
into $w$ smaller subproblems using the properties of the Kronecker
product.
This reduces the computational complexity from ${\cal O}(d^w)$
to ${\cal O}(dw)$, which  is why it is called a {\em reduced} method).
However, it still requires substantial storage for both the ${d^w}^2$ matrix
and 
the inverse matrix before to be performed the product to the ($d^w$)-dimensional vector 
of the empirical probabilities. 
We therefore attempt to limit the storage overhead by
performing $w$ inverse matrix products iteratively.
Our proposed {\em castell} algorithm updates
the multi-dimensional empirical probabilities incrementally
by producing each inverse matrix, which requires a storage size of $d^w$
in total. 

Finally, with regard to the domain sparsity issue, 
we propose the {\em hybrid} and {\em truncated} schemes. 
In the {\em hybrid} scheme, 
we combine the baseline scheme with {\sf RR-Independent}
to cover a broad range of dimensionalities.
We calculate the upper bounds for the estimation error in both schemes
(Theorem~\ref{th.rrind} for {\sf RR-Independent}, and
Theorem~\ref{th.avd} for {\sf RR-Ind-Joint}), which
suggest thresholds for the features of high-dimensional data
(the dimensionality, the number of respondents, and a privacy budget)
that enable adoption of the most appropriate scheme.
The {\em truncated} scheme 
truncates the joint probability of $w$ attributes
at the upper limits estimated by the joint probability for the lower ($w-1$)-dimensional
data.

\subsection{RR-Ind-Joint}\label{sec.ij}
\subsubsection{Randomization}

The randomization process is the same as for {\sf RR-Independent}.
That is, suppose that $m$ attributes are independently randomized
to give $m$ randomization matrices $P^1, \ldots, P^m$, respectively.
After the $n$ respondents perform the randomization
processes to their respective answers $X = (X^1\ldots, X^m)$ independently,
giving $Y^j = {\sf RR}_{P^j}(X^j)$ for $j = 1,\ldots, m$,
a central server observes the perturbed records 
$Y = (Y^1,\ldots, Y^m)$. 
Here, $X$ and $Y$ are $m$-dimensional data for $n$ records over
the full domain $\Omega$ defined by the
Cartesian products of $m$ domains as 
$\Omega = \Omega_1\times \cdots \times \Omega_m$,
where each domain size is $d_i = |\Omega_i|$ for $i = 1,\ldots, m$.

When multiple attributes are randomized independently, 
can we find a joint randomization matrix that yields
the same randomization and allows us to estimate the joint distribution?
To answer this question, we leverage the independence
of the attribute randomization. 
Specifically,
two events 
$(a_1 \rightarrow a_2)$ and 
$(b_1 \rightarrow b_2)$ for attribute $A$ and $B$
are independent if and only if
\[
Pr(a_1 \rightarrow a_2 \land b_1 \rightarrow b_2)
= Pr(a_1 \rightarrow a_2) Pr(b_1 \rightarrow b_2).
\]
This makes the aggregation of multiple randomizations 
simple and practical.
We now present the following theorem for obtaining
the aggregated randomization matrix. 
\begin{Theorem}\label{th.otimes}  Let $P^i$ and $P^j$ be 
  nonsingular $(d_i^2)$ and $(d_j^2)$ randomization matrices
  for attributes $A^i$ and $A^j$, respectively.
  The matrix $P^i \otimes P^j$ 
  is a non-singular $(d^i d^j)^2$ randomization matrix
  for $(A^i, A^j)$.
\end{Theorem}

By recursively applying Theorem~\ref{th.otimes}
to every pair of attributes, 
it is straightforward to generalize it. 

\begin{Corollary}\label{col.rand}
  Let $P^{i_1}, \ldots, P^{i_w}$ be non-singular randomization matrices
  for $w$ attributes $A^{i_1},\ldots, A^{i_w}$.
  A matrix defined by $P^{i_1} \otimes \cdots \otimes P^{i_w}$
  is a non-singular
  randomization matrix for $|\Omega^{i_1}| \times \cdots \times |\Omega^{i_2}|$
  where $|\Omega^{i}|$ is a domain of attribute $A^i$. 
\end{Corollary}

A differential private randomization matrix with
$p = \frac{e^{\epsilon}}{e^{\epsilon} + d -1}$, 
$q = \frac{1}{e^{\epsilon} + d -1}$ becomes
singular only when $\epsilon = 0$ and $p = 1/d$.
Because it is not hard to avoid the trivial case of $\epsilon = 0$,
we can confirm that there exists a non-singular accumulated
randomization matrix for any given set of
(non-singular) randomization matrices.

\subsubsection{Estimation}\label{sec.est}

We now consider the estimation of the $w$-way joint probability distribution
of set of attributes 
$S=  \{A^{i_1}, \ldots, A^{i_w}\} \subset \{A^1, \ldots, A^{m}\}$
from the independently RRs $Y^1, \ldots, Y^m$.
Let $w$ be the size of subset $S$, i.e.,  $|S| = w \le m$. 

Given an aggregated randomization matrix $P^{i_1}\otimes P^{i_w}$,
the $w$-way joint probability is given as
\[
\hat\Pi^S = ((P^{i_1}\otimes P^{i_w})^T)^{-1} \lambda^S,
\]
where $\lambda$ is a $(d_{i_1}\cdot d_{i_w})$-dimensional vector
for the empirical distribution of $Y^{i_1},\ldots, Y^{i_w}$.

\subsubsection{Inverse Matrix}

The dominant cost in estimating the joint probability is for
the matrix inversion. If Strassen's algorithm~\cite{Strassen},
known as the best performing algorithm,
is used, 
the inversion cost is ${\cal O}(|\Omega|^{2.807})$.
We therefore must be able reduce the cost of  matrix inversion
for both computation and storage reasons. 
  
Let us consider the inverses of aggregated matrix
of $(d_A^2)$ and $(d_B^2)$ matrices $P_A$ and $P_B$, 
respectively.
From the fundamental property of the Kronecker product
that $(A\otimes B)^{-1} = A^{-1}\otimes B^{-1}$, 
we can compute the inverse matrix as follows,
\begin{eqnarray}
  \hat\Pi^{AB} &=& \left(P_A \otimes P_B\right)^{-1} \lambda^{AB} \label{eq.naive} \\
  &=& \left({P_A}^{-1} \otimes {P_B}^{-1}\right) \lambda^{AB} \label{eq.reduced}
\end{eqnarray}
where $\lambda^{AB}$ is a $(d_A \times d_B)$-dimensional vector. 
(For simplicity, we omit the initial transpose of $P$ hereafter.)
We call Eqs.~(\ref{eq.naive}), and (\ref{eq.reduced}) as 
{\em na\"ive}, and {\em reduced}, respectively. 
The {\em reduced} method divides the computational cost
of the matrix inversion into 
those for lower-dimensional
$d_A^2$  and $d_B^2$.
However, it still requires the storage of the aggregated 
matrix, which is $(d_A\times d_B)^2$ one.
The aggregated matrix becomes too large to store realistically
when $|\Omega| = d_1 \times d_w > 5000$.

Note that we are proposing the {\em castell} method that 
solves the inverse matrix while incurring only lightweight computation costs
manageable storage requirements.
Recall Eq.~(\ref{eq.reduced}), which can be written as
\begin{eqnarray}
  &=& (P_A^{-1} \otimes P_B^{-1}) \lambda^{AB} \nonumber \\
  &=& \left(\begin{array}{cc}
    a_{11}P_B^{-1} & a_{12}P_B^{-1} \\
    a_{22}P_B^{-1} & a_{22}P_B^{-1} 
  \end{array}\right)
  \left(\begin{array}{c}
    \lambda_{a_1b_1} \\
    \lambda_{a_1b_2} \\
    \lambda_{a_2b_1} \\
      \lambda_{a_2b_2} \\
  \end{array}\right) \nonumber \\
  &=& \left(\begin{array}{c}
    a_{11}P_B^{-1} \lambda_{a_1} + a_{12}P_B^{-1} \lambda_{a_2} \\
    a_{21}P_B^{-1} \lambda_{a_1} + a_{22}P_B^{-1} \lambda_{a_2} \\
    \end{array}\right) \label{eq.expand} \\
  &=& \left(\begin{array}{cc}
    a_{11} & a_{21} \\
    a_{21} & a_{22} \\
  \end{array}\right)
  \left(\begin{array}{c}
    P_B^{-1}\lambda_{a_1} \\
    P_B^{-1}\lambda_{a_2} \\
  \end{array}\right)  \nonumber \\
  &=& P_A^{-1} P_B^{-1} \lambda^{AB}, \label{eq.vector}
\end{eqnarray}
where $(a_{ij})$ is an element of $P_A^{-1}$ and
$\lambda_{a_i}$ of Eq.~(\ref{eq.expand}) is $d_B$-dimension vector $(\lambda_{a_ib_1} \, \lambda_{a_ib_{d_B}})^T$
for $i = 1, \ldots, d_B$. 
Note that we assume $d_A = d_B = 2$ here for simplicity. 
Eq.~(\ref{eq.vector}) is $(d_A \times d_B)$-dimension vector 
$\hat\Pi^{AB} = \left(\begin{array}{c} 
  \Pi^{AB}_{a_1} \\ 
  \Pi^{AB}_{a_2}
\end{array}\right)$.
If we rearrange the vector as 
$d_A \times d_B$ matrix,  
Eq.~(\ref{eq.vector}) can be written simply as
\(
  (\Pi^{AB}_{a_1} \, \Pi^{AB}_{a_2}) 
  = P_A^{-1} \left( P_B^{-1} {\Lambda^{AB}}^T \right)^T,
\)
where $\lambda^{AB}$ in Eq.~(\ref{eq.reduced}) is replaced by
$\Lambda^{AB}$, which is a $(d_A \times d_B)$ matrix
obtained by rearrangement of the elements of  the $\lambda^{AB}$ 
empirical distribution over $(Y_A, Y_B)$. 
This is the basic idea in the {\em castell} method for matrix inversion.
The computation for the {\em castell} inverse is as lightweight as 
that for the {\em reduced} method 
because it only requires the inversion of each $d_i^2$ matrix.
The matrix size does not increase as it does for  multi-dimensional data,
and
the storage size stays constant, irrespective of the number of
dimensions being processed. 
It requires storage only for the
empirical distribution $\Lambda^{AB} (= \lambda^{AB})$, which
is proportional to $|\Omega| = d_1\times\cdots\times d_w$.
Therefore, the {\em castell} inversion is efficient
in terms of both computational and communicational costs. 

\begin{figure}[tb]\centering
  \includegraphics[width=0.7\linewidth]{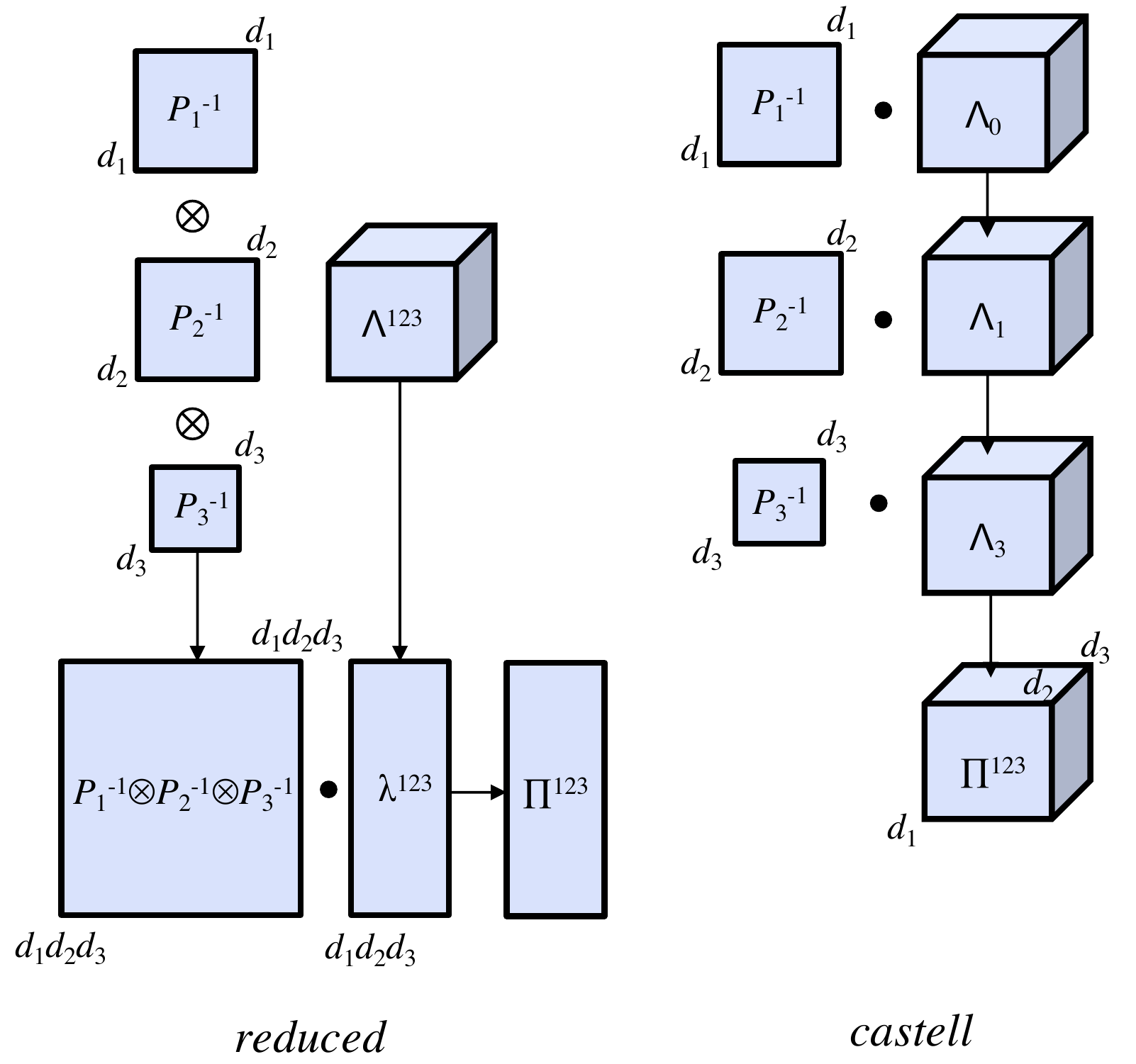}
  \caption{Flows in the joint probability estimation
    from three randomization matrices $P_1, P_2$, and $P_3$
    for the {\em reduced} and {\em castell} algorithms}
  \label{fig-castell}
\end{figure}

To make the difference between the {\em reduced} and {\em castell} algorithms
clear, 
we illustrate the two estimation steps in Fig.~\ref{fig-castell}.
Given three randomization matrices $P_1$, $P_2$ and $P_3$
having $d_1^2$, $d_2^2$ and $d_3^2$ dimensions, respectively, 
the 3-way joint probabilities $\Pi^{123}$ are estimated. 
Note that the estimated joint probabilities $\Pi^{123}$
in the vector for the {\em reduced} method are
identical to those for the ($d_1 \times d_2 \times d_3$)-dimensional
data $\Pi^{123}$ (i.e., 
the two estimations are mathematically equivalent).

To extend the basic idea of {\em castell} inversion to $w$-dimensional data,
we introduce a new transposition for multi-dimensional data.
\begin{Def}
Let $A$ be $d_1\times d_2 \times \cdots \times d_w$ multi-dimensional data
\(
(a_{i_1\cdots i_w})
\)
for $1 \le i_j \le d_j$. 
Then, $i$-th transposition of $A$ is a
$d_i \times (d_1\cdots d_{i-1} d_{i+1} \cdots d_w)$ matrix
denoted by $A^{T_i}$ 
such that 
\begin{eqnarray}
  A^{T_i} &=&
  \label{eq.castell} 
  \left(\begin{array}{ccc}
    a_{\cdots 1 \cdots 1} & \cdots & a_{\cdots 1\cdots d_w} \\
    \vdots & \ddots & \vdots \\
    a_{\cdots d_i\cdots 1} & \cdots & a_{\cdots {d_i} \cdots d_w}  \\
  \end{array}\right). 
\end{eqnarray}
The inverse of $i$-th transposition, 
denoted by $A^{T_i^{-1}}$, 
is  a $d_1\times \cdots \times d_w$ multi-dimensional data such that
$(A^{T_i})^{T_{i}^{-1}} = A$.  
\end{Def}
Note that $A^{T_i}$ is a matrix (2-dimensional data)
and $A^{T_i^{-1}}$ is a multi-dimensional data. 
The rows of the matrix in Eq.~(\ref{eq.castell}) are ordered according to
the $i$-th attribute
and the columns can be arranged arbitrarily.
We now introduce a simple method for arranging the columns via
a permutation $\sigma$ of the sequence of
dimension identities $\langle 1,2,\ldots, w \rangle$. 
For example, consider the 3-dimensional data comprising  {\em race}, {\em sex}, and {\em income},
with domain sizes $d_1 = 5$, $d_2 = 2$, and $d_3 = 2$, respectively.
Let $\Lambda$ be the empirical distribution for $S = \{A_{race}, A_{sex}, A_{income}\}$
(comprising
$5\times 2\times 2 = 20$ elements) as
\[
\left(
\left(\begin{array}{cc}
  1 & 6 \\
  2 & 7 \\
  3 & 8 \\
  4 & 9 \\
  5 & 10 \\
  \end{array}\right), 
\left(\begin{array}{cc}
  11 & 16 \\
  12 & 17 \\
  13 & 18 \\
  14 & 19 \\
  15 & 20 \\
  \end{array}\right)
\right).
\]
By letting $\sigma$ be a permutation of the sequence $\langle 1, 2, 3 \rangle$,
where $\sigma(1) = 3, \sigma(2) = 1$, and $\sigma(3) = 2$
for the $(i = 2)$-th transposition, we have
$d_2 \times d_3\cdot d_1 = 2 \times 10$ matrix
\[
\Lambda^{T_2} =
\left(\begin{array}{cccccccccc}
  1 & 11 & 2 & 12 & 3 & 13& 4 & 14 & 5  & 15 \\
  6 & 16 & 7 & 17 & 8 & 18& 9 & 19 & 10 & 20 \\
  \end{array}\right)
\]
and the inverse transposition defined by
$\sigma^{-1}(1) = d - i + 2 = 2$, $\sigma^{-1}(2) = 3$, and $\sigma^{-1}(3) = 1$
gives a $d_1 \times d_2 \times d_3$ multi-dimensional data.

A general {\em castell} inversion of matrix is defined as follows.
Let $P_1, \ldots, P_w$ be randomization matrices.
Given an empirical distribution $\Lambda^S$ of $d_1\times \cdots \times d_s$
dimensionality, the $w$-way joint probability is estimated as
\[
\hat\Pi^S =
          {P_1}^{-1}
          \left(
          \cdots
          \left(P_{w-1}^{-1}
          \left(P_w^{-1} {\Lambda^S}^{T_w}\right)^{T_w^{-1} T_{w-1}}
          \right)^{\cdots} 
          \right)^{T_1}. 
\]
Note that the inverse of transposition $T_i^{-1}$ is inserted
for every $i$-th product. 
This leaves the order of dimensions in the empirical distribution $\Lambda$
unchanged when cascading the $w$ products. 
That is, the estimate $\hat\Pi$ always remains 
a $(d_1\times d_2 \times \cdots \times d_w)$ dimensional data. 


To implement this approach of computing the estimates incrementally, 
we present the Algorithm~\ref{alg.ijoint},
which is a procedure for estimating
the joint probabilities of the $w$ attributes from independently
RRs $Y^1 \ldots Y^w$.

\begin{algorithm}[tb]\centering
  \caption{Estimation {\sf RR-Ind-Joint}, {\em Castell}}\label{alg.ijoint}
  \begin{algorithmic}[1]
    \State $P_1,\ldots, P_m \gets $ randomization matrices.
    \State $Y^i \gets {\sf RR}_{P_i}(X^i)$ for $i = 1,\ldots, m$. 
    \State $S \subset \{A^1,\ldots, A^m\}$ such that $|S| = w$. 
    \State $\hat{\Lambda}^{S} \gets$ 
    $(d_1\times \cdots \times d_w)$-dimension data of 
    empirical distribution for attributes $(Y^1,\ldots, Y^m)$. 
    \State $\Lambda_w \gets \hat\Lambda^S$. 
    \ForAll {$i = w,\ldots, 1$}
    \State $\Lambda_{i-1} \gets (P_{i}^{-1} \Lambda_i^{T_{i}})^{T_{i}^{-1}}$
    \EndFor
    \State \Return $\hat{\Pi}^{S} \gets \Lambda_0$
  \end{algorithmic}
\end{algorithm}

We assume the use of Strassen's algorithm~\cite{Strassen} 
for the primitive matrix inversions and
a domain size for the $w$ attributes of
$d_1 = \cdots = d_w = d$, for simplicity. 
\begin{Theorem}\label{th.complexity}
  Let $\Lambda$ be a $(d^w)$-dimensional data representing
  the empirical distribution
  of $w$ attributes. 
  Algorithm~\ref{alg.ijoint} runs in ${\cal O}(wd^{2.807})$ time
  and requires $d^w$ of storage. 
\end{Theorem}
Table~\ref{tbl.invcost} summarizes the computation and storage costs
for three matrix inversion algorithms.

\begin{table}[tb]\centering
  \caption{Costs for matrix inversion}\label{tbl.invcost}
  \begin{tabular}{r|c|c|c} \hline
    method & inverse & computation cost & storage cost \\ \hline
    1. {\em na\"ive} & $(P_1\otimes P_2 \otimes P_3)^{-1}$ & ${\cal O}(d^{w 2.807})$ & $d^{2w}$ \\
    2. {\em reduced} & $P_1^{-1}\otimes P_2^{-1} \otimes P_3^{-1}$ & ${\cal O}(wd^{2.807})$ & $d^{2w}$ \\
    3. {\em castell} & $P_1^{-1} P_2^{-1} P_3^{-1}$ & ${\cal O}(wd^{2.807})$ & $d^{w}$ \\ \hline
  \end{tabular}
\end{table}

\subsection{Hybrid Scheme}\label{sec.hybrid}

We now consider a hybrid scheme positioned
between {\sf RR-Independent} and {\sf RR-Ind-Joint}.
As we will show shortly in Sections~\ref{sec.error.ind} and
\ref{sec.error.joint},
{\sf RR-Ind-Joint} is efficient when $w$ is small and $n$ is large,
whereas estimation via {\sf RR-Independent} is stable, simple, and
less dependent on the dimensionality $w$.
It is therefore useful to consider a hybrid of these two schemes
for estimating
probabilities in a more general environment. 

The optimal algorithm will depend on the given data, for which
several parameters are involved. 
Fortunately, 
our analysis shows that the estimation accuracy
is monotonic (increasing/decreasing) with respect to
the parameters;
$n$ (a number of respondents)
$\epsilon$ (a privacy budget), and 
$w$ (dimensionality)
for both schemes. 
An optimal estimation algorithm can therefore be found by
switching between {\sf RR-Independent} and {\sf RR-Ind-Joint}
at predetermined threshold $n^*$, $\epsilon^*$, and $w^*$. 
Table~\ref{tbl.crossover} shows an example of
the thresholds estimated for the Adult dataset
($n = 32,561$ and $d = 16$ (education)). 
(The detailed analysis is given in Appendix~\ref{sec.thresholds}.) 

\begin{table}[tb]\centering
  \caption{Thresholds for selecting the optimal algorithm as either
    {\sf RR-Ind-Joint} or {\sf RR-Independent}}\label{tbl.crossover}
    \begin{tabular}{c|r|c} \hline
      threshold & Adult dataset & scheme \\ \hline
      $n > n^*$ & 121,000 &  \\ 
      $\epsilon > \epsilon^*$ & 0.473 & {\sf RR-Ind-Joint} \\
      $w <  w^*$ & 1.374 & \\ \hline 
      otherwise  & & {\sf RR-Independent} \\ \hline
    \end{tabular}
\end{table}

\subsection{Truncated Scheme}\label{sec.truncated}

An estimation method that uses the inverse matrix
could generate an invalid probability, such as a negative value,
or a value greater than $1.0$.
In addition to restricting these trivial invalid values,
we develop a heuristic for regulating the estimated error growth,
based on 
the relationship between the joint and the marginal probabilities as
\begin{eqnarray*}
  Pr(A,B) &=& Pr(A|B)Pr(B) \le Pr(B), \\
  &=& Pr(B|A)Pr(A) \le Pr(A), \\
  &\le& \min(Pr(A), Pr(B)).
\end{eqnarray*}
We leverage this relationship to give the generalized inequality
\begin{equation}
  0 \le Pr(S) \le \min_{S'\subset S, |S'| = |S|+1} Pr(S').
\end{equation}
This represents a limitation on the valid elements of the estimated probabilities.
Using the limits estimated for $w-1$ dimensionality,
we can truncate a too-high probability at the $w$ level,
as shown in Algorithm~\ref{alg.truncated}. 

\begin{algorithm}[tb]\centering
  \caption{Estimate improvement {\em Truncated}}\label{alg.truncated}
  \begin{algorithmic}[1]
    \State $S = \{A^1,\ldots, A^w\}$ a subset of set of $m$ attributes.
    \State $\hat{\Lambda}^{S} \gets$ 
    $\Omega = (d_1\times \cdots \times d_w)$-dimension data of 
    empirical distribution for attributes $(Y^1,\ldots, Y^m)$. 
    \State $\hat\Pi^S \gets $ {\em Castell} $(\hat\Lambda^{S})$
    \State $\Pi_0 \gets \hat\Pi^S$ where all minus values are replaced with $0$.
    \ForAll {$i = 1,\ldots, w$}
    \State $S_i \gets S - \{A^i\}$
    \State $\Pi_{i}(a) \gets $ $\min
    (\Pi_{i-1}(a), \mbox{{\em Castell}}(\Lambda^{S_i}(a)))$ for $a \in \Omega$.
    \EndFor
    \State \Return $\Pi_{w}$
  \end{algorithmic}
\end{algorithm}

\subsection{Privacy}

The privacy of the \textsf{RR-ind-joint} scheme is the same as that of \textsf{RR-Ind}.
Consider a simple RR that gives a response $x$ with a probability of 
$p = \frac{e^{\epsilon}}{e^{\epsilon}+d -1}$
and gives a randomly chosen value in $\Omega_A$ as a response
with a probability of $q = (1-p)/(d - 1) = \frac{1}{e^{\epsilon}+d -1}$

The LDP holds for all independent RR attributes as follows.

\begin{Theorem}\label{th.ldp} \textsf{RR-Ind-Joint} satisfies 
$(\epsilon, 0)$-LDP
for attribute $A$.
With $m$ attributes $A^1,\ldots, A^m$, 
\textsf{RR-Ind-Joint}
satisfies $(m\epsilon,0)$-LDP. 
\end{Theorem}

Because LDP guarantees that it is unable to infer the true value 
from a randomized one,
it does not provide DP~\cite{Dwork2014}.
LDP is related to attribute inference attack~\cite{Fang2020}
and DP prevents membership inference attack~\cite{Shokri2017}. 
According to Yeom et al.~\cite{Yeom2018},
attribute inference is at least as difficult as
membership inference.
Therefore, we think that a multi-dimensional data randomized to meet
the LDP guarantee implies it will also meet the DP guarantee. 


\subsection{Estimation Error ({\sf RR-Independent})}\label{sec.error.ind}

We evaluate the accuracy loss 
for the estimated joint probability
in terms of the mean absolute error (MAE), the mean absolute error (MAE)
and 
the average variation distance (AVD)\footnote{
  Ren et al.~\cite{LoPub} suggested the average variant distance,
  which is essentially equivalent to the AVD.
}. 
MAE is defined as
\(
MAE = 1/{d^2} \sum_{x \in |A|\times |B|} |\Pi^{AB}(x) - \hat\Pi^{AB}(x)|.
\)
%
%
AVD was suggested by \cite{LoPub} and \cite{WAE}.
Let $C$ be a subset of attributes 
$\{X^{i_1},\ldots, X^{i_w}\}$.
The AVD between
the real joint probability distributions $\Pi^C$ and 
the estimated distributions $\hat\Pi^C$ is defined as
\[
\mbox{\em AVD}  = \frac{1}{|A|} \sum_{C \in A}
\sup_{(c \in C} | \Pi^{c} - \hat\Pi^{c} |,
\]
where $A$ is a power set of attributes such that
$C$ has $w$ distinct attributes.

First, we show a bound for the MSE of \textsf{RR-Independent}.
\begin{Theorem}\label{th.rrind.mse}
  Let $A$ and $B$ be two attributes with
  Cramer's V statistics $V$ and the same number of values
  in both domains, i.e., $d = |A| = |B|$.
  The MSE of \textsf{RR-Ind} is less than $V^2/d$. 
\end{Theorem}
Taking the squared root of both sides, we estimate that the MAE for \textsf{RR-Independent}
is proportional to $V/\sqrt{d}$ 

Next, we consider an upper bound for the estimation error when
$w$-way joint probability is estimated by {\sf RR-Independent}.
Assume that the marginal probability $\pi(a)$ for $a \in A^i$ is proportional
to the domain size $|\Omega_i| = d_i$, for all $i \le w$.
The $w$-way joint probability is then
\[\Pi^S_{\sf RR-Ind} (a_1,\ldots, a_w) =
\hat\pi^1(a_1)\cdots \hat\pi(a_w) = \frac{1}{d_1}\cdots\frac{1}{d_w}.
\]
We can now identify a range of possible values for
the joint probability, as follows.
\begin{Lemma}\label{le.marginal}
  Let $\Pi$ be a real $w$-way joint probability of $w$ attributes with
  marginal probabilities $\pi^1,\ldots, \pi^w$. Then, for any $a_1,\ldots, a_w$
  of $w$ attributes, 
  \[
  0 \le \Pi(a_1,\ldots, a_w) \le \min( \pi^1(a_1),\ldots, \pi^w(a_w))
  \]
  holds. 
\end{Lemma}
This means that the estimated probability must belong within
the interval $[0, \min(\pi^1,\ldots, \pi^w)]$.
We can derive an upper bound for the estimation error in {\sf RR-Independent} as follows.
\begin{Theorem}\label{th.rrind}
  Let $\hat\pi^i$ be 
  the estimated marginal probability of the $i$-th attribute,
  which follows a uniform distribution of $1/d_i$, where $d_i$ is 
  the size of the $i$-th domain
  for $i = 1,\ldots, w$.
  A $w$-way joint probability of $w$ attributes
  is estimated by {\sf RR-Independent} with an error less than 
  \begin{equation}\label{eq.avd-rrind}
    \max(d_*^{-w}, 1/d_* - d_*^{-w}),
  \end{equation}
  error where $d_* = \max_{i \le w} d_i$. 
\end{Theorem}

\subsection{Estimation Error ({\sf RR-Ind-Joint})}\label{sec.error.joint}

The MAE of \textsf{RR-Ind-Joint} does not depend on $V$ 
because it estimates the joint probability of attributes
via inversion of the randomization matrix. 
\textsf{RR-Ind-Joint} has no estimation error 
provided all randomization matrices for the attributes are
non-singular. 

The estimation of \textsf{RR-Ind-Joint}
suffers a rounding error in the empirical probability distribution $\lambda^{AB}(Y)$. 
The observed probability of $(a,b)$ for $Y$ is the fraction of
respondents who send $(a,b)$ out of the $n$ respondents. The precision of empirical probability $\lambda^{AB}(Y)$ is therefore $1/n$.
We consider a model of empirical probability having the form,
\(
\hat\lambda = \lambda + \Delta\lambda,
\)
where $\lambda$ is the real joint probability
of the randomization matrix and $\Delta\lambda$ is the rounding error.
For example, suppose that we observe
the empirical probability of a set of $n = 10$ perturbed 
records as
\[
\hat\lambda^{AB} = \left(\begin{array}{c}3/10\\ 3/10\\ 1/10\\ 3/10
\end{array}\right)
 = \left(\begin{array}{c}0.2875\\ 0.2625\\ 0.1625\\ 0.2875
\end{array}\right)
+ \left(\begin{array}{r} 0.0125\\ 0.0375\\ -0.0625\\ 0.0125
\end{array}\right),
\]
where 
the empirical probabilities are the sums of
the expected values, determined by a randomization mechanism 
($P^A$, $P^B$ and $f^X$ (see Appendix~\ref{sec.exp}))
and 
the rounding error $\Delta\lambda$.  

We consider the rounding error as a uniform distribution over
$[-1/n, 1/n]$, for which 
$E[\Delta\lambda] = 0$ and  $E[|\Delta\lambda|] = 1/2n$ holds.
Note that the rounding errors are within the range
\(
-1/10 < -0.0625 < 0.0375 < 1/10.
\)
Using this model, 
the estimation of the joint probability is
\[
\hat\Pi = P^{-1} \hat\lambda = P^{-1}(\lambda - \Delta\lambda)
= \Pi - P^{-1}\Delta\lambda.
\]
The last term in the above formula is the source of the estimation error. 
It is a linear combination of $d^2$ uniform distributions
and can be approximated as a normal distribution whose
mean increases with $1/n$. 

\begin{Lemma}\label{le.adj}
  Let $P$ be a randomization matrix for a set of $d$ elements
  with $p_{ii} = p = e^{\epsilon}/(e^{\epsilon} + d -1)$ and
  $p_{ij} = 1/(e^{\epsilon} + d -1)$ for $i \ne j \in \{1,\ldots,d\}$. 
  An element of $P^{-1}$ is at most $1/p$. 
\end{Lemma}

\begin{Lemma}\label{le.inv}
  Let $X_1$ and $X_2$ be attributes of $n$ records 
  with domains of size $d_1$ and $d_2$, respectively. 
  An independently randomized
  matrix $P$ with $\epsilon$-DP
  has a rounding error such as
  \[
  \max P^{-1}\Delta\lambda 
  < \left(1 + \frac{\max(d_1,d_2) -1}{e^{\epsilon}}\right)^2 \frac{d_1 d_2}{n}
  \]
\end{Lemma}

\begin{Theorem}[upper bound of estimation error]\label{th.avd}
  A $w$-way joint probability distribution of $n$ records
  with domain sizes $d_1, \ldots, d_w$, respectively, 
  is estimated from an independently randomized matrix
  with $\epsilon$-DP in {\sf RR-Ind-Joint}
  with an error not exceeding
  \begin{equation}\label{eq.avd}
  \left(1 + \frac{d-1}{e^{\epsilon/w}}\right)^w\frac{d^w}{n},
  \end{equation}
  where $d = \max(d_1,\ldots, d_w)$. 
\end{Theorem}
Note that the estimation error is asymptotically linearly related to
the size of the full domain $|\Omega| = d_1 \times \cdot \times d_w < d^w$.

\section{Evaluation}\label{sec.eval}
\subsection{Data}

We evaluate the utility loss in RR processing and estimating
using four major open-source datasets and 
a synthetic dataset (see Appendix~\ref{sec.synthetic}).

Table~\ref{tbl.datasets} shows the specifications for the open-source datasets 
that are required to evaluate the performance of the proposed schemes.
We chose major open-source datasets that comprised
multi-dimensional data records.
Each dataset has $n$ records (rows) of $w$ attributes (columns). 
Each attribute has a domain $\Omega$ of possible values.
The full domain for the $w$-dimensional data is denoted by
the Cartesian product of all attributes 
$\Omega = \Omega_1 \times \cdots \times \Omega_w$.
We denote the size of the full domain by $|\Omega|$.
Generally, $|\Omega|$ increases exponentially with data dimensionality $w$.
For example, the US Census dataset has 68 categorical attributes with several 
domain sizes ranging from 2 (Sex, iKorean) to 18 (iYewarsch).
The full domain is $1.7 \times 10^{44}$. 
Depending on the dataset, we randomly choose 20 -- 50 combinations
of $w$ attributes to form $A$ 
and take the average of the estimation errors (distances)
for the $w$-way joint distributions.

\begin{table}[tb]\centering
  \caption{Dataset specifications}\label{tbl.datasets}
  \begin{tabular}{lp{2cm}rrr} \hline
    dataset & description & \# records $n$ & \# att. $w$ & domain size $\Omega_i$ \\ \hline
    Adult & UCI census income data~\cite{Adult}		& 32,561 	& 8 	& 1814400 \\
    Census & US Census Data (1990)~\cite{Uscensus}	& 2,458,285	& 68	& 1.711505e+44 \\
    Credit & German Credit Data (2000)~\cite{Credit}			& 1,000		& 13	& 34,560,000 \\
    Nursery & Enrollment data in 1980's Nursery school~\cite{Nursery}	& 12,960	& 9	& 64,800 \\ \hline
  \end{tabular}
\end{table}

\subsection{Results (Open-source Data)}\label{sec-results}

Table~\ref{tbl.adult} shows the MAE for two attribute values
in the Adult dataset: namely,
$n = 32,561$ and privacy budget $\epsilon = 1$. 

\begin{table}[tb]\centering\small
\caption{Example MAE for the Adult dataset}\label{tbl.adult}
\begin{tabular}{r|cc|cc|cc} \hline
  \mbox{} & sex & income & sex & race & edu. & occupa. \\ \hline
  $d$   & 2     &  2     & 2   & 5    & 16        & 15 \\
  $V$   & 
\multicolumn{2}{|c}{0.216} & 
\multicolumn{2}{|c}{0.118} &
\multicolumn{2}{|c}{0.187} \\
{\sf RR-Ind} & 
\multicolumn{2}{|c}{$1.88 \times 10^{-3}$} & 
\multicolumn{2}{|c}{$1.10 \times 10^{-4}$} &
\multicolumn{2}{|c}{$2.15 \times 10^{-5}$} \\
{\sf RR-Ind-Joint} & 
\multicolumn{2}{|c}{$7.07 \times 10^{-10}$} &
\multicolumn{2}{|c}{$2.26 \times 10^{-9}$} &
\multicolumn{2}{|c}{$6.28 \times 10^{-7}$} \\ \hline
\end{tabular}
\end{table}

\begin{figure}[tb]\centering
    \includegraphics[width=\wdd \linewidth]{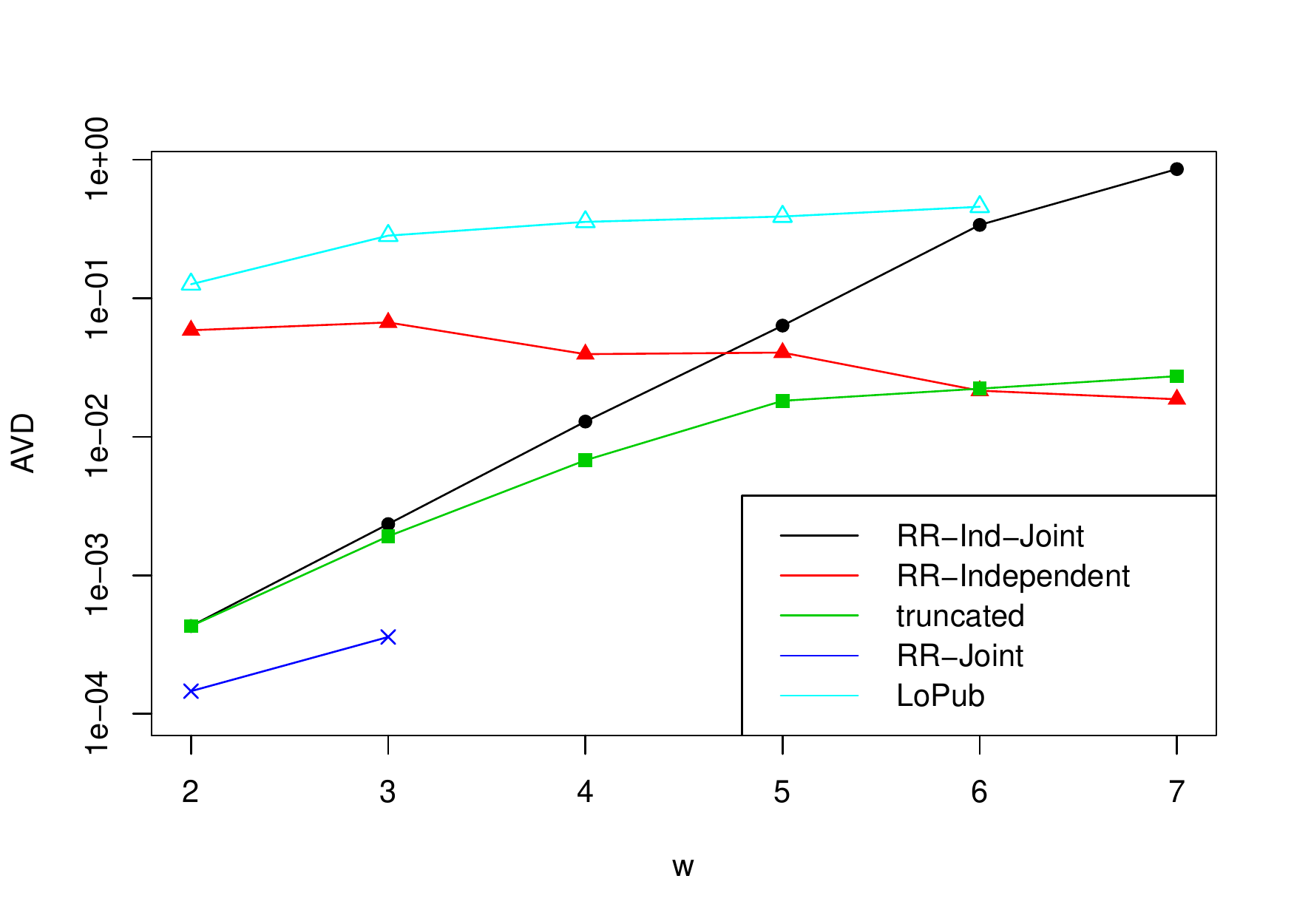}
    \caption{AVDs for several algorithms
      with respect to the dimensionality $w$}\label{fig-lopub}
\end{figure}

Fig.~\ref{fig-lopub} shows the AVD between 
the real and the estimated joint probability
distributions with respect to dimensionality $w$.
We estimate the $w$-way joint probability via LoPub~\cite{LoPub}
(Lasso regression with the 64 bits for Bloom filter and 5 hash functions)
and via the proposed {\sf RR-Ind-Joint} method. 

The AVDs for {\sf RR-Ind-Joint} are distributed around
$4\time 10^{-4}$ to $1\time 10^{-2}$ for $w \le 4$. 
There are quite small in comparison to LoPub. 
The accuracy is very high in comparison with any recent
multi-dimensional LDP schemes such as LoCop~\cite{LoCop}
and Wasserstein autoencoder (WAE)~\cite{WAE}
(According to Fig.~5~\cite{WAE}, the AVDs for LoCop
are close to those for LoPub and the AVDs of WAE are
almost half of those for LoPub and LoCop. The estimation results
of WAE are in the range $0.05$ to $0.09$. )

Fig.~\ref{fig-lopub} also shows the AVDs of the multi-dimensional
RR schemes {\sf RR-Independent} and {\sf RR-Joint}~\cite{MDRR}.
Because of the exponential nature of 
computational and capacity costs, the estimating via {\sf RR-Joint}
with the dimensionality of more than 3 was not feasible.
The AVDs for {\sf RR-Ind-Joint} are much better than those for 
{\sf RR-Independent}. 
Note that we have plotted the AVDs using a logarithmic scale. 
Table~\ref{tbl-comp} shows that
the {\em hybrid} and {\em truncated} schemes
outperform the conventional works and
the AVDs for 6-way joint probabilities of {\em hybrid} and {\em truncated}
schemes are $0.0155$ and $0.0099$, which is $0.03$ of that in 
LoPub with the same condition. 

\begin{figure}[tb]\centering
    \includegraphics[width=\wdd \linewidth]{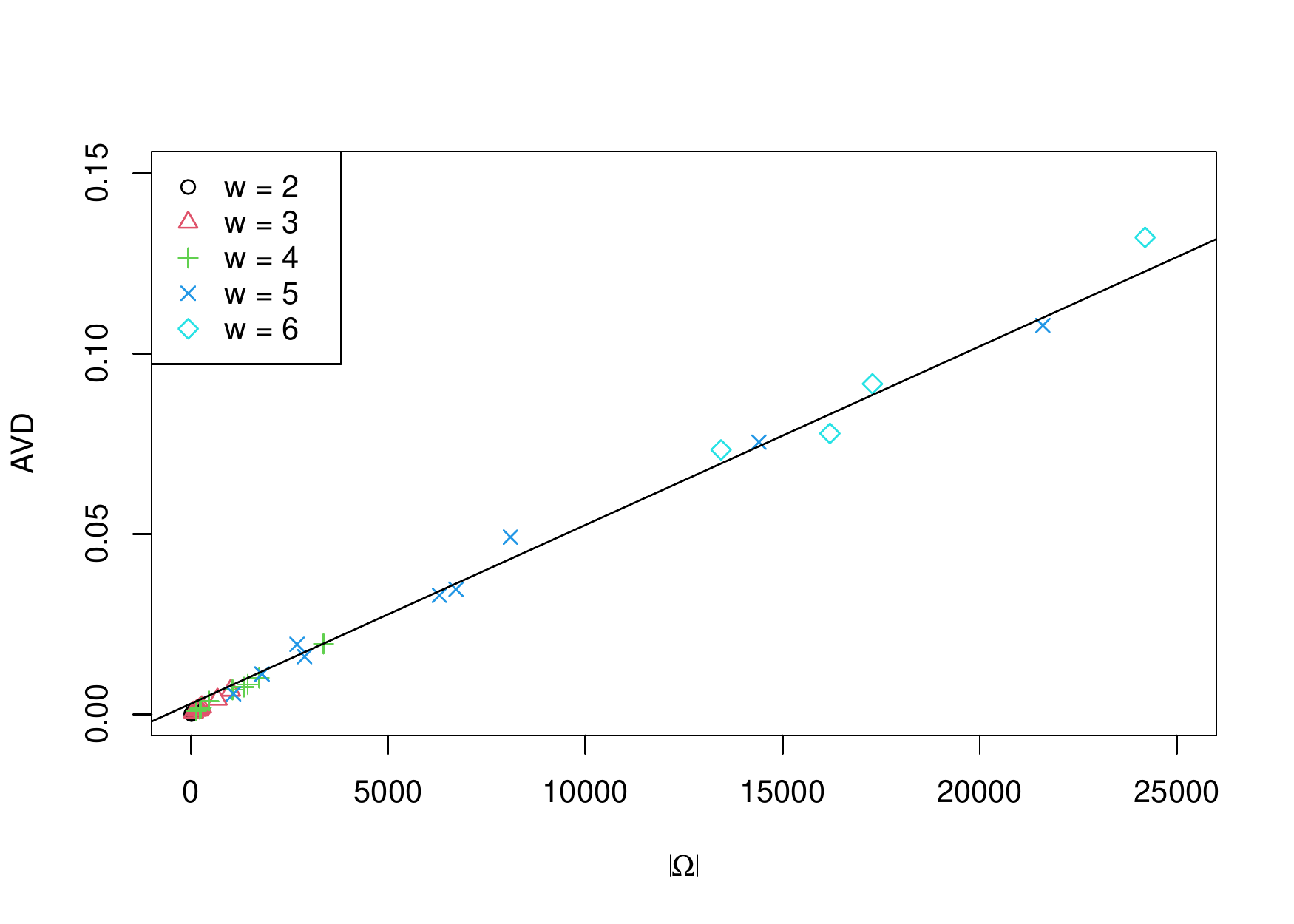}
    \caption{AVDs for {\sf RR-Ind-Joint}
      with respect to the domain size $|\Omega|$}\label{fig-omega}
\end{figure}

\begin{table}[tb]\centering
  \caption{AVDs of various schemes (the Adult dataset, 
    $n = 32,561, \epsilon = 4, |\Omega| = 16800$) } \label{tbl-comp}
    \begin{tabular}{l|c|c|c|c|c|c}
    \hline
    schemes $\backslash\, w$ & 2 & 3 & 4 & 5 & 6 & mean \\ \hline
    {\sf RR-Ind-Joint} & 0.0004 & 0.0023 & 0.0129 & 0.0635 & 0.3384 & 0.0835 \\ 
    {\sf \scriptsize RR-Independent} & 0.0588 & 0.0669 & 0.0395 & 0.0405 & 0.0215 & 	0.0455 \\ 
    {\em hybrid} & 0.0004 & 0.0023 & 0.0129 & 0.0405 & 0.0215 & 	0.0155 \\ 
    {\em truncated} & 0.0004 & 0.0019 & 0.0068 & 0.0182 & 0.0223 & 	0.0099 \\ 
    {\sf RR-Joint}~\cite{MDRR}
    	& 0.0001 & 0.0004 & ~ & ~ & ~ & 			0.0003 \\ 
    LoPub~\cite{LoPub} 
    	& 0.1262 & 0.2832 & 0.3560 & 0.3891 & 0.4576 & 	0.3224 \\ \hline
    \end{tabular}
\end{table}


The AVD increases exponentially as the dimensionality $w$ increases and
increases linearly with the full domain size
$|\Omega| = |\Omega_1 \times \cdots \times \Omega_w|$,
where $\Omega_i$ is the domain of $i$-th attribute.
The estimation error is related to the full domain size $|\Omega|$.
Fig.~\ref{fig-omega}
shows a scatter plot of AVD against the domain size $|\Omega|$.
It shows that
the domain size $|\Omega|$ varies with the dimensionality $w = 2,\ldots,5$
and that the AVD is linear with respect to $|\Omega|$. 
In the figure, the maximum domain size $|\Omega| = 16,800$
is given by the product of 
$|\Omega_{race}| = 5$,
$|\Omega_{education}| = 16$,
$|\Omega_{occupation}| = 15$,
$|\Omega_{marital-status}| = 7$, and
$|\Omega_{income50k}| = 2$.
Fitting the AVD to a linear function, we have
a simple prediction
\begin{equation}\label{eq.avd.omega}
\hat{\mbox{\em AVD}}(|\Omega|) = 2.944\cdot 10^{-3} + 4.953\cdot 10^{-6} |\Omega|.
\end{equation}
If we assume a maximum error as $\mbox{\em AVD}^* = 0.5$,
solving Eq.~(\ref{eq.avd.omega}) gives
the maximum domain size $|\Omega^*| = 100,365$.

\begin{figure*}[tb]\centering
  \begin{subfigure}[t]{\wdq \linewidth}\centering
    \includegraphics[width=\linewidth]{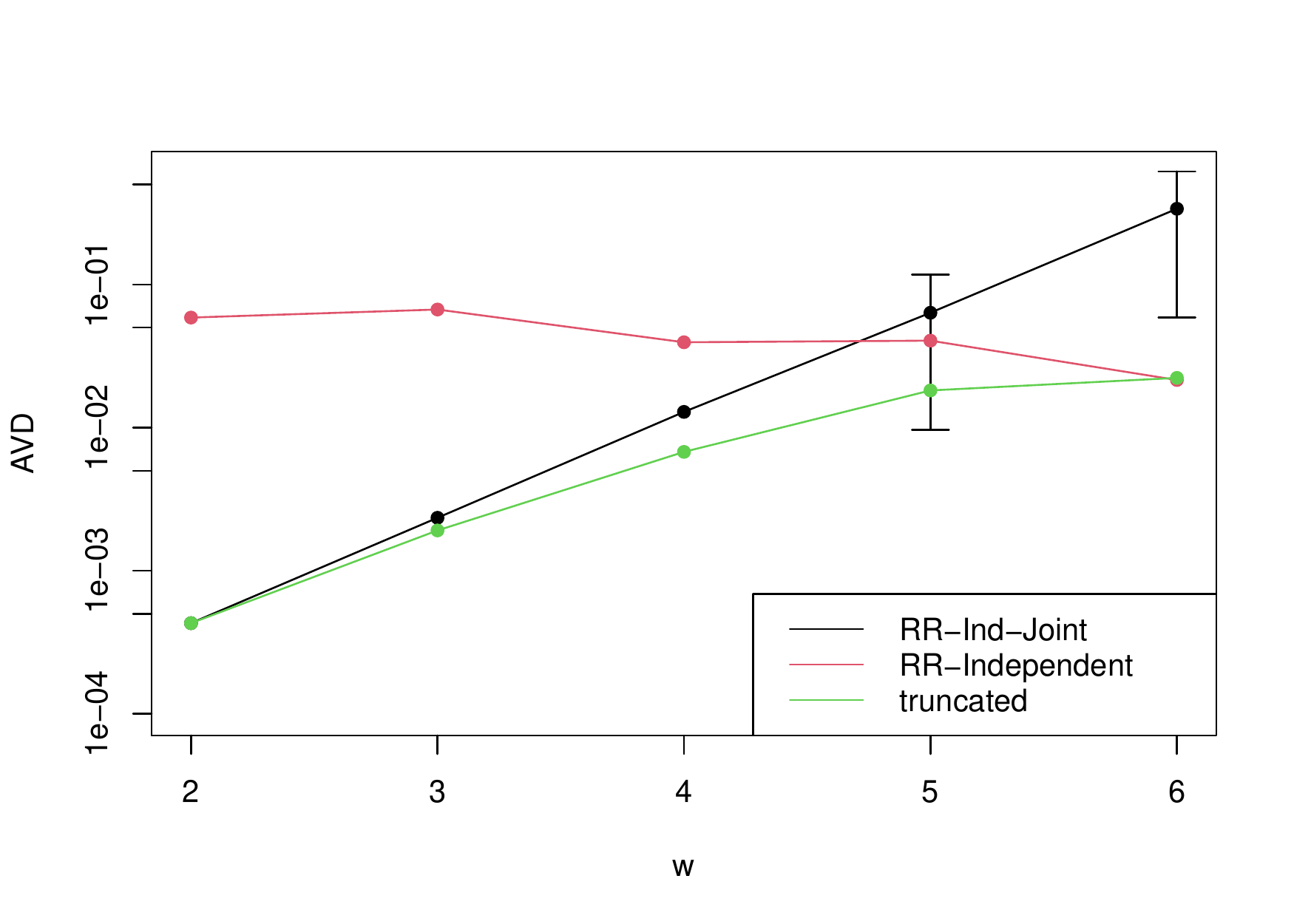}
    \caption{Adult}
  \end{subfigure}
  \hfill
  \begin{subfigure}[t]{\wdq \linewidth}\centering
    \includegraphics[width=\linewidth]{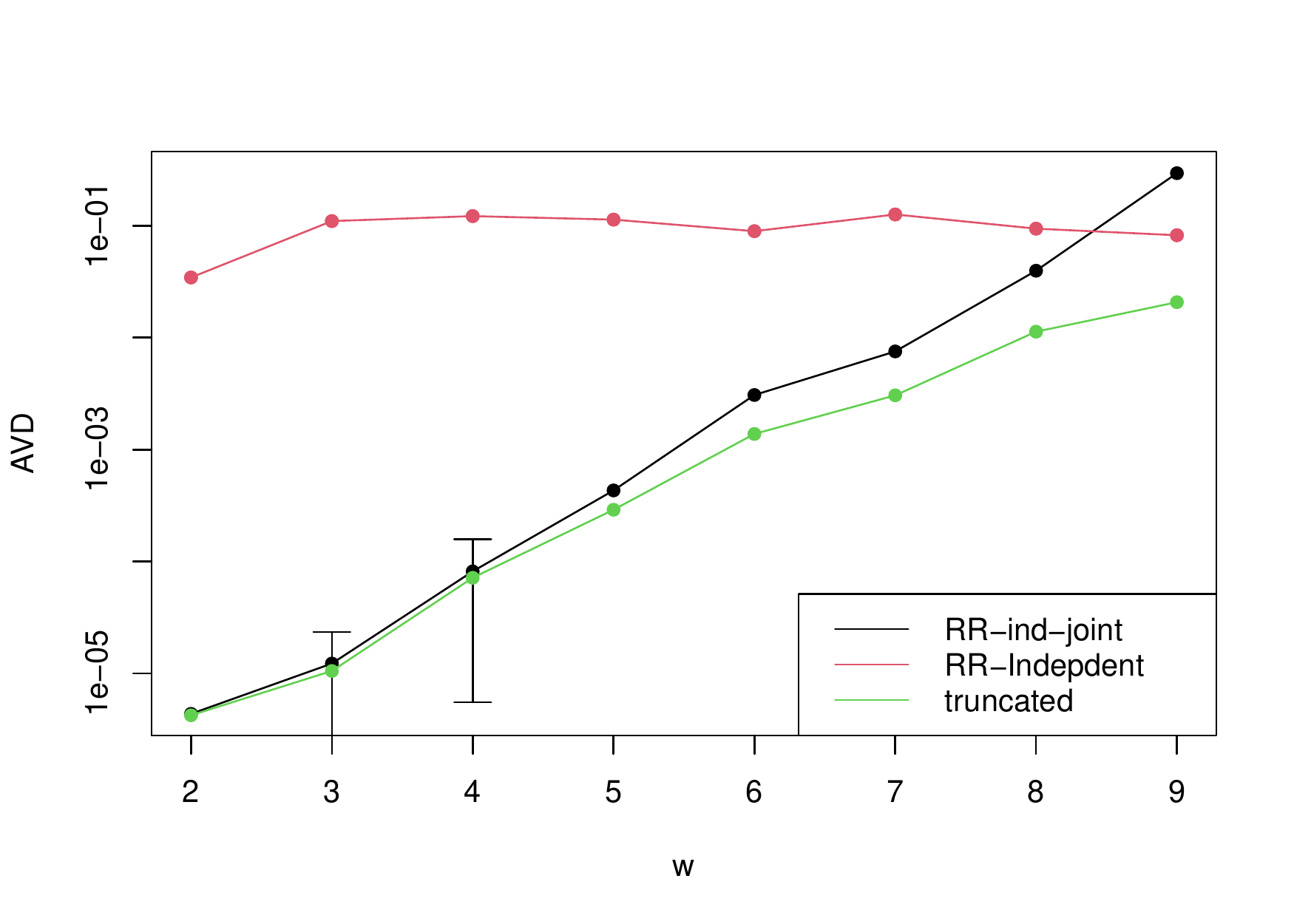}
    \caption{US Census}
  \end{subfigure}
  \hfill
  \begin{subfigure}[t]{\wdq \linewidth}\centering
    \includegraphics[width=\linewidth]{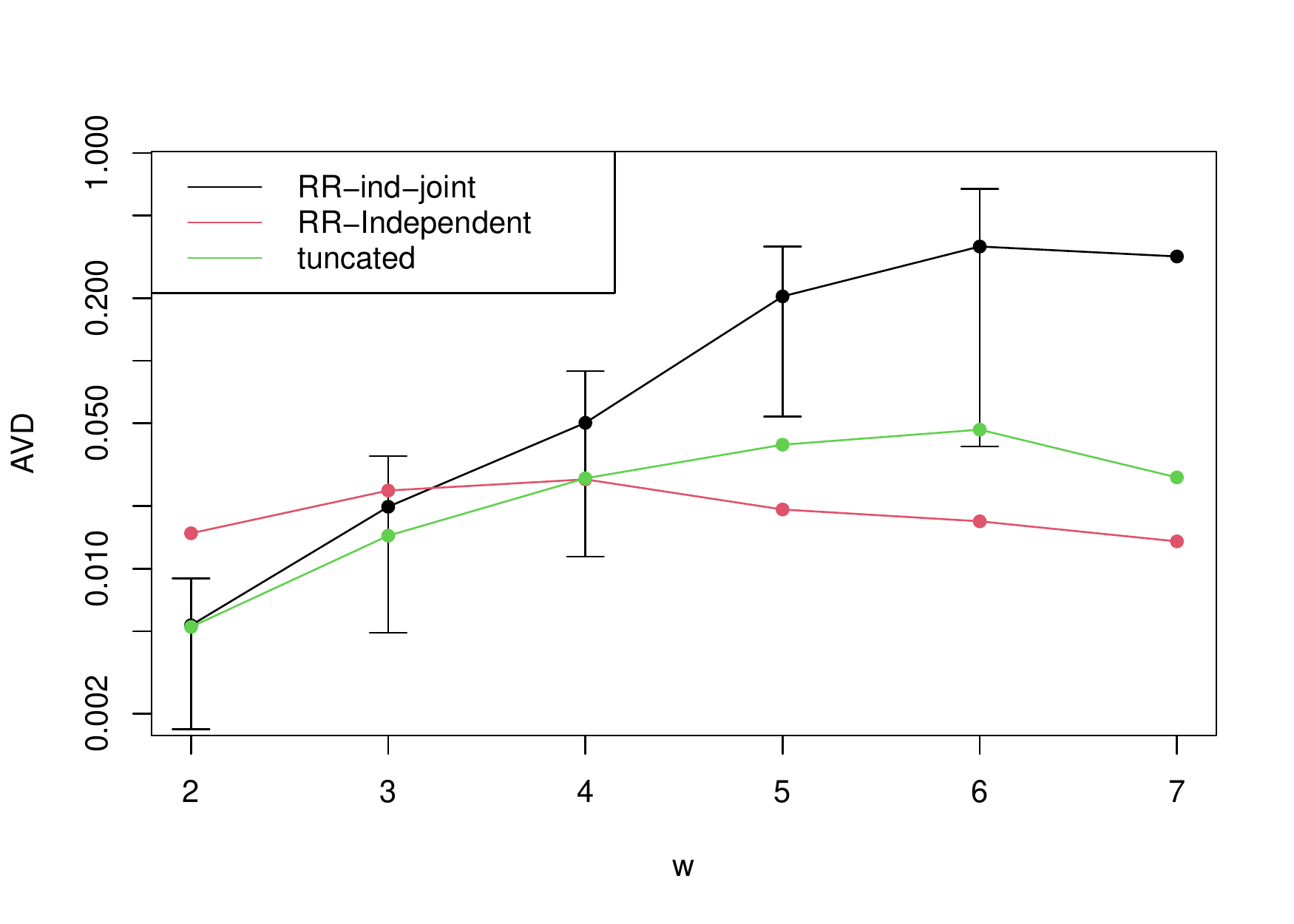}
    \caption{Credit}
  \end{subfigure}
  \hfill
  \begin{subfigure}[t]{\wdq \linewidth}\centering
    \includegraphics[width=\linewidth]{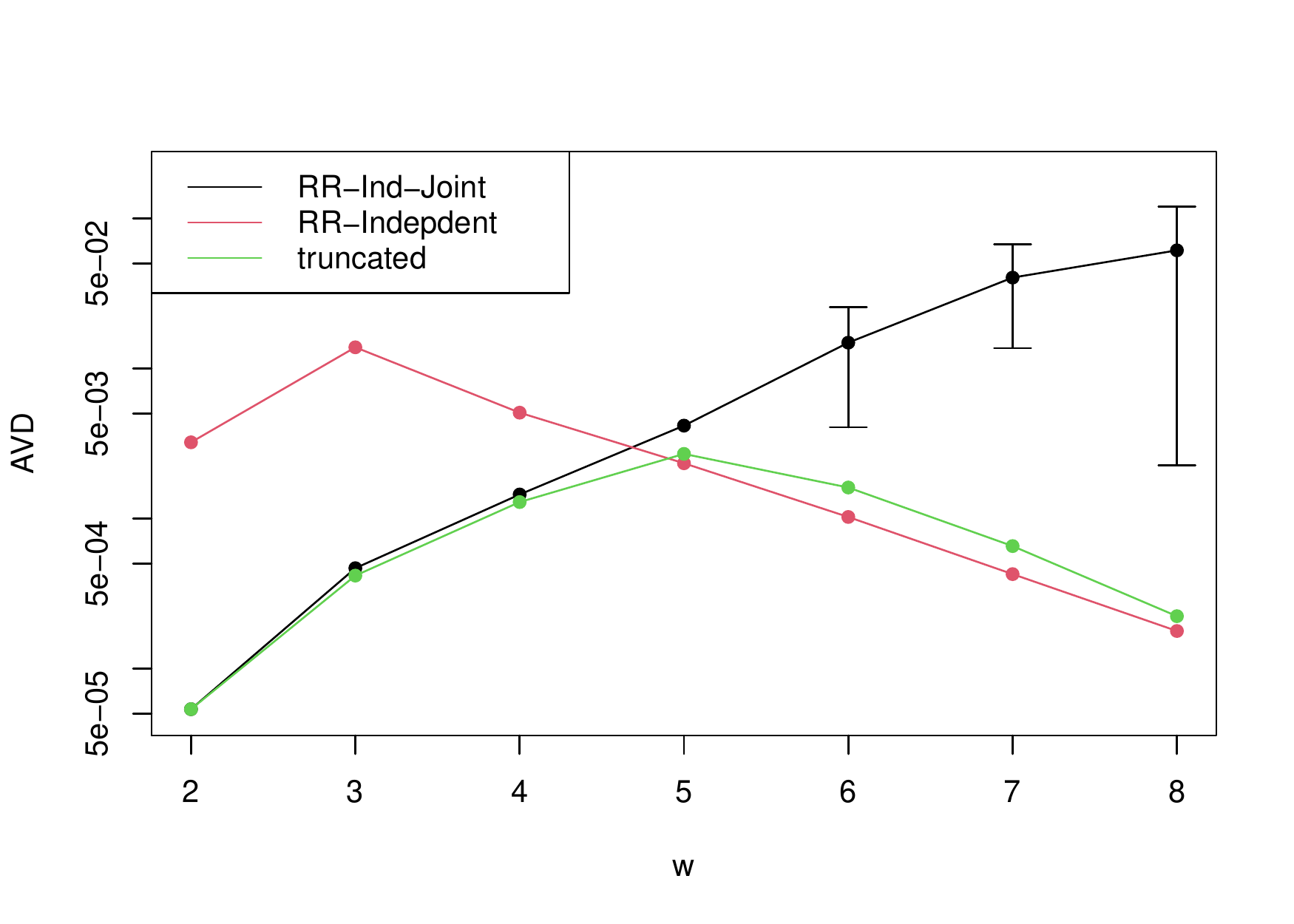}
    \caption{Nursery}
  \end{subfigure}
  \caption{AVDs between real and estimated values for $w$-way joint probabilities
    with respect to dimensionality $w$}\label{fig-adw}
\end{figure*}

Fig.~\ref{fig-adw} shows the AVDs of $w$-way joint probability distributions
estimated for {\sf RR-Ind-Joint}.
All datasets show similar behavior, in that the AVDs increase
exponentially with $w$, as $d^w$. 
This is consistent with the upper bound given by Eq.~(\ref{eq.avd}). 
The standard deviation, shown as the 68\% confidence interval of $\pm \sigma$,
also grows with $w$ (excessively long intervals that have negative values
are not shown).

In Fig.~\ref{fig-adw},
the AVDs of estimated by {\sf RR-Independent} are shown in red. 
The AVDs estimated by {\sf RR-Independent} increase with $w$ ($-p^w$) when $w$ is small 
and turns to be decreasing ($\min(p_i)$). 
Compared to the {\sf RR-Ind-Joint} case, the estimated joint probabilities
are distributed stably. 
We can therefore conclude that
{\sf RR-Ind-Joint} estimates joint probabilities more accurately
than {\sf RR-Independent} for small dimensionality $w$.
However, if $w$ is sufficiently large, {\sf RR-Independent} is
more accurate than {\sf RR-Ind-Joint}. 
There is always a crossover point $w^*$ within which
{\sf RR-Ind-Joint} estimates accurately for all datasets.  
For example, using the Adult dataset,
we would prefer {\sf RR-Ind-Joint} for estimating $w$-way joint probabilities
if $w \le 4$, with {\sf RR-Independent} being preferred otherwise.
The crossover points for the other datasets are
$w^* = 8, 3$, and $4$ for US census, Credit and Nursery, respectively.


\begin{figure*}[tb]\centering
  \begin{subfigure}[t]{\wdq \linewidth}\centering
    \includegraphics[width=\linewidth]{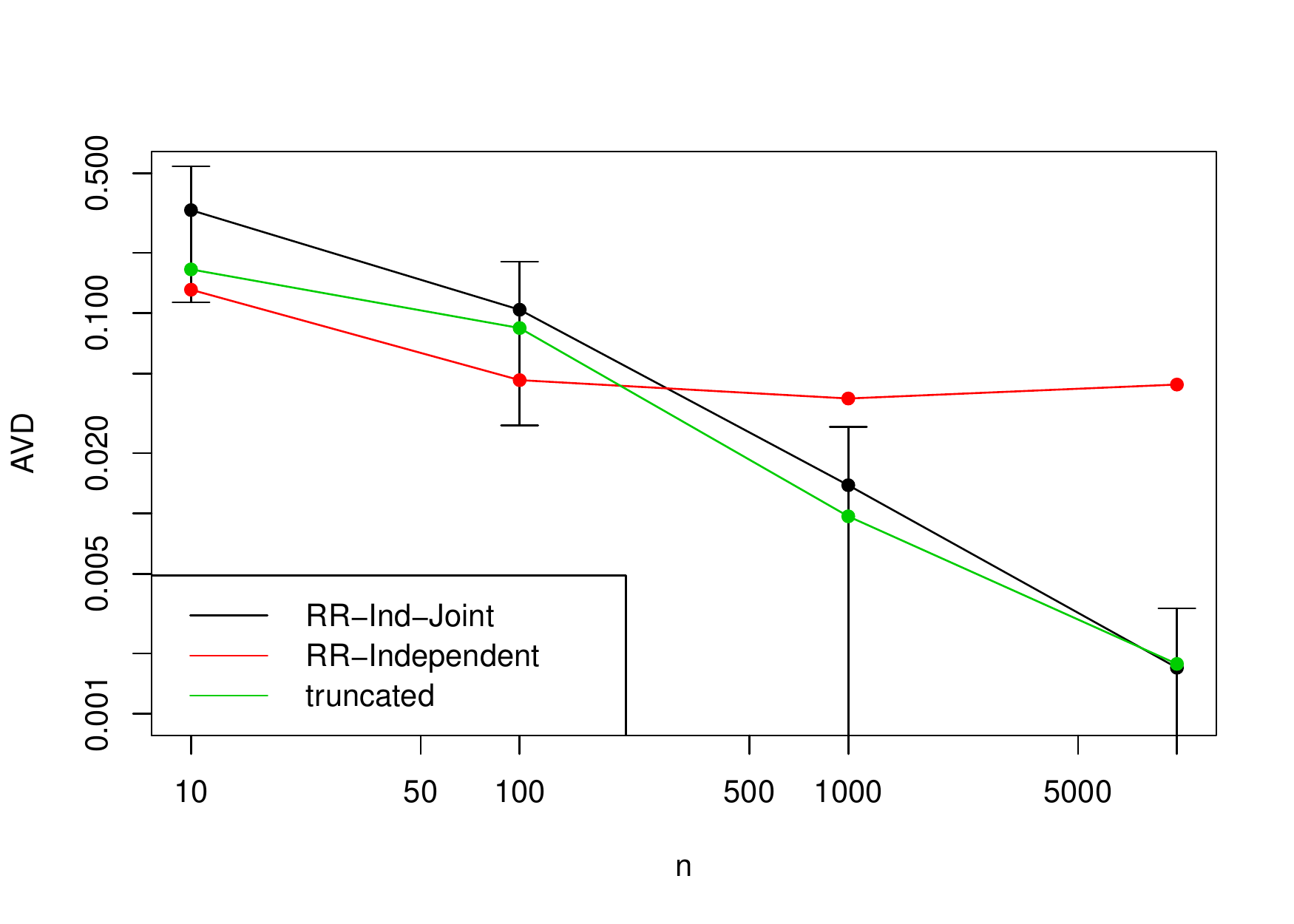}
    \caption{Adult}
  \end{subfigure}
  \hfill
  \begin{subfigure}[t]{\wdq \linewidth}\centering
    \includegraphics[width=\linewidth]{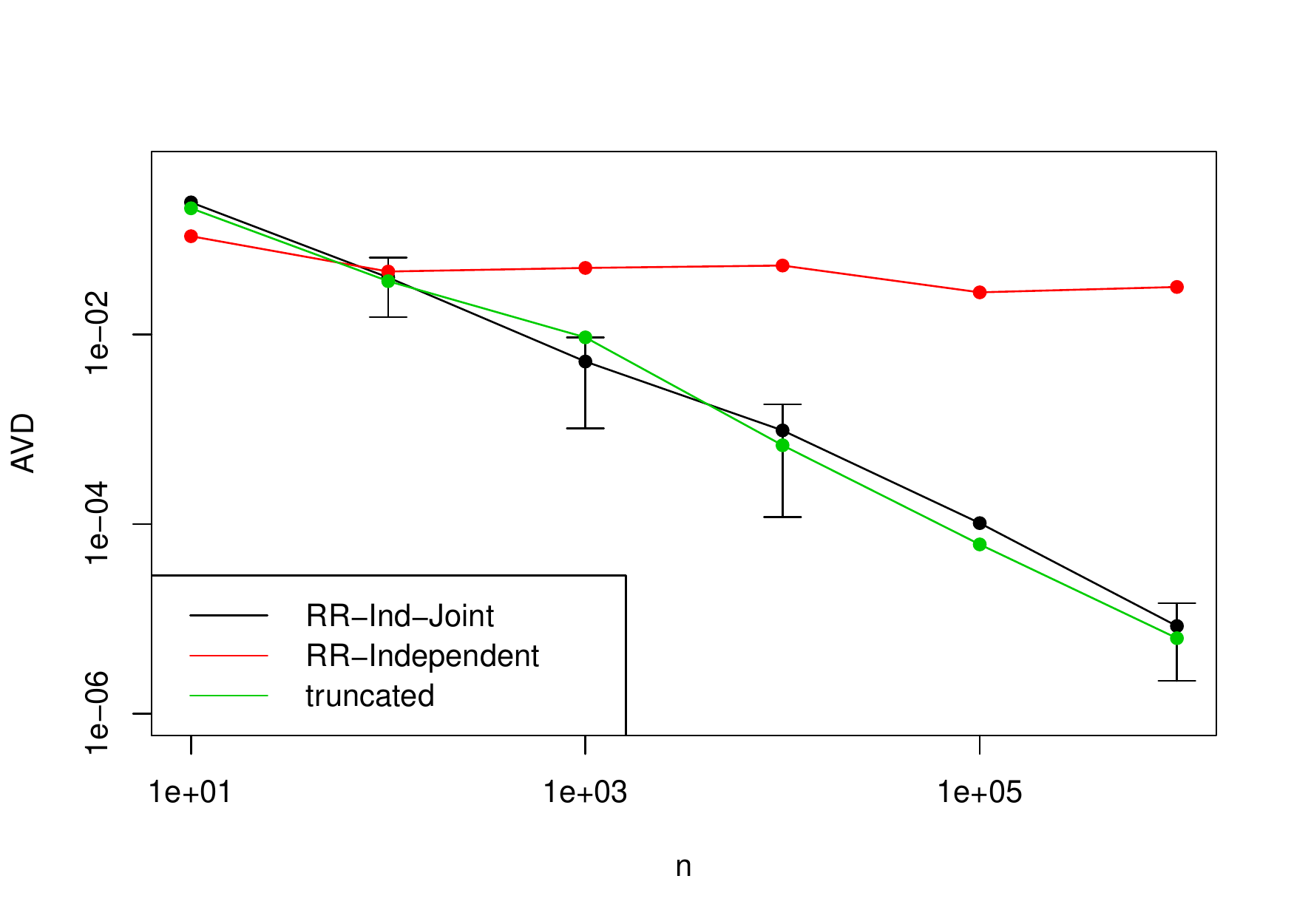}
    \caption{US Census}
  \end{subfigure}
  \hfill
  \begin{subfigure}[t]{\wdq \linewidth}\centering
    \includegraphics[width=\linewidth]{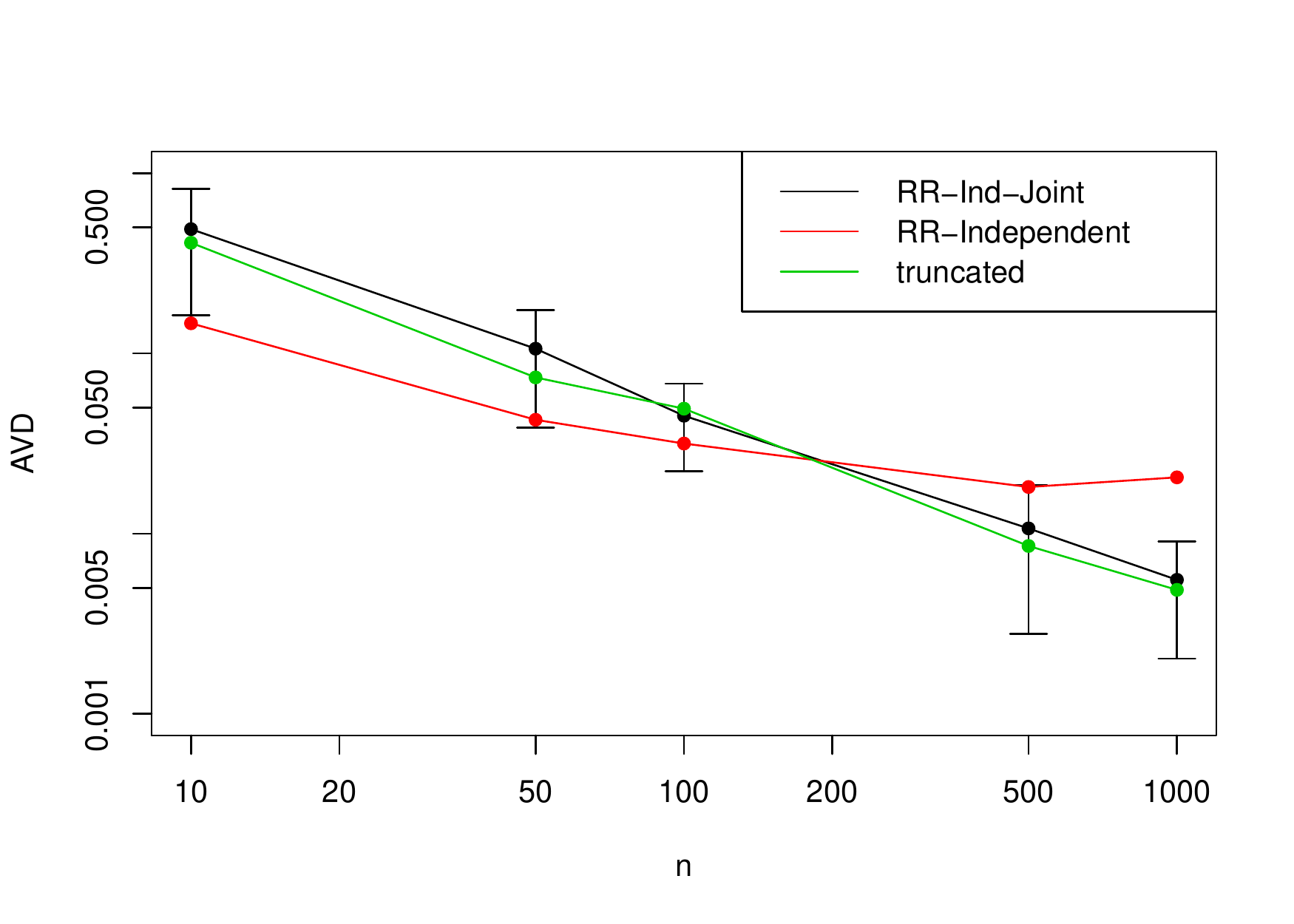}
    \caption{Credit}
  \end{subfigure}
  \hfill
  \begin{subfigure}[t]{\wdq \linewidth}\centering
    \includegraphics[width=\linewidth]{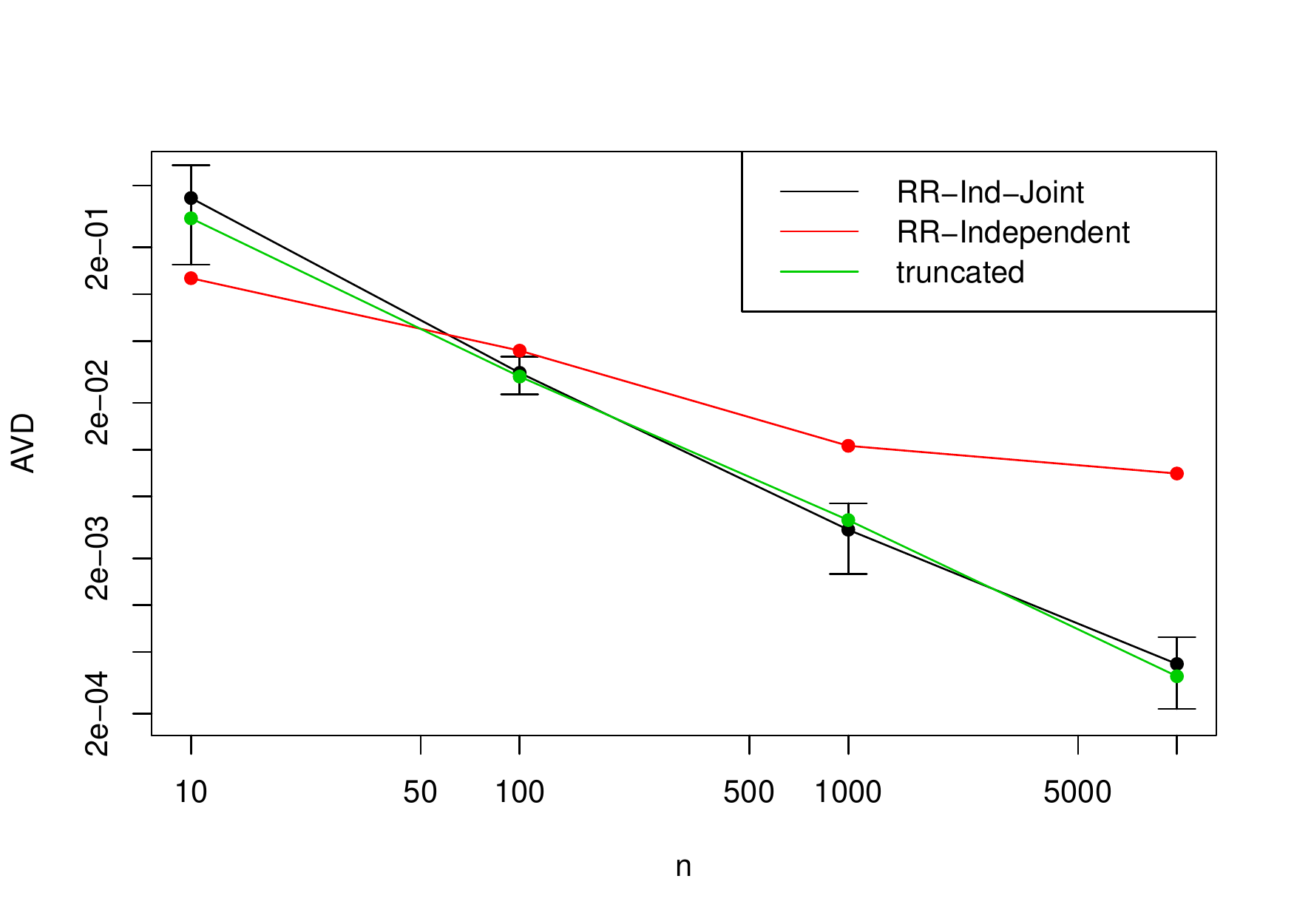}
    \caption{Nursery}
  \end{subfigure}
  \caption{AVD between real and the estimated $(w = 2)$ way joint probabilities
    with respect to the number of respondents $n$}\label{fig-adn}
\end{figure*}

The upper bound for the AVD using {\sf RR-Ind-Joint} in Eq.~(\ref{eq.avd}) 
indicates that 
the AVD and number of respondents $n$ are inversely proportional. 
Fig.~\ref{fig-adn} shows 
the distributions of AVDs with respect to the number of respondents (records) $n$.
We see that {\sf RR-Ind-Joint} estimates probabilities more accurately than
{\sf RR-Independent} when $n$ is large for all datasets. 
This result suggests that {\sf RR-Ind-Joint} is the most appropriate when $n > 500$, 
in most cases. 


\begin{figure*}[tb]\centering
  \begin{subfigure}[t]{\wdq \linewidth}\centering
    \includegraphics[width=\linewidth]{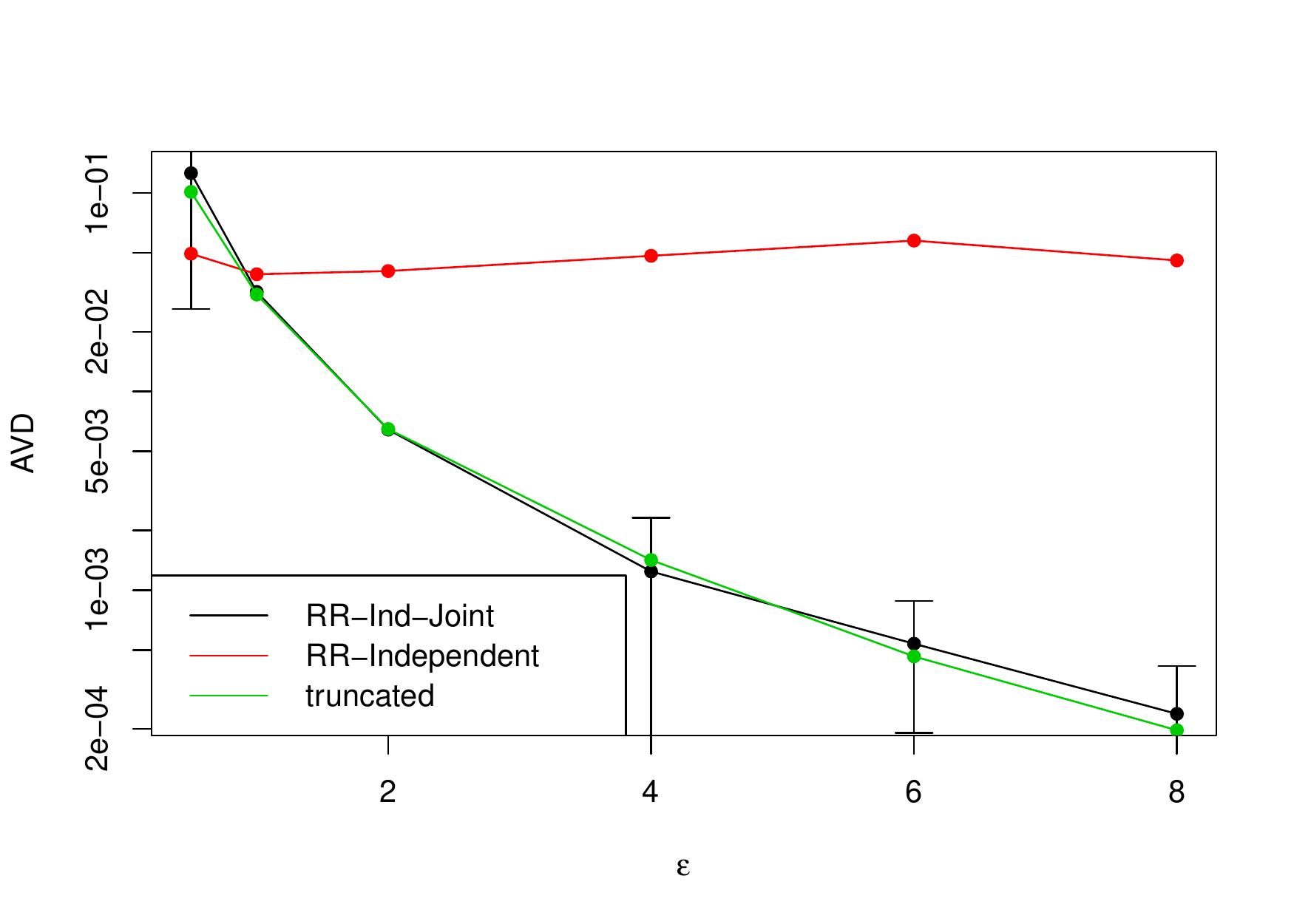}
    \caption{Adult}
  \end{subfigure}
  \hfill
  \begin{subfigure}[t]{\wdq \linewidth}\centering
    \includegraphics[width=\linewidth]{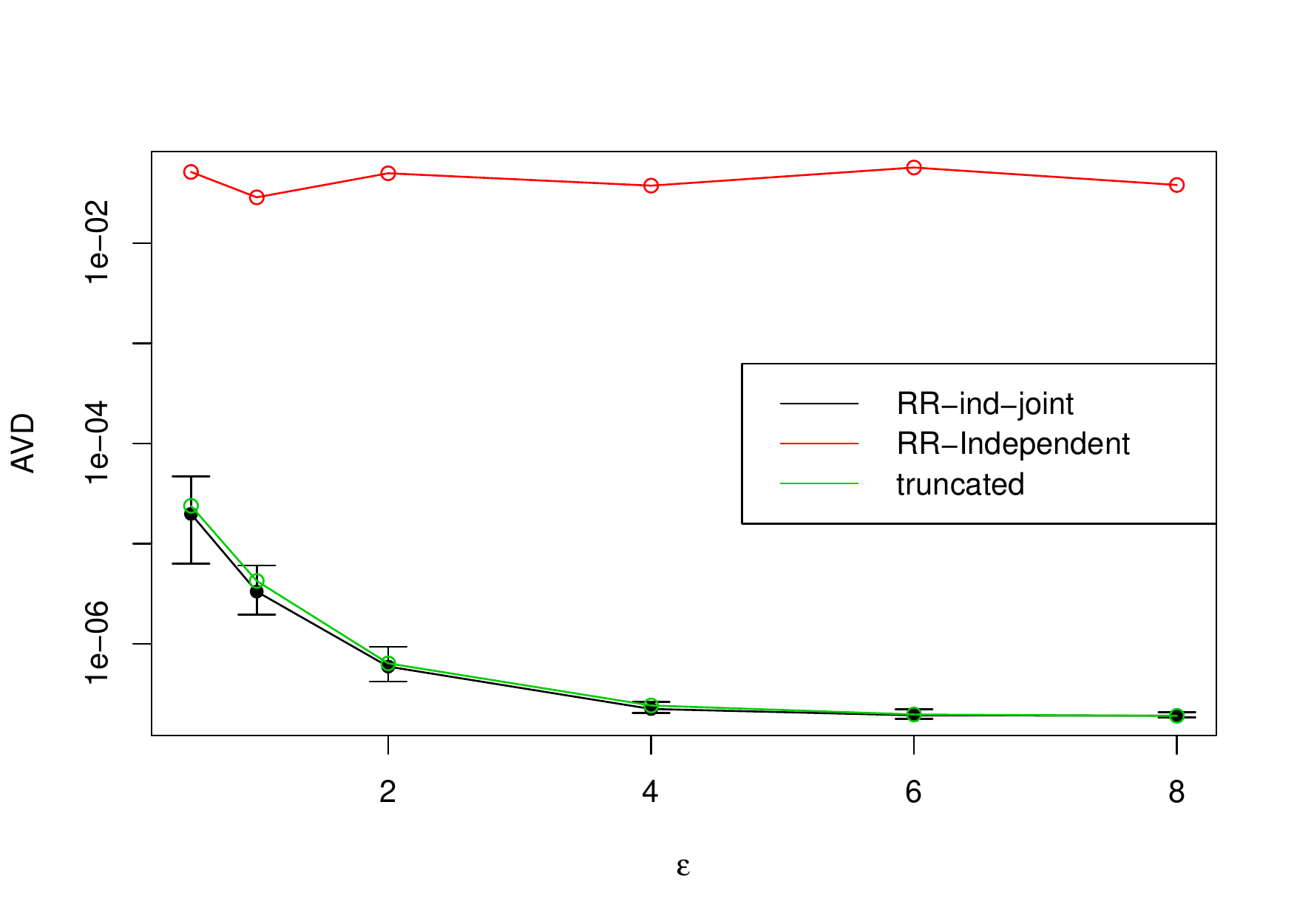}
    \caption{US Census}
  \end{subfigure}
  \hfill
  \begin{subfigure}[t]{\wdq \linewidth}\centering
    \includegraphics[width=\linewidth]{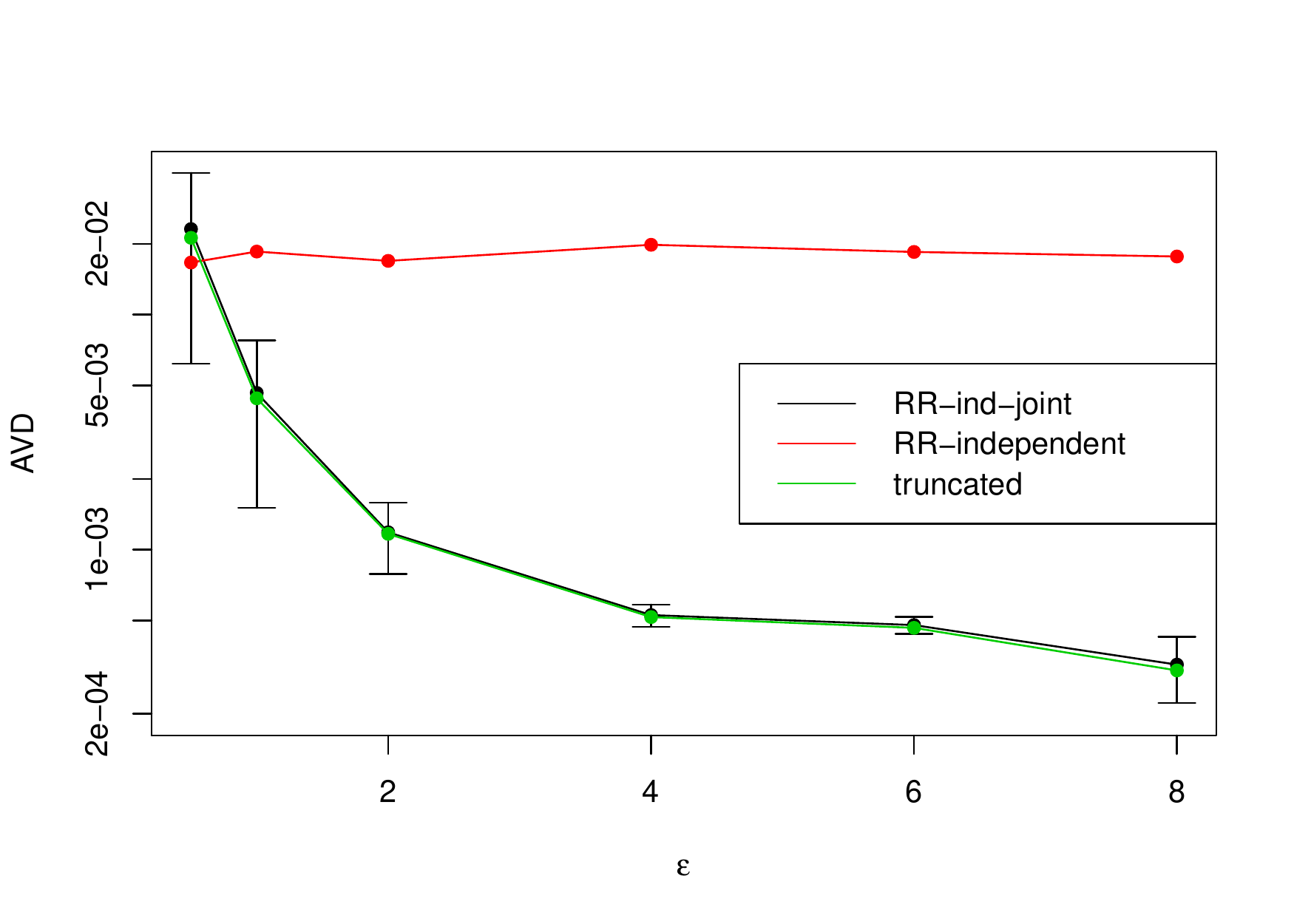}
    \caption{Credit}
  \end{subfigure}
  \hfill
  \begin{subfigure}[t]{\wdq \linewidth}\centering
    \includegraphics[width=\linewidth]{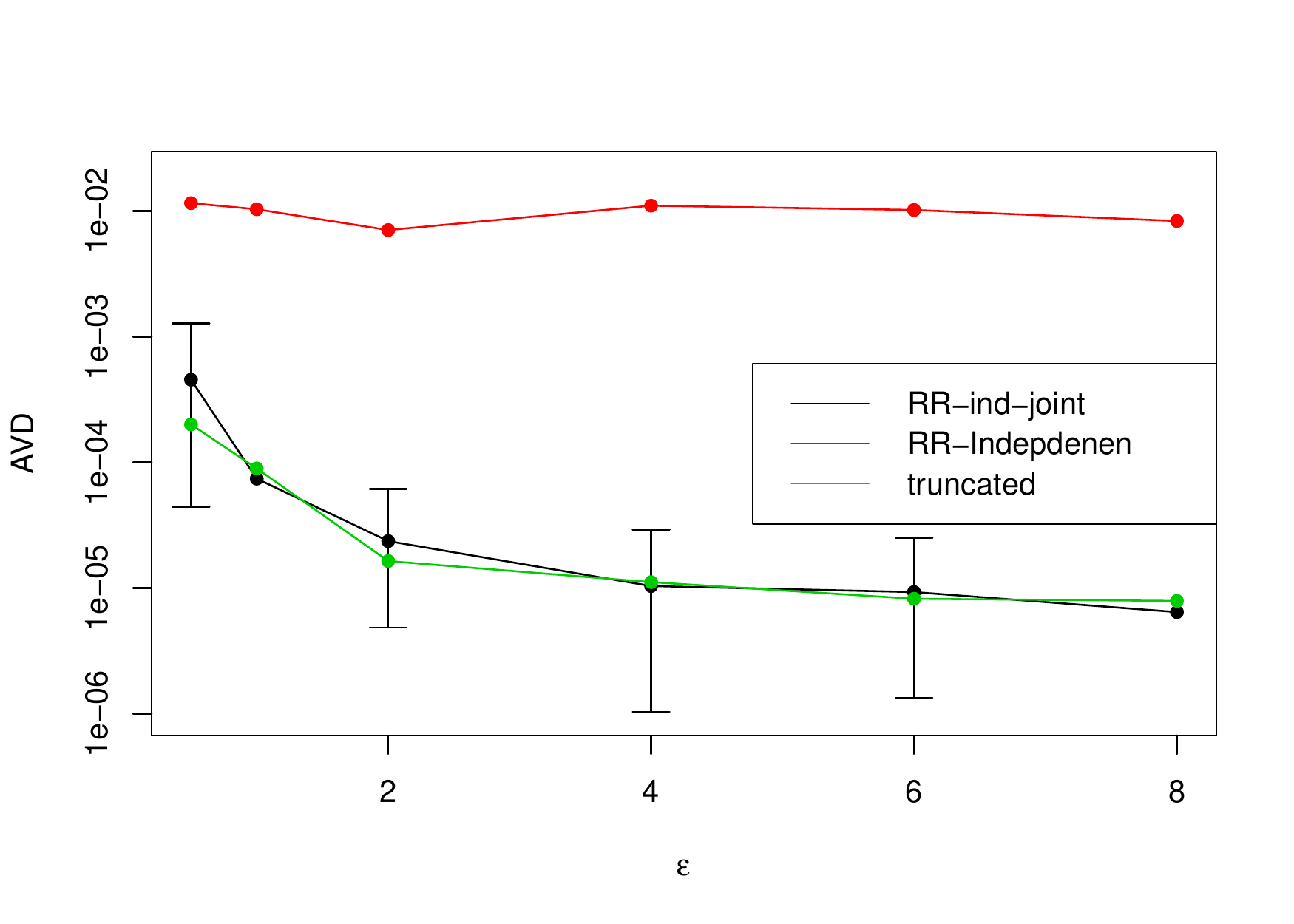}
    \caption{Nursery}
  \end{subfigure}
  \caption{AVDs between real and the estimated  $2$-way joint probabilities
    with regard to privacy budget $\epsilon$}\label{fig-ade}
\end{figure*}

Estimation error depends on privacy budget $\epsilon$.
The AVDs of {\sf RR-Ind-Joint} decrease as $\epsilon$ increases
because Eq.~(\ref{eq.avd}) converges to $d^w/n$ when $\epsilon$ becomes large. 
In contrast, the estimation error for {\sf RR-Independent} are
independent of $\epsilon$
because the error incurred by the independence of attributes dominates in this case.
Fig.~\ref{fig-ade} shows the AVDs of $2$-way joint probabilities 
using {\sf RR-Ind-Joint} and {\sf RR-Independent}.
The datasets were perturbed with privacy budget $\epsilon$ ranging from $0.5$ to $8$. 
Note that the AVDs for {\sf RR-Ind-Joint} decrease 
as the privacy budget $\epsilon$ increases, 
whereas the AVDs for {\sf RR-Independent} are stable. 
Except for cases where $\epsilon = 0.5$ for the Adult and Credit datasets,
{\sf RR-Ind-Joint} is indicated as the preferred option for estimation.

\subsection{Processing Time}

\begin{figure}[tb]\centering
  \includegraphics[width=\wdd\linewidth]{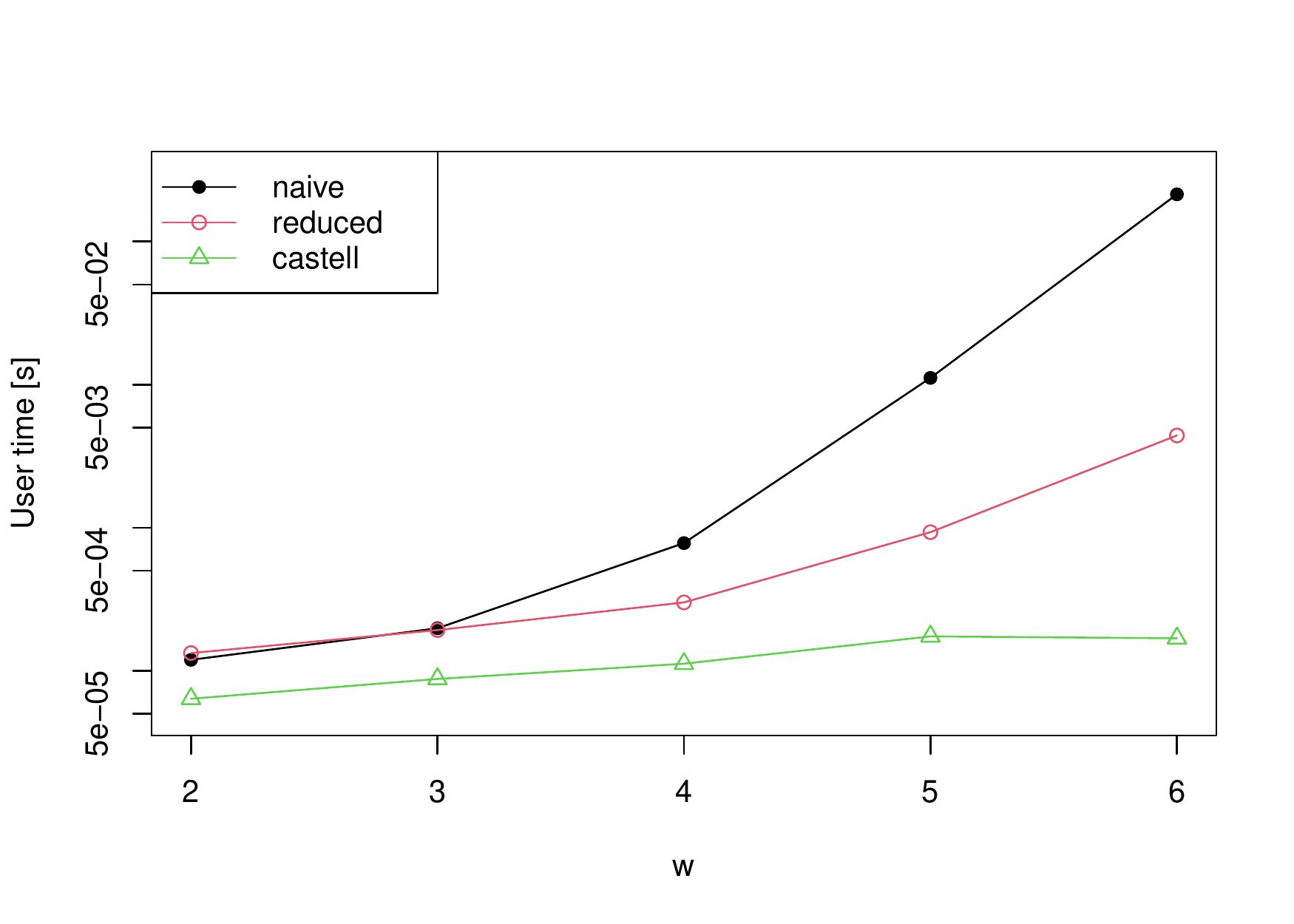}
  \caption{Comparison of processing time for estimations
    as a function of dimensionality $w$}\label{fig-time}
\end{figure}

{\sf RR-Ind-Joint} is scalable with respect to the dimensionality $w$ of
the joint probability estimation.
In Algorithm~\ref{alg.ijoint}, the inverses are computed
for each of $w$ randomization matrices having $(d_j,d_j)$ dimensions
for $j = 1,\ldots w$ rather than inverting a
$(d_1\times \cdots d_w, d_1\times \cdots d_w)$-dimension matrix,
which would require both a large computational capability and a large
amount of storage.
We show the reduction of computation cost in Fig.~\ref{fig-time},
as
the processing time for computing
the inversion of a $(3^w,3^w)$ matrix with $p = 0.5$
for $w = 2,\ldots, 6$ dimensionality.
The measurements were repeated 100 times 
in R version 4.0.0, running on a 2.3-GHz Intel Core i9, with 32 GB DDR4. 
The figure shows that the processing time for  $(P\otimes \cdots \otimes P)^{-1}$
computations
(labeled as ``na\"ive'') increases exponentially with $w$,
whereas the processing time for $(P^{-1}) \otimes \cdots \otimes (P^{-1})$
computations 
(labeled as ``reduced'') increases more slowly. 
The computation time at $w = 6$ is $0.007$ seconds.
The {\em castell} inversion algorithm not only reduces the storage
requirements for the matrix but also helps minimize the computation time.

\section{Discussion}
\subsection{Limitations}

Although our scheme scales up high-dimension data,
it still requires storage for the $d^w$ multi-dimensional contingency table
for empirical distribution.
The cross-tabulation for counting the combination of all attributes
is available for R ({\tt table}) and python (Pandas {\tt crosstab} method)
and is out of scope of this work.
But, it is used inside of Algorithm~\ref{alg.ijoint} (for empirical distribution)
and may have the limitation. 

The upper bounds of estimation loss 
 (Theorem~\ref{th.otimes} and \ref{th.rrind})
assume that $d = d_1 = \cdots = d_m$ (domain size) for simplicity.
In practice, it does not hold generally and hence the bounds are
loose when the variance of domain size is large
($d$ ranges from $2$ to $16$ and has median $6.5$ in Adult data).
This insufficient accuracy would incur the error of thresholds in {\em hybrid} scheme.
For example, Table.~\ref{tbl.crossover} provides the 
thresholds estimated by the upper bounds, suggesting 
{\sf RR-Ind-Joint} algorithm as preferable for $w < w^* = 1.374$.
However, according to the experimental results in Fig.~\ref{fig-adw}, 
{\sf RR-Ind-Joint} outperforms for $w < 5$. 
For an alternative estimation of thresholds, 
a sampling-based analysis should be considered. 

In the evaluation in Section~\ref{sec.eval}, we dropped numerical
attributes such as Age, Capital-gain/loss, Hour-per-week, and Country (42 values).
Although numerical attributes can be converted to categorical ones,
it is not trivial to identify the optimal number of bins. 
An automated and adaptive algorithm for the optimal granularity for
conversion to categorical attributes is one of the future works. 



\subsection{Extensions}

The accurate joint probability estimation could follow
an accurate synthetic data.
For example,
several synthetic algorithms have been proposed in
LoPub~\cite{LoPub}, LoCop~\cite{LoCop} and WAE~\cite{WAE}. 
LoPub performs random sampling of clusters of attributes
and assigns values according to the estimated
conditional distributions. 
LoCop uses the inverse cumulative distribution 
function for the multivariate Gaussian copula. 
The WAE generates a random vector from
Gaussian distribution at the latent layer and
feeds them into the decoder of the autoencoder. 
We will explore the best synthetic algorithms
and evaluate the accuracy for major machine
learning algorithms 
as one of the future studies. 
  
{\sf RR-Ind-Joint} is very accurate for low dimensionality.
Hence, privacy-preserving key-value data is
one of its potential applications.
With an appropriate conversion of numerical values
to categorical date, we can apply RR to key-value
data with LDP guarantee and estimate accurate
the joint probability distribution that reveals
the correlations between keys and values. 



\section{Related Works}

\subsection{Differential Private Data Publishing}

DP~\cite{Dwork2014}
has been used for privacy protection in data publishing. 
LDP~\cite{LDP}
was proposed to
eliminate the assumption of trust in a central server
and applied to many use cases
including
crowdsourcing participants~\cite{Duchi-SFCS2013}, and
heavy hitter detections~\cite{Bassily-STOC2015}, \cite{Qin-CCS2016}. 
There were significant studies 
for useful building blocks and properties;
a compositional theorem~\cite{Kairouz2015}, 
an optimized local hashing (OLE)~\cite{Wang-Usenix2017}, 
a sampling-based approach~\cite{Zhang2018}, 
post-processing for improving utility~\cite{Wang-NDSS2020}
and 
on the optimal data-independent noise distribution~\cite{Jordi-IS2013}. 

For the works related to our goal, 
the attempts for multi-dimensional data publishing
with DP or LDP are classified into
some categories; 
the (Laplacian or Gaussian) Noise-based:
\cite{Ding-SIGMOD2011},
\cite{Qardaji2013}, \cite{Qardaji2014},
\cite{Zhang-PrivBayes17},
\cite{Chen-KDD2015},
\cite{Day-AsiaCCS2015},
\cite{Zhang-CCS2018},
\cite{Xu-IFS2017}, 
the RR-based:
\cite{Fanti-PETS2016},
\cite{Wang-EDBT2016},
\cite{MDRR},\cite{LoCop},\cite{WAE}, 
the  Key-Value based (2-dimensional data) schemes:
\cite{Harmony},
\cite{PrivKV},
\cite{Fang2020}.

\subsection{Multi-Dimensional LDP schemes}
\subsubsection{Nose-based schemes}

The first attempt to add Laplace  noise to high-dimensional data
was done by Ding et al.~\cite{Ding-SIGMOD2011}.
They injected DP noise into an initial subset of cuboids
and then compute the remaining cuboids from the initial subset. 
To improve accuracy of $w$-way marginal estimation, 
several studies have been done. 
{\em PriView} due to Qaraji et al.~\cite{Qardaji2013}, \cite{Qardaji2014} uses an entropy maximization.
{\em PrivBayes} proposed by Zhang et al.~\cite{Zhang-PrivBayes17} uses a Bayesian network
to iteratively learn a set of low-dimensional conditional probabilities.
Chen et al.~\cite{Chen-KDD2015}
uses a junction tree algorithm to find the optimal mechanism
based on sampling-based testing to explore pairwise dependencies of attributes.
DPSense proposed by Day and Li~\cite{Day-AsiaCCS2015} controls sensitivity 
with a threshold of counts and proves the optimization of the thresholds. 
CALM proposed by Zhang et al.~\cite{Zhang-CCS2018}
partitions the set of users into some groups and assigns them to one group,
and then aggregates to obtain a noisy marginal table and
performs reconstruction steps. 
DPPro studied by Xu et al.~\cite{Xu-IFS2017} uses a random projection
to maximize utility and to preserve pairwise distances between attributes.
They add Gaussian noise to intermediate vectors to maximize the utility
and proves DP. 
Arcolezi et al.~\cite{Arcolezi-CIKM2021} proposed sampling techniques
for saving privacy budget and shows the 9-way MSE of the Adult datasets. 
Cormode et al.~\cite{Cormode-SIGMOD2018} provided the utility guarantee
and evaluated the estimate using open-source trajectory datasets. 

Most of these studies aimed to satisfy DP rather than LDP
and focused on the optimality of subsets of attributes to minimize
estimation error.
Hence, no sufficient evaluation of estimation accuracy with respect to $w$
were made. It is hard to compare our work for scalability. 

\subsubsection{RR-based schemes}

RR~\cite{RR2} based multi-dimensional studies were inspired after 
RAPPOR~\cite{RAPPOR} successfully encoded data as a Bloom Filter
and then performs RR of each bit of filter. 
Soon after RAPPOR,
Fanti et al.~\cite{Fanti-PETS2016} proposed 2-dimensional joint
probabilities using the Expectation Maximization (EM) algorithm.
However, it incurs a considerable computational cost for higher
dimension data. 
Wang et al.~\cite{Wang-EDBT2016} theoretically derive
a mathematical model of the mean squared error of RR and Laplace
mechanism and show frequency estimation.

Recently, 
some advanced works~\cite{MDRR}, \cite{LoPub}, \cite{LoCop}, and \cite{WAE}
using RR for high-dimensional data were made. 
Domingo-Ferrer and Soria-Comas proposed some RR-based schemes
toward the dimensionality issues. Their study, reviewed
detailed in Section~\ref{sec.mdrr}, 
 provides insights but has some drawbacks that motivate this work.
Ren et al.~\cite{LoPub} proposed an LDP scheme called LoPub,
estimating multi-dimensional joint probability distributions.
They perturb a multi-dimensional data encoded binary vectors
using Bloom filter and combine a Lasso regression with an EM
to estimate the joint probabilities accurately. 
They also show a method for synthetic data that preserves the utility of
the original data in the sense that classification accuracies
for some machine learning algorithms are preserved as the original.
LoCop~\cite{LoCop} improves LoPub by introducing multivariate
Gaussian copula to estimate cross-attribute dependencies.
To improve the accuracy of LoPub and LoCop, Jiang et al. \cite{WAE}
combines the Wasserstein Autoencoder (WAE) and the federated
learning to propose DP-FED-WAE framework. With an LDP algorithm
called SignDS for saving privacy budget, 
they reported that DP-FED-WAE outperforms LoPub and LoCop. 
They show the comparison of joint probability estimation
accuracy in terms of $w$-way joint probabilities.
We showed that {\sf RR-Ind-Joint} outperforms the state-of-art schemes
in Fig.~\ref{fig-lopub} and Table~\ref{tbl-comp}. 

\subsubsection{Key-Value schemes}

Key-value data has been used for several services and
its privacy-preserving has a high demand.
As for dimensionality, the dimension is fixed as $w = 2$, but
some studies deal with dependency between key and value similar
to our study. 

Nguyen et al.~\cite{Harmony} proposed {\em Harmony} in which
for any numerical data is encoded as binary data
according to the value. 
Using {\em Harmony} as building block, 
Ye et al.~\cite{PrivKV} proposed an LDP protocol designed for key-value data.
Their scheme 
perturbs the key jointly with the encoded value 
using a variation of RR. 
The associations between key and value are preserved from
the randomized pairs with the privacy of input being guaranteed in a specified privacy budget.
Note that PrivKV randomizes key and value jointly with probability depending on
value. 
Fang et al.~\cite{Fang2020} study the local model poisoning attack to
LDP schemes. Under assumption that attacker manipulates the value on the compromised device,
they report some defense techniques have limited success in open data experiments.

\section{Conclusion}\label{sec.conc}

In this paper, we have studied the randomization of a multi-dimensional
data, where 
independent randomization of attributes
would seem to address the curse of dimensionality.
However, a na\"ive approach to independent randomization can suffer from
three main drawbacks. 
First, 
the combination of domains grows exponentially. 
Second, the approach masks any association among attributes.
Third, there can be too few records to cover the combined domain
(the domain sparsity). 

Our proposed multi-dimensional RR scheme {\sf RR-Ind-Joint}
randomizes all attributes {\em independently},
and estimates {\em jointly} the multi-dimensional 
joint probability distributions, while addressing these issues. 

Using an accumulated arbitrary number of independent randomization matrices,
we can estimate the joint probability with high accuracy. 
We have proposed a {\em castell} algorithm, which 
inverts its aggregated randomization matrix efficiently, and  
requires only lightweight computation costs (linear with respect to dimensionality $w$)
and manageable storage costs ($d^w$, which is the same as for the 
cross-tabulation matrix). 
We develop {\em upper bounds of the estimation errors} for
two primitive RR schemes ({\sf RR-Ind-Joint} and {\sf RR-Independent})
and propose a hybrid RR scheme that
switches efficiently between them,
depending on the values of the relevant parameters 
(dimensionality, the privacy budget, and the number of respondents). 
Our experimental results using open-source datasets
show that the proposed scheme can deal
with a wide range of datasets 
and can estimate joint probabilities to practical levels of accuracy.

We plan to tighten the upper bounds for the estimation
errors. The estimated crossover points given in Table~\ref{tbl.crossover}
are too large to serve as practical threshold values, such as
were observed in Figs.~\ref{fig-adw}, \ref{fig-adn}, and \ref{fig-ade}. 
In our current experiments, we ignore some continuous attributes that
should be studied in the future. 

\subsection*{Acknowledgment}

\bibliography{bib}

\begin{thebibliography}{9}

\bibitem{jama2020} Pedro F. Saint-Maurice, et al.
``Association of Daily Step Count and Step Intensity With Mortality Among US Adults,''
 JAMA, 323(12) pp.1151-1160, 2020.

\bibitem{Shen2017} H. Shen, M. Zhang, and J. Shen, 
  ``Efficient privacy-preserving cubedata aggregation scheme for smart grids,''
  IEEE Trans. Inf. Forensics and Security, vol. 12, no. 6, pp. 1369-1381, 2017.

\bibitem{MDRR} J. Domingo-Ferrer and J. Soria-Comas, 
  ``Multi-Dimensional Randomized Response,'' 
  in IEEE Transactions on Knowledge and Data Engineering, 2022, 
  \url{doi: 10.1109/TKDE.2020.3045759}.

\bibitem{LoPub} X. Ren et al., ``$\textsf{LoPub}$ : 
  High-Dimensional Crowdsourced Data Publication With Local Differential Privacy,''
  in IEEE Transactions on Information Forensics and Security, 
  vol. 13, no. 9, pp. 2151-2166, Sept. 2018, \url{doi: 10.1109/TIFS.2018.2812146}.


\bibitem{LoCop}
  Teng Wang, Xinyu Yang, Xuebin Ren, Wei Yu, and Shusen Yang,
  ``Locally private high-dimensional crowdsourced data
  release based on copula functions,'' 
  {\em IEEE Transactions on Services Computing}, pp. 1-1, 2019.

\bibitem{WAE} Xue Jiang, Xuebing Zhou,  Jens Grossklags, 
  ``Privacy-Preserving High-dimensional Data Collection 
  with Federated Generative Autoencoder'',
  Proceedings on Privacy Enhancing Technologies,
  pp. 481-500, 2022. \url{DOI: 10.2478/popets-2022-0024}

\bibitem{WAE2018}
  Ilya Tolstikhin, Olivier Bousquet, Sylvain Gelly, and Bernhard
  Scholkopf,
  ``Wasserstein auto-encoders,'' In International
  Conference on Learning Representations (ICLR 2018), Vancouver,
  BC, Canada, 2018. 

\bibitem{cramerV} H. Cram\'er, {\it Mathematical Methods of Statistics},
  Princeton University Press, 1946. 

\bibitem{PrivKV}Q. Ye, H. Hu, X. Meng, H. Zheng, ``PrivKV : Key-Value Data Collection with Local Differential Privacy'', IEEE S\&P, pp. 294-308, 2019. 

\bibitem{McSherry}C. Dwark, F. McSherry, K. Nissim, A. Smith, ``Calibrating noise to sensitivity in private data analysis,'' TCC, Vol. 3876, p. 265-284, 2006. 

\bibitem{LDP}J. C. Duchi, M. I. Jordan, M. J. Wainwright, ``Local privacy and statistical minimax rates,'' FOCS, pp. 429-438, 2013. 

\bibitem{RR1}P. Kairouz, S. Oh, and P. Viswanat, ``Extremal mechanisms for local differential privacy'', NIPS, pp. 2879-2887, 2014. 

\bibitem{RR2}S. L. Warner, ``Randomized response: A survey technique for eliminating evasive answer bias'', Journal of the American Statistical Association, pp. 63-69, 1965. 

\bibitem{Harmony}T. T. Nguy\^{e}n, X. Xiao, Y. Yang, S. C. Hui, H. Shin, J. Shin, ``Collecting and analyzing data from smart device users with local differential privacy'', \url{arXiv:1606.05053}, 2016. 

\bibitem{McScherry} F. McSherry, ``Privacy integrated queries: An extensible platform for privacy-preserving data analysis'', SIGMOD, pp. 19-30, 2009. 

\bibitem{RAPPOR} \'{U}lfar Erlingsson, Vasyl Pihur, Aleksandra Korolova, ``RAPPOR: Randomized Aggregatable Privacy-Preserving Ordinal Response'', ACM Conference on Computer and Communications Security, pp.1054-1067, 2014.

\bibitem{MS} ``Learning with Privacy at Scale'' \url{https://machinelearning.apple.com/2017/12/06/learning-with-privacy-at-scale.html} (accessed on 2019).

\bibitem{Kairouz2015} Kairouz, P., Oh, S.,  Viswanath, P.,
  ``The Composition Theorem for Differential Privacy'' 
  Proceedings of the 32nd International Conference on Machine Learning, 37, pp. 1376-1385, 2015. 

\bibitem{Dwork2014} C. Dwork and A. Roth, 
  ``The algorithmic foundations of differential privacy'', 
  Found. Trends Theor. Comput. Sci. 9, 3-4, 211-407, 2014.

\bibitem{Strassen} V. Strassen, ``Gaussian elimination is not optimal'',
  {\it Numerische Mathematik}, 13(4), pp. 354-356, 1969. 

\bibitem{Adult}
  K. Bache and M. Lichman, UCI Machine Learning Repository, 2013.
  \url{https://archive.ics.uci.edu/ml/datasets/adult}
\bibitem{Uscensus}
  Meek, Thiesson, and Heckerman, 
  ``The Learning Curve Method Applied to Clustering'',
  The Journal of Machine Learning Research, 2011. 
  \url{https://archive.ics.uci.edu/ml/datasets/US+Census+Data+(1990)}
\bibitem{Credit}
  Hans Hofmann, 
  ``Statlog (German Credit Data) Data Set'', 
  \url{https://archive.ics.uci.edu/ml/datasets/Statlog+%28German+Credit+Data%29}
\bibitem{Nursery}
  \url{https://archive.ics.uci.edu/ml/datasets/Nursery}

\bibitem{Yeom2018}  S. Yeom, I. Giacomelli, M. Fredrikson and S. Jha, 
  ``Privacy Risk in Machine Learning: Analyzing the Connection to Overfitting,''
  2018 IEEE 31st Computer Security Foundations Symposium (CSF),
  pp. 268-282, 2018. 

\bibitem{Shokri2017} R. Shokri, M. Stronati, C. Song and V. Shmatikov, 
  ``Membership Inference Attacks Against Machine Learning Models,''
  2017 IEEE Symposium on Security and Privacy (SP), pp. 3-18, 2017.

\bibitem{Fang2020}
  M. Fang, X. Cao, J. Jia and N. Gong,
  ``Local Model Poisoning Attacks to {Byzantine-Robust} Federated Learning,''
  29th USENIX Security Symposium (USENIX Security 20), 
  pp. 1605--1622, 2020.


  

\bibitem{Wang-EDBT2016} 
  Y. Wang, X. Wu and D. Hu., ``Using randomized response for
  differential privacy preserving data collection,''
  In Proceedings of the EDBT/ICDT 2016 Joint Conference, 2016. 

\bibitem{Wang-Usenix2017}
  T. Wang, J. Blocki, N. Li and S. Jha,
  ``Locally differentially private
  protocols for frequency estimation,''
  In Proceedings of the 26th USENIX Security Symposium, ACM, pp. 729-745,
  2017.

\bibitem{Zhang2018}
  X. Zhang, L. Chen, K. Jin and X. Meng. Private high-dimensional
  data publication with junction tree. Journal of Computer Research
  and Development 55 (2018) 2794-2809

\bibitem{Zhang-PrivBayes17}
  J. Zhang, G. Cormode, C.M. Procopiuc, D. Srivastava and X. Xiao,
  ``PrivBayes: Private data release via Bayesian networks,'' ACM
  Transactions on Database Systems (TODS), 42(4), pp.1-4, 2017.

\bibitem{Zhang-CCS2018}
  Z. Zhang, T. Wang, N. Li, S. He and J. Chen,
  ``CALM: Consistent adaptive local marginal for
  marginal release under local differential privacy,''
  In Proceedings of the 2018 ACM SIGSAC Conference on
  Computer and Communications Security (CCS'18), ACM, 
  pp. 212-229, 2018. 

\bibitem{Fanti-PETS2016}
  Giulia Fanti, Vasyl Pihur, and Ulfar Erlingsson,
  ``Building a Rappor with the unknown: Privacy-preserving learning of
  associations and data dictionaries,''
  Proceedings on Privacy Enhancing Technologies, 3:1-21, 2016.

\bibitem{Duchi-SFCS2013}
  J. C. Duchi, M. I. Jordan, and M. J. Wainwright,
  ``Local privacy and statistical minimax rates,''
  in Proc. IEEE 54th Annu. Symp. Foundations Comput. Sci., pp. 429-438, 2013.

\bibitem{Qardaji2014}
  W. Qardaji, W. Yang, and N. Li,
  ``PriView: Practical differentially
  private release of marginal contingency tables,''
  in Proc. ACM SIGMOD Int. Conf. Manage. Data, pp. 1435-1446, 2014.

\bibitem{Ding-SIGMOD2011}
  B. Ding, M. Winslett, J. Han, and Z. Li,
  ``Differentially private data cubes:
  Optimizing noise sources and consistency,'' in Proc. ACM
  SIGMOD Int. Conf. Manage. Data, pp. 217-228, 2011.

\bibitem{Chen-KDD2015}
  R. Chen, Q. Xiao, Y. Zhang, and J. Xu,
  ``Differentially private high-dimensional
  data publication via sampling-based inference,'' in
  Proc. 21th ACM SIGKDD Int. Conf. Knowl. Discovery Data Mining,
  pp. 129-138, 2015. 

\bibitem{Xu-IFS2017}
  C. Xu, J. Ren, Y. Zhang, Z. Qin, and K. Ren,
  ``DPPro: Differentially private high-dimensional
  data release via random projection,'' IEEE Trans. Inf. Forensics Security,
  vol. 12, no. 12, pp. 3081-3093, 2017.

\bibitem{Qardaji2013}
  Wahbeh H. Qardaji and Weining Yang and Ninghui Li, 
  ``Understanding Hierarchical Methods for Differentially Private Histograms,''
  in Proc. VLDB Endow., Vol. 6, pp. 1954-1965, 2013. 

\bibitem{Day-AsiaCCS2015}
  W. Day and N. Li,
  ``Differentially private publishing of high-dimensional
  data using sensitivity control,''
  in Proc. ASIACCS, pp. 451-462, 2015.

\bibitem{Bassily-STOC2015}
  R. Bassily and A. SmithLi,
  ``Local, private, efficient protocols for succinct histograms,''
  in Proc. ACM STOC, pp. 127-135, 2015. 

\bibitem{Qin-CCS2016}
  Z. Qin, Y. Yang, T. Yu, I. Khalil, X. Xiao, and K. Ren,
  ``Heavy hitter
  estimation over set-valued data with local differential privacy,''
  in Proc. ACM CCS, pp. 192-203, 2016.

\bibitem{Jordi-IS2013}
  Jordi Soria-Comas, Josep Domingo-Ferrer,
  ``Optimal data-independent noise for differential privacy,''
  Information Sciences,
  Volume 250, pp. 200-214, 2013.

\bibitem{Wang-ICDE2019}
  Wang, Ning and Xiao, Xiaokui and Yang, Yin and Zhao, Jun and Hui, Siu Cheung and Shin, Hyejin and Shin, Junbum and Yu, Ge, 
  ``Collecting and Analyzing Multidimensional Data with Local Differential Privacy,''
  2019 IEEE 35th International Conference on Data Engineering (ICDE),
  pp. 638-649, 2019. 

\bibitem{Wang-NDSS2020}
  Wang T, Lopuhaa-Zwakenberg M, Li Z, Skoric B, Li N.,
  ``Locally Differentially Private Frequency Estimation with Consistency,''
  In NDSS 2020, 2020.

\bibitem{Arcolezi-CIKM2021} 
  Heber H. Arcolezi, Jean-Francois Couchot, Bechara Al Bouna, and Xiaokui Xiao,
  ``Random Sampling Plus Fake Data: Multidimensional Frequency Estimates With Local Differential Privacy,''
  In Proceedings of the 30th 
  ACM International Conference on Information \& Knowledge Management
  (CIKM '21), ACM, pp. 47-57, 2021. 

\bibitem{Cormode-SIGMOD2018} 
  Graham Cormode, Tejas Kulkarni, and Divesh Srivastava,
  ``Marginal Release Under Local Differential Privacy,''
  In Proceedings of the 2018 International Conference on Management of Data (SIGMOD '18), ACM, pp.131-146, 2018. 

\end{thebibliography}

\appendix
\subsection{Proofs}

\begin{ProofOf}{Theorem~\ref{th.otimes}}
  First, we show that the matrix has a corresponding
  conditional probability. 
  Let $u$ and $v$ be tuples of attributes $A^i$ and $A^j$
  such that $u = (y^i,y^j)$ and $v = (x^i, x^j)$.
  From the premise of the randomization matrix for attributes $A^i$ and $A^j$,
  $p^i_{x^i y^i} = Pr[Y^i = y^i| X^i = x^i]$ and
  $p^j_{x^j y^j} = Pr[Y^j = y^j| X^j = x^j]$ hold. 
  According to the definition of the Kronecker product, we obtain the 
  $(d^i d^j)\times (d^i d^j)$ matrix as
  \[
  P^i \otimes P^j = \left( \begin{array}{ccc}
    p_{11}P^j & \cdots & p_{1d_i} P^j \\
    \vdots  & \ddots & \vdots \\
    p_{d_i1}P^j & \cdots & p_{d_id_i} P^j
  \end{array} \right),
  \]
  where element $p_{uv}$ is $p^i_{y^i x^i}\cdot p^j_{y^j x^j}$,
  which is equal to the joint probability of $u$ and $v$
  because the two randomizations are independent.
  Second, we show it satisfies the conditions for probability.
  If $p_{11}+\cdots p_{1d_i} = 1$ and $p_{11}+\cdots p_{1d_j} = 1$,
  the sum of the Kronecker product
  \(
  p_{11}p_{11} +\cdots + p_{1d_i}p_{1d_j} 
  = p_{11}(p_{11}+\cdots+ p_{1d_j}) +\cdots + p_{1d_i}(p_{11}+\cdots+ p_{1d_j})
  = p_{11}(1) +\cdots + p_{1d_i}(1) = 1
  \)
  holds. 
  Finally, we show that the matrix can be inverted. 
  Because of the property of Kronecker products,
  $P^i \otimes P^j$ is non-singular if and only if
  $P^i$ and  $P^j$ are non-singular. 
  Hence, we have the theorem. 
\end{ProofOf}

\begin{ProofOf}{Theorem~\ref{th.complexity}}
  A permutation $\sigma_i$ takes $w$ time.
  The time for the transposition and its inversion are negligible
  because these can be predetermined from the data structure. 
  An inversion of a $d^2$ matrix takes ${\cal O}(d^{2.807})$ time.
  Therefore, repeating these costs $w$ times, the total
  processing time is ${\cal O}(wd^{2.807})$ and 
  the storage requirement of $d^w$ is constant. 
\end{ProofOf}

\begin{ProofOf}{Theorem~\ref{th.ldp}}
  For any $x, x' \in |A|$ such that $x \ne x'$, and any $y \in |A|$
  \[
  \frac{Pr[RR(x) = y]}{Pr[RR(x') = y]}
    = \frac{p}{q} = e^{\epsilon}
    \]
    Because the $m$ attributes are perturbed independently,
    the sequential decomposition theorem~\cite{McSherry} states that
    \textsf{RR-Ind-Joint} satisfies 
    $(m \epsilon,0)$-LDP. 
\end{ProofOf}

\begin{ProofOf}{Theorem~\ref{th.rrind.mse}}
  The definition of V statistics is $V = \sqrt{\chi^2/n(d-1)}$.
  Squaring and dividing both sides by $d$, we have
  \begin{eqnarray*}
  V^2/d &=& \frac{\chi^2/n}{d(d-1)} 
  \le \frac{1}{nd^2} \sum^{d^2}_{i = 1} \frac{(o_i - e_i)^2}{e_i} \\
  &=& \frac{1}{d^2} \sum_{a \in |A|, b \in |B|} \frac{(o_{(a,b)}/n -\hat\lambda_a\hat\lambda_b)^2}{\hat\lambda_a\hat\lambda_b} \\
  &\le& \frac{1}{d^2} \sum_{a \in |A|, b \in |B|} (\Pi^{AB}(a,b) -\hat\lambda_a\hat\lambda_b)^2 \\
  &=& MSE(\Pi^{AB}).
  \end{eqnarray*}
  Note that the expected value $e_i$ is the mean of the binomial distribution
  of $p = \Pi^{AB}_{RR Ind}$ with $n$ trials, i.e., 
  $np = n\hat\lambda^A(a)\hat\lambda^B(b)$. 
  The final inequality holds when $\hat\lambda_a\hat\lambda_b \le 1.0$.
\end{ProofOf}

\begin{ProofOf}{Lemma~\ref{le.marginal}}
  Suppose there exists an $i$-th attribute and value $a_i$
  for which $\pi^i(a_i) < \Pi(a_1,\ldots,a_w)$.
  This immediately contradicts 
  the marginal probability given as
  $\sum_{a \in \Pi^i} \Pi(a_1,\ldots,a,\ldots, a_w) > \pi^i(a_i)$. 
  $\Pi(a_1,\ldots,a_w)$ must therefore be less than the minimum for $\pi^j$. 
\end{ProofOf}

\begin{ProofOf}{Theorem~\ref{th.rrind}}
  Given sufficient records and an accurate randomization matrix,
  a marginal distribution can be estimated without error. We consider this as
  $\hat\pi = \pi$. 
  With the premise that $\pi^i \sim 1/d_i$ and Lemma~\ref{le.marginal} holds, 
  the estimated probability is
  \begin{eqnarray*}
    0 &\le& \Pi^S_{\sf RR-Ind} (a_1,\ldots, a_w) = \frac{1}{d_1 \cdots d_w} \le \frac{1}{(\max d_i)^w} \\
    &\le&
    \min( \pi^1(a_1),\ldots, \pi^w(a_w)) = \frac{1}{\max{d_i}}.
  \end{eqnarray*}
      Therefore, the longer interval either
      $[0, 1/\max(d_i)^w]$ or $[1/(\max d_i)^w, 1/\max{d_i}]$
        is an upper bound on the estimation error. 
\end{ProofOf}

\begin{ProofOf}{Lemma~\ref{le.adj}}
  The adjugate matrix of $P$ shows that 
  the largest elements of the inverse are along the diagonal
  and are less than $1/p$ for any $d \ge 2$.
\end{ProofOf}

\begin{ProofOf}{Lemma~\ref{le.inv}}
  The inverse of the product is given as $P^{-1} = P_1^{-1}\otimes P_2^{-1}$,
  whose largest elements are along the diagonal and are at most
  $1/p_1 p_2$ from Lemma~\ref{le.adj}.
  $\Delta\lambda$ is a $(d_1\times d_2)$-dimensional vector of
  uniform random values within $[-1/n, 1/n]$.
  The rounding error is the inner product of $P^{-1}$ and
  $\Delta\lambda$ such that
  \[
  \max P^{-1}\cdot \Delta\lambda \le
  (1/p_1p_2 \cdots 1/p_1p_2)\cdot
  \left(\begin{array}{c} 1/n\\ \vdots \\ 1/n \end{array}\right)
  = \frac{d_1 d_2}{p_1 p_2 n}
  \]
\end{ProofOf}

\begin{ProofOf}{Theorem~\ref{th.avd}}
  With $\epsilon' = \epsilon/w$, $w$ attributes are
  randomized independently. From Lemma~\ref{le.inv}
  and $\epsilon'$,
  we have
  \[
  \max P^{-1}\Delta\lambda < \frac{d_1\cdots d_w}{p_1 \cdots p_w n}
  < \frac{d^w}{p^w n} = 
  \left(\frac{e^{\epsilon/w}+d-1}{d^{\epsilon/w}}\right)^w \frac{d^w}{n}
  \]
  where $d = \max(d_1,\ldots, d_w)$ and $p = \min(p_1,\ldots, p_w)$. 
\end{ProofOf}

\subsection{Example}\label{sec.exp}
  
Consider a dataset $X$ on $n = 10$ parties with two attributes $A$ and $B$,
where domain $\Omega_A = \{a_1, a_2\}$ and $\Omega_B = \{b_1, b_2\}$. The empirical
(true) joint probability distribution of $X$ is 
\begin{eqnarray*}
  \Pi_{AB}(a_1, b_1) &=& 4/10, \\
  \Pi_{AB}(a_2, b_1) &=& 2/10, \\
  \Pi_{AB}(a_1, b_2) &=& 0,    \\
  \Pi_{AB}(a_2, b_2) &=& 4/10.
\end{eqnarray*}
This yields marginal distributions $\pi_A = (0.4, 0.6)$ and
$\pi_B = (0.6, 0.4)$.
We express the frequencies of $X$ as a $2\times 2$ matrix
\(
f^X = \left(\begin{array}{cc}
  4 & 0 \\
  2 & 4
\end{array}\right),
\)
which indicates 
frequencies of $(a_1,b_1), (a_2,b_1), (a_1,b_2), (a_2,b_2)$ for $X$, 
respectively.

With $\epsilon = \log(3)$ and $p_A = p_B = 3/4$, 
we have randomization matrices for $A$ and $B$ as
\[
P^A = \left(\begin{array}{cc}
p_A & q_A \\
q_A & p_A \\
\end{array}\right)
= \left(\begin{array}{cc}
3/4 & 1/4 \\
1/4 & 3/4 \\
\end{array}\right)
= P^B, 
\]
where $p = \frac{e^{\log 3}}{e^{\log 3} + d - 1} = 3/4$ and $q = 1-p$. 
The respondents randomize their two responses $x_i^A$, and $x_i^B$ independently.
Suppose that the randomized $Y^A = \textsf{RR}_{P^A}(X^A)$ 
and $Y^B = \textsf{RR}_{P^B}(X^B)$ are observed as
\(
f^Y = \left(\begin{array}{cc}
  3 & 1 \\
  3 & 3
\end{array}\right),
\)
for which the empirical probabilities of $Y$ 
are $\lambda^A = (0.4,0.6)$ and $\lambda^B = (0.6, 0.4)$.
Note that the $V$ statistics for $A$ and $B$ gives $V_{AB}(Y) = 0.25$, which
is reduced from $V_{AB}(X) = 0.66$ for the original dataset.
Here, the correlation between $A$ and $B$ has been partially lost by
the independent randomizations. 

\textsf{RR-Independent} estimates the joint probabilities as
the product of the estimated marginal distributions 
$\hat\pi^A = {P^A}^{-1} \lambda^A = (0.3, 0.7)$ 
and $\hat\pi^B = {P^B}^{-1} \lambda^B = (0.7, 0.3)$, giving
\[
\hat\Pi^{AB}_{\sf RR\mathchar`-Ind} = \left(\begin{array}{cc}
0.21 & 0.09 \\
0.49 & 0.21
\end{array}\right),
\]
which estimates $\Pi^{AB}$ with
MSE$ = 0.041$ and AVD = $0.29$.
The value for $V$ statistics is nearly 0. 

\textsf{RR-Ind-Joint} treats the two independent randomization matrices
as a single accumulated matrix 
$P^A \otimes P^B$
\[
= \left(\begin{array}{cccc}
  p_a p_b & p_a q_b & q_a p_b & q_a q_b \\
  p_a q_b & p_a p_b & q_a q_b & q_a p_b \\
  q_a p_b & q_a q_b & p_a p_b & p_a q_b \\
  q_a q_b & q_a p_b & p_a q_b & p_a p_b
\end{array}\right)
= 
\frac{1}{16}
\left(\begin{array}{cccc}
9 & 3 & 3 & 1 \\
3 & 9 & 1 & 3 \\
3 & 1 & 9 & 3 \\
1 & 3 & 3 & 9
\end{array}\right).
\]
Given the observed the empirical distributions for $Y$, 
we estimate the joint probabilities as
\begin{eqnarray*}
\hat\Pi^{AB}_{\sf RR\mathchar`-Ind\mathchar`-Joint} &=& (P^A \otimes P^B)^{-1} \Lambda^{AB} \\
&=& {P^A}^{-1} \left( {P^B}^{-1} {\Lambda^{AB}}^T \right)^T \\
&=& {P^A}^{-1} 
	\left(
	\left( \begin{array}{cc} 1.5 & -0.5 \\ -0.5 & 1.5 \end{array} \right)
        \left(\begin{array}{cc} 0.3 & 0.1 \\ 0.3 & 0.3 \end{array} \right)^T
        \right)^T \\
&=& \left(\begin{array}{cc} 1.5 & -0.5 \\ -0.5 & 1.5 \end{array} \right)
        \left(\begin{array}{cc} 0.4 & 0.3 \\ 0. & 0.3 \end{array} \right)^T \\
&=& \left(\begin{array}{cc} 0.45 & -0.15 \\ 0.25 & 0.45 \end{array} \right).
\end{eqnarray*}
The estimation error is 
MSE $= 0.0075$ and AVD $= 0.15$. 

Using the limit on valid probabilities estimated via
the $(w-1)$-way joint (marginal) probabilities
$\hat\pi^A = (0.3, 0.7)$
and $\hat\pi^B = (0.6, 0.4)$,
we obtain the revised probability for the {\em truncated} algorithm
\begin{eqnarray*}
  \hat\Pi^{AB}_{truncated} &=& \left(\begin{array}{cc}
    \min(0.45,0.3,0.7) & \min(0, 0.3,0.3) \\
    \min(0.25,0.7,0.7) & \min(0.45, 0.7,0.3) \\
  \end{array}\right) \\
  &=& \left(\begin{array}{cc}
    0.3  & 0 \\
    0.25 & 0.3 \\
  \end{array}\right),
\end{eqnarray*}
which improves accuracy as MSE $= 0.15$ and  AVD $= 0.1$.

The privacy of the independent randomization is assured by
$\epsilon = \ln(p/q) = \ln(3)$.
With two attributes, the privacy budget is $2\epsilon = 2 \ln 3$ in total.
The same privacy is assured by {\sf RR-Joint} with
$p' = \frac{e^{2\epsilon}}{e^{2\epsilon} + d_1 d_2 -1} = 9/12$, 
and $q' = 1/12$ and is expressed as 
\[
\frac{1}{12}\left(\begin{array}{cccc}
  9 & 1 & 1 & 1 \\
  1 & 9 & 1 & 1 \\
  1 & 1 & 9 & 1 \\
  1 & 1 & 1 & 9 \\
\end{array}\right)
\]
for which $(2 \ln 3,0)$-LDP holds.

\subsection{Thresholds analysis for hybrid scheme}\label{sec.thresholds}

Experimental results in Fig.~\ref{fig-adw}, \ref{fig-adn} and \ref{fig-ade}
suggest that there are crossover points between two estimations. 

By combining 
the upper bound of estimation error of {\sf RR-Ind-Joint} in Eq.~(\ref{eq.avd})
with that of {\sf RR-Independent} in Eq.~(\ref{eq.avd-rrind}),
we identify the thresholds for number of respondents $n^*$ beyond which
AVD$_{\sf RR-Ind-Joint}$ is less than AVD$_{\sf RR-Ind}$ as,
\[
n^* \ge
\left(1 + \frac{d-1}{e^{\epsilon/w}}\right)^w
\frac{d^w}{\max(d^{-w}, 1/d - d^{-w})},
\]
where $d$ is the maximum domain size for $w$ attributes. 
It implies that {\sf RR-Ind-Joint} shall be chosen
when there are enough records $n$ according to
the domain size $d = |\Omega|$ and dimension $w$.
We see that (c) Credit and (d) Nursery has higher
cross-points in Fig.~\ref{fig-adn} due to the lack of records $n$. 

Similarly, we have the threshold for privacy budget $\epsilon^*$
beyond which {\sf RR-Ind-Joint} estimates better than
{\sf RR-Independent} as, 
\[
\epsilon^* \ge w\log \frac{d-1}
        {\left(\frac{1}{d} - \frac{1}{d^w} \right)^{1/w} \frac{n^{1/w}}{d} -1}.
\]
Note that the threshold for the privacy budget is linear to dimension $w$
here because the sequential composition of $w$ randomizations results
$w\epsilon$-differential privacy. 
If a required privacy budget is $\epsilon < \epsilon^*$, we
can use {\sf RR-Ind-Joint} to estimate joint probability. 

Finally, suppose that AVD of {\sf RR-Ind-Joint} is smaller
than that of {\sf RR-Independent}.
Then, by noticing $\exp(\epsilon/w) \rightarrow 1$ as $w$ becomes
large enough, 
\[
\frac{d^{2w}}{n}
 <
 \left( 1 +  \frac{d-1}{e^{\epsilon/w}} \right)^w \frac{d^w}{n}
 \le \frac{1}{d} - \frac{1}{d^w}
 \le \frac{1}{d}
\]
holds.
By solving it for $w$, we have the threshold for
dimension $w^*$ as
\[
w^* \le \frac{\log n - \log d}{\log d^2}.
\]

The threshold values enable us to combine two
estimation algorithms as efficient hybrid scheme.
We use {\sf RR-Ind-Joint} for small dimension
joint probability estimation and switch to
{\sf RR-Independent} for high dimensional cases.
With observation of fundamental features of data,
$n$, $m$ and $d$, we estimate joint probability
for arbitrary dimension $w$ from independently
randomized data with $(w\epsilon,0)$-LDP guarantee.

\subsection{Evaluation with Synthetic Data}\label{sec.synthetic}
\subsubsection{Methodology}

Our synthesized dataset has two attributes $A$ and $B$
with marginal probabilities $\lambda^A = \lambda^B$
distributed as $Pr(A = a) = c/a$ for $a = 2,\ldots, d$ and a constant 
$c = 1/(\sum_{a = 2}^d 1/a)$. 
The domain of attribute $A$ is denoted by $|A| = \{c/2,\ldots, c/d\}$,
where $d$ is the number of unique values in attribute $A$. 
The correlation between attributes is controlled to give
values for Cramer's V statistics $v = V_{AB} \in [0,1]$.

\begin{figure*}[tb]\centering
  \begin{subfigure}[t]{\wdq \linewidth}\centering
    \includegraphics[width=\linewidth]{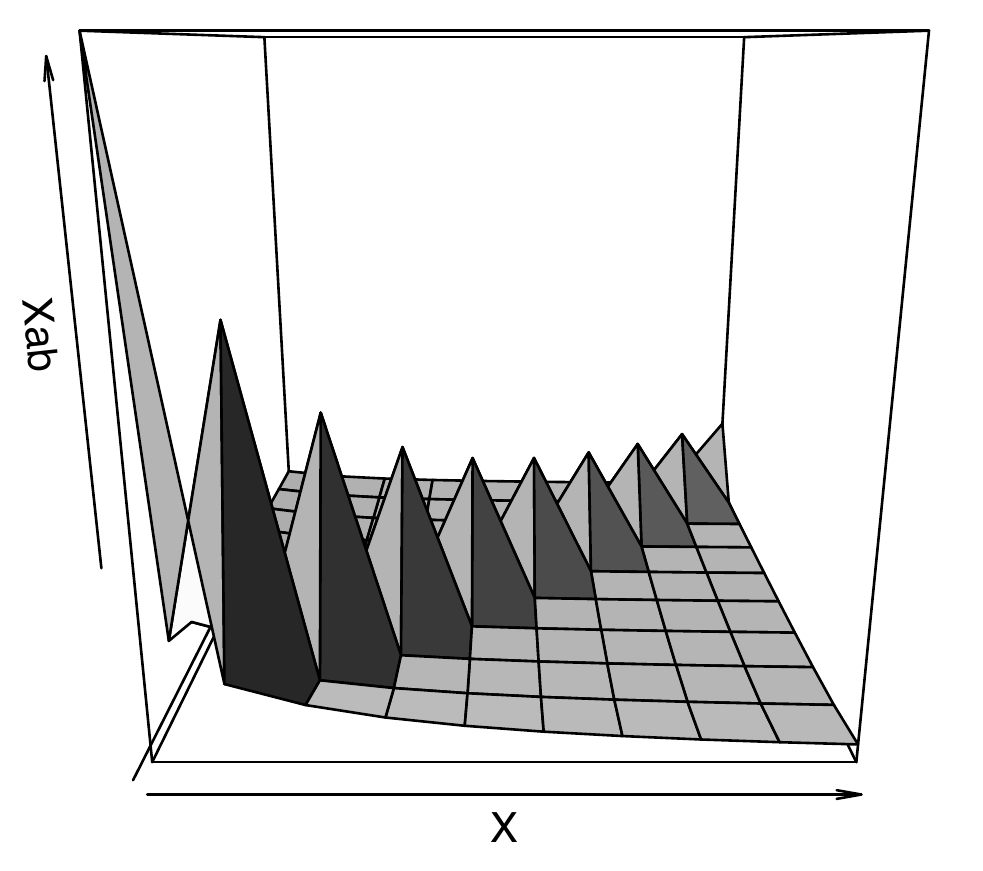}
    \caption{Real data $\Pi^{AB}(X)$}\label{fig-3d-xab}
  \end{subfigure}
  \hfill
  \begin{subfigure}[t]{\wdq \linewidth}\centering
    \includegraphics[width=\linewidth]{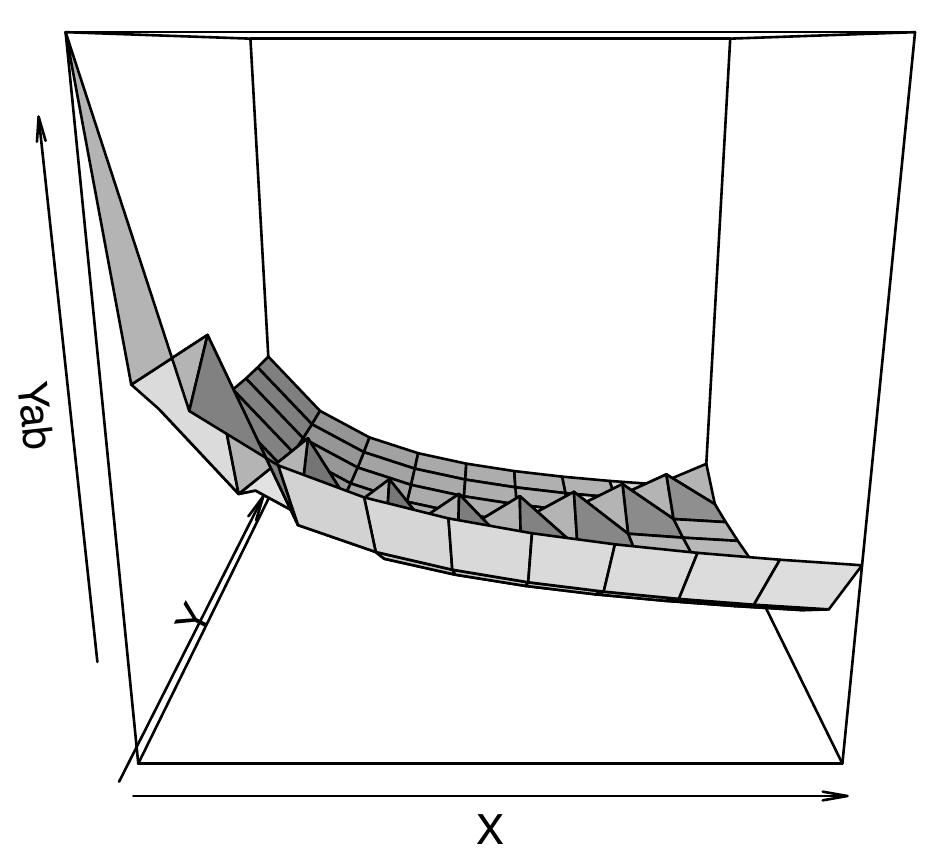}
    \caption{Perturbed $\lambda^{AB}(Y)$}\label{fig-3d-yab}
  \end{subfigure}
  \hfill
  \begin{subfigure}[t]{\wdq \linewidth}\centering
    \includegraphics[width=\linewidth]{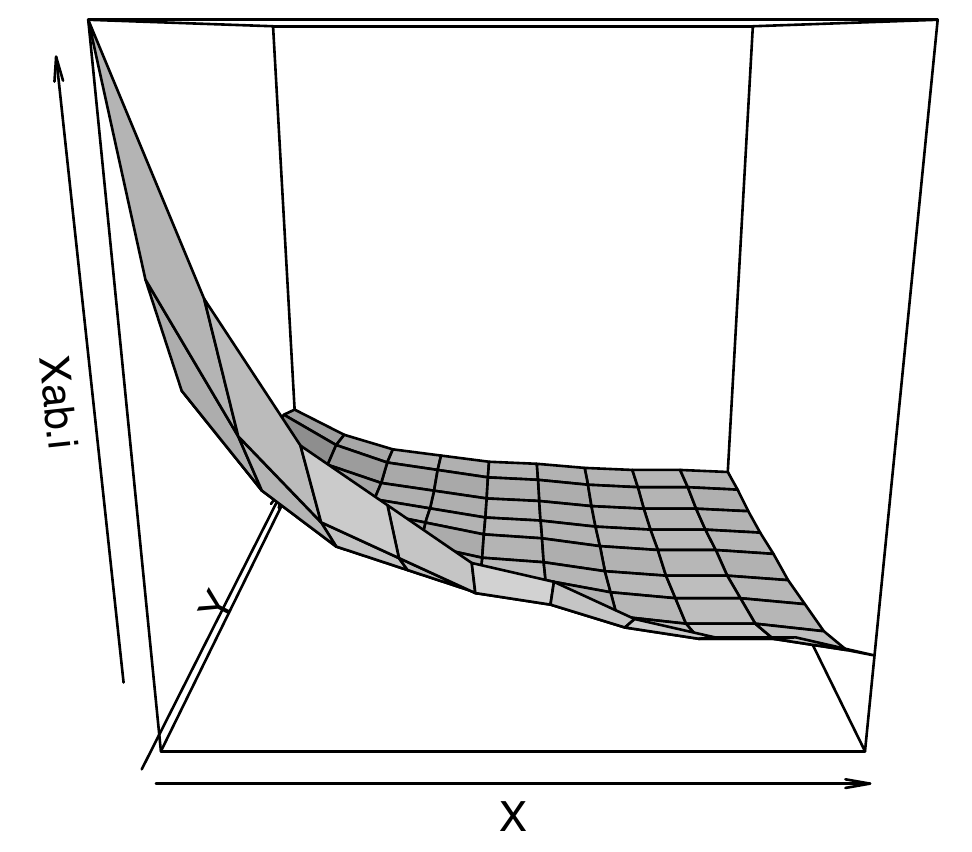}
    \caption{Estimated $\hat\Pi^{AB}_{\sf RR Ind}(X)$}\label{fig-3d-xab-ind}
  \end{subfigure}
  \hfill
  \begin{subfigure}[t]{\wdq \linewidth}\centering
    \includegraphics[width=\linewidth]{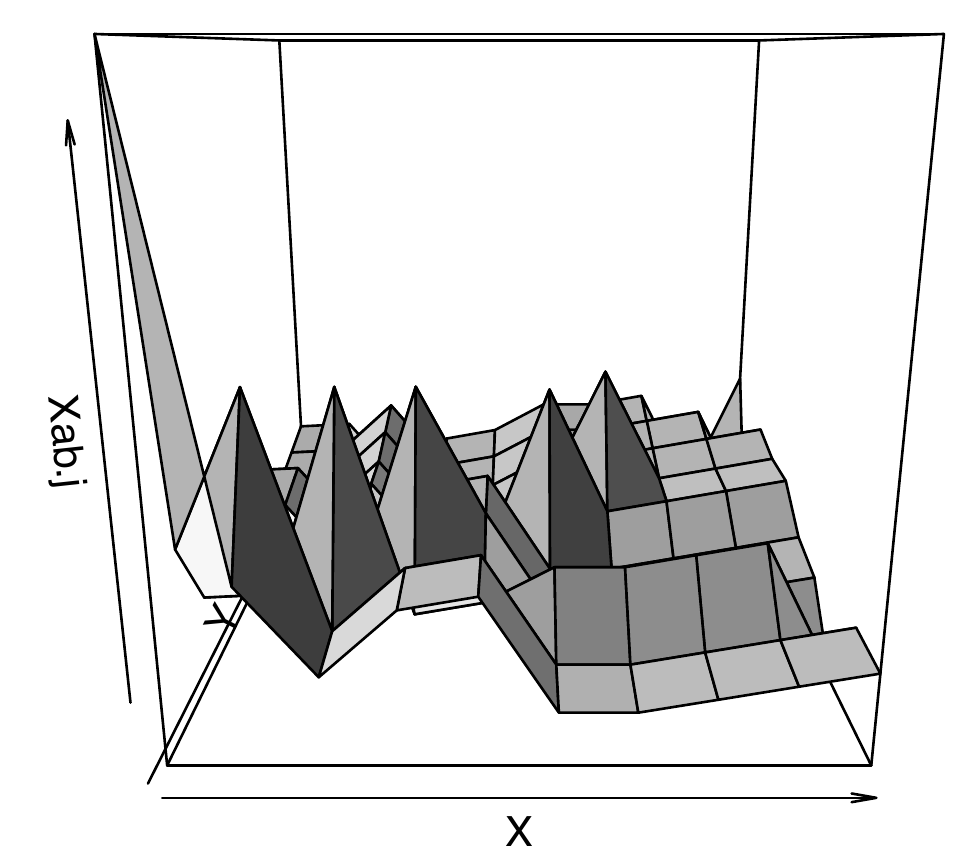}
    \caption{Estimated $\hat\Pi^{AB}_{\sf RR Ind Joint}(X)$}\label{fig-3d-xab-joint}
  \end{subfigure}  
  \caption{Example of a 2-way joint probability distribution,
    with estimated distributions
    for {\sf RR-Independent} and {\sf RR-Ind-Joint}}\label{fig-3d}
\end{figure*}

Fig.~\ref{fig-3d}
shows the joint probability distributions 
 $A$ and $B$ with
$n = 1000, d = 10, v = 0.5$,
for the synthetic data $\Pi^{AB}(X)$ (\ref{fig-3d-xab}),
the perturbed data $Y = \textsf{RR}(X)$ with $\epsilon = 1$ 
$\lambda^{AB}(Y)$ (\ref{fig-3d-yab}) ,
the estimated probability by \textsf{RR-Ind} $\hat\Pi^{AB}_{\sf RR Ind}(X)$
(\ref{fig-3d-xab-ind})
and
the estimated  \textsf{RR-Ind-Joint} probability $\hat\Pi^{AB}_{\sf RR Ind Joint}(X)$
(\ref{fig-3d-xab-joint}). 
Note that the joint probability of the given data $X$
with Cramer's V of $v = 0.5$
has a strong correlation along the diagonal elements in the Cartesian
product $|A|\times |B|$,
which is distributed widely in the perturbed data $Y$.
\textsf{RR-Ind} fails to estimate the strong correlation between
the two attributes in $\hat\Pi^{AB}_{\sf RR Ind}(X)$. 
In contrast, \textsf{RR-Ind-Joint} estimates the joint probabilities
more accurately (see Fig.~\ref{fig-3d-xab-joint}). 
The estimated probabilities are not exactly the same as those for the original $X$
because the precision of the empirical distribution $\lambda^{AB}$
depends on environmental parameters, e.g., the number
of individuals $n$, the size of the attribute domain (the number
of unique values) $d$, the privacy budget $\epsilon$ and
the correlation between two attributes. 
We evaluate the accuracy loss in terms of these parameters. 

\subsubsection{Results (Synthetic Data)}

\begin{figure*}[tb]\centering
  \begin{subfigure}[t]{\wdq\linewidth}\centering
    \includegraphics[width=\linewidth]{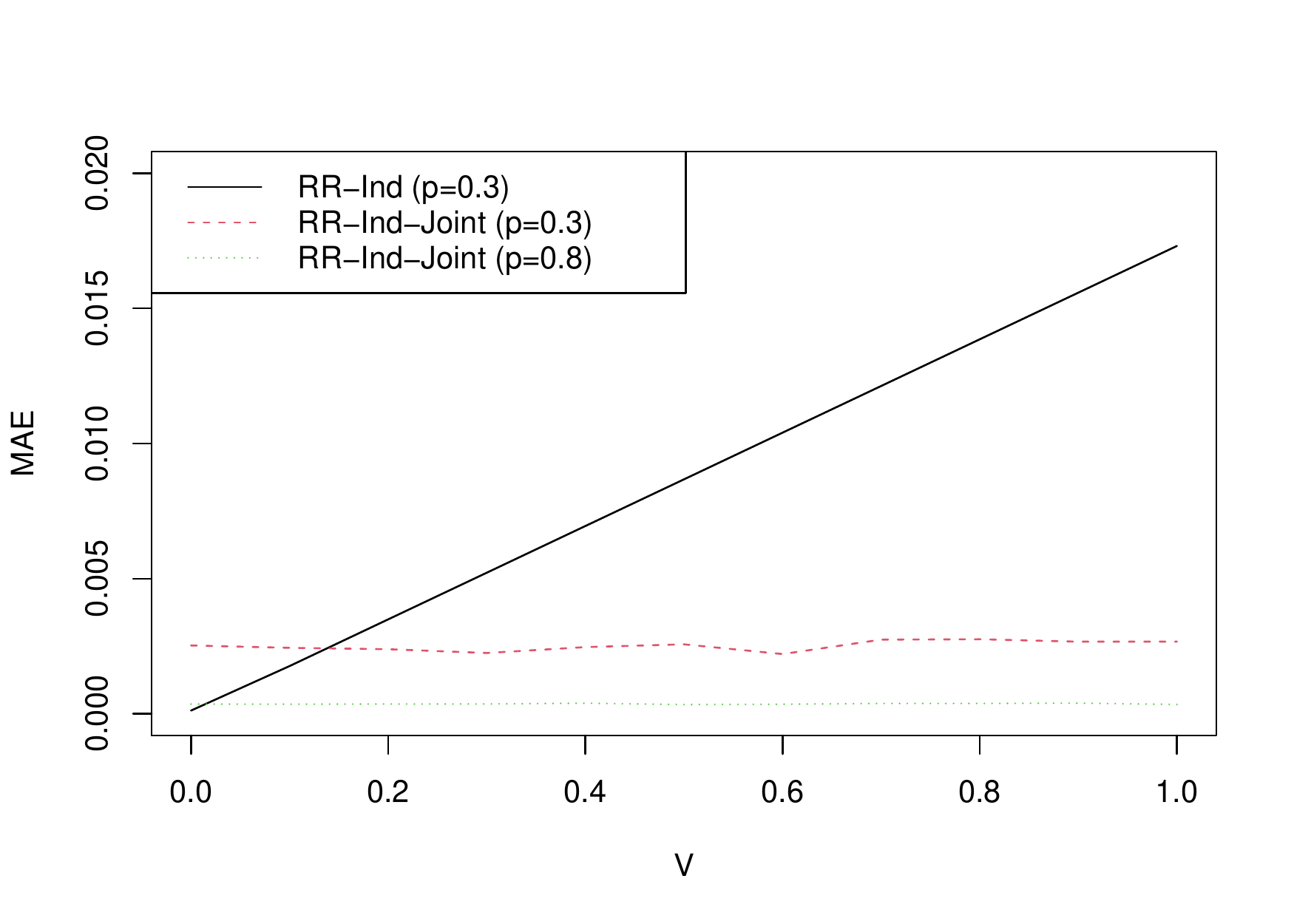}
    \caption{Correlation $v$ }\label{fig-v}
  \end{subfigure}
  \hfill
  \begin{subfigure}[t]{\wdq\linewidth}\centering
    \includegraphics[width=\linewidth]{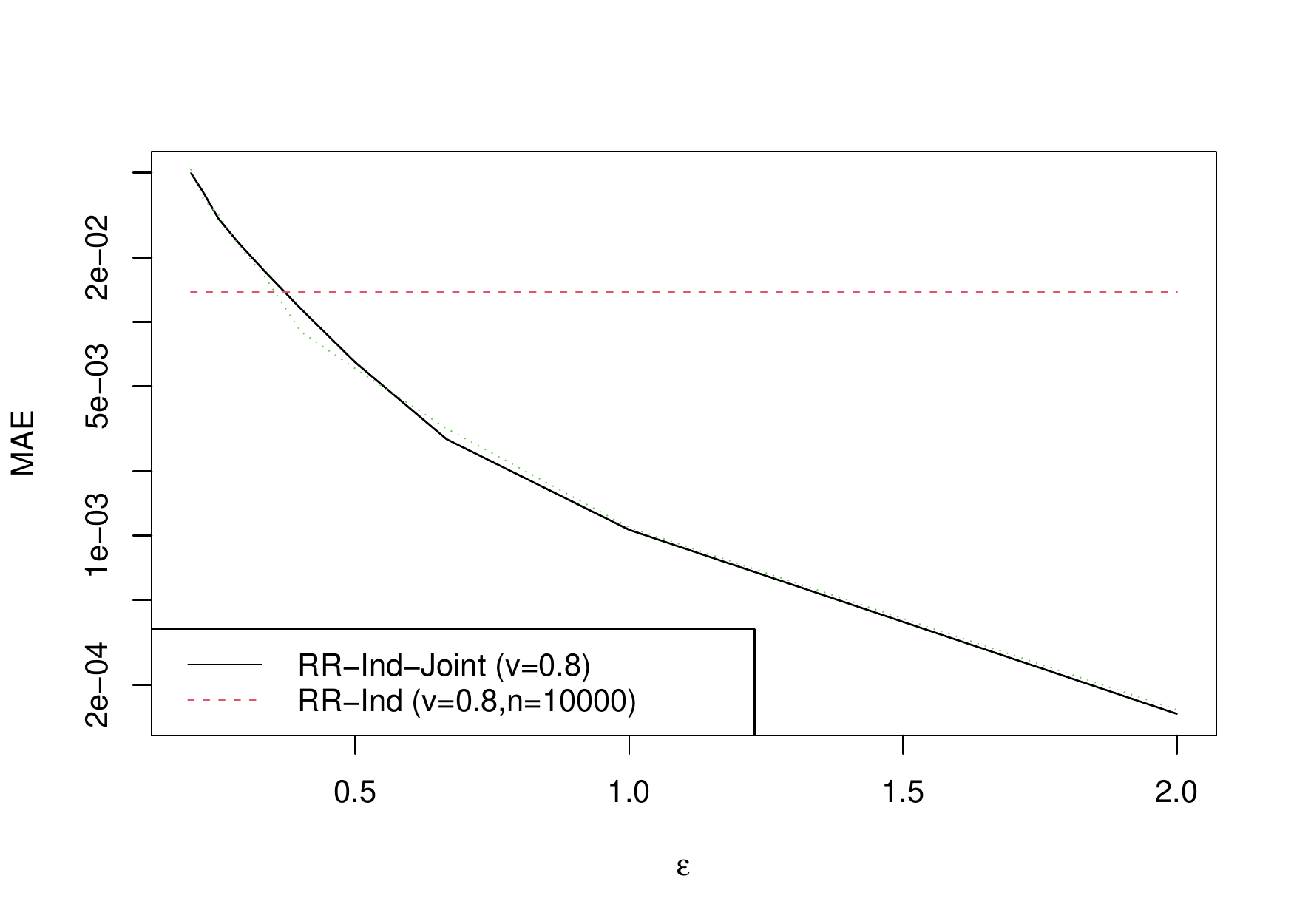}
    \caption{Privacy budget $\epsilon$ }\label{fig-eps}
  \end{subfigure}
  \hfill
  \begin{subfigure}[t]{\wdq\linewidth}\centering
    \includegraphics[width=\linewidth]{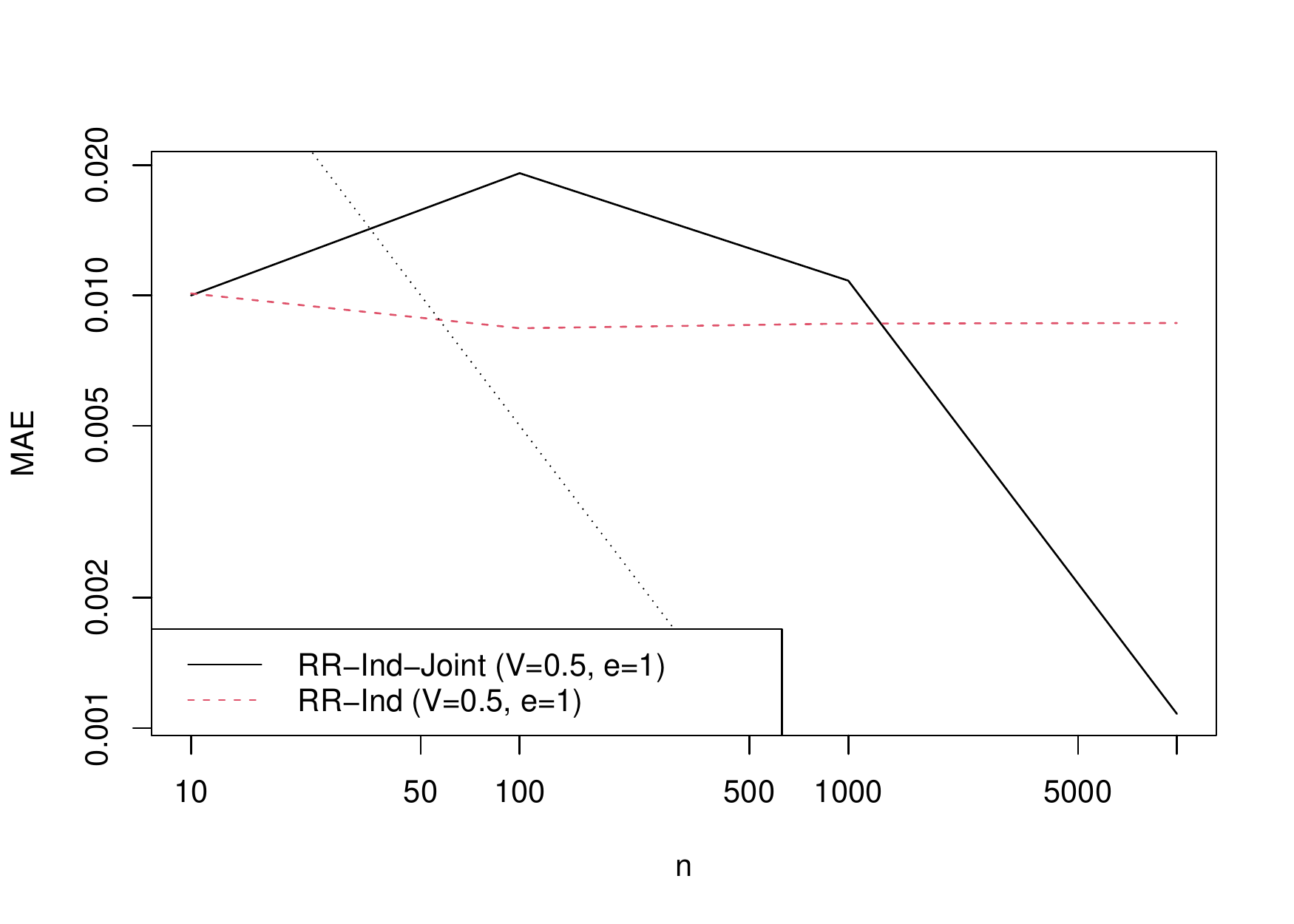}
    \caption{Number of respondents $n$ }\label{fig-n}
  \end{subfigure}
  \hfill
  \begin{subfigure}[t]{\wdq\linewidth}\centering
    \includegraphics[width=\linewidth]{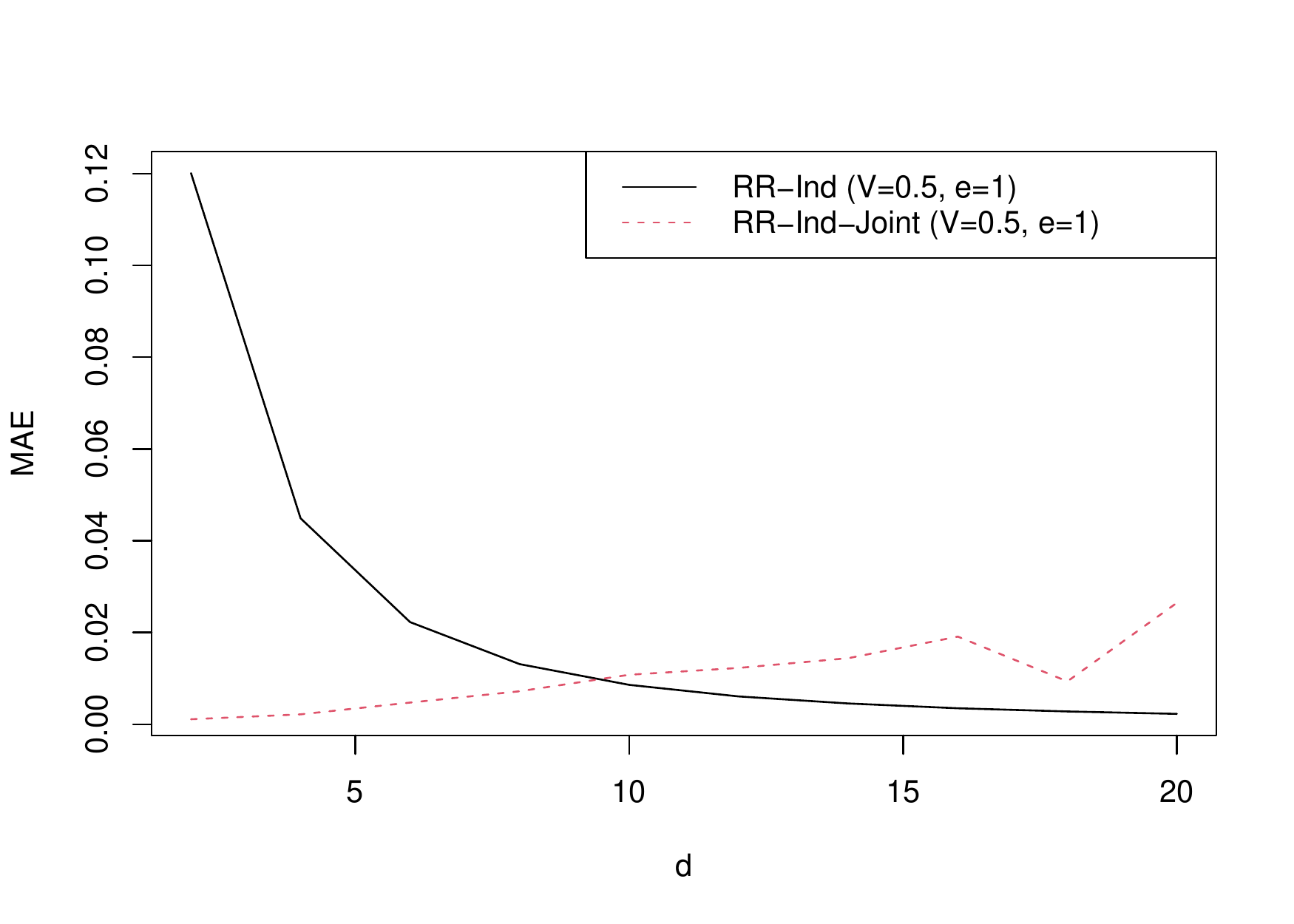}
    \caption{Domain size $d (= |A|)$ }\label{fig-d}
  \end{subfigure}
  \caption{Synthetic-data MAEs}\label{fig-4}
\end{figure*}

Figures~\ref{fig-4} shows synthetic-data MAE values for four sets of
parameter values, namely, 
Cramer's V statistics $v \in [0,1]$,
Privacy budget $\epsilon = 0.1, \ldots, 2$,
Number of individuals $n = 10, 100, 1000, 10000$, and
Domain sizes (the number of unique values in
attribute) $d (=|\Omega_A| = |\Omega_B|) = 2, \ldots, 20$.
   
Figure~\ref{fig-v} shows that the estimation error for \textsf{RR-Independent}
depends on the correlation between attributes.
The MAE is proportional to $V$ with two extreme cases:
0 when $A$ and $B$ are independent ($V = 0$) and
the highest value when $A$ completely depends on $B$ ($V = 1$).
\textsf{RR-Independent} estimates the joint probability, via
the product of two marginal probabilities, as
$\hat\Pi^{AB}(a,b) = \hat\sigma^A(a) \hat\sigma^B(b)$,
under the assumption of independent attributes, for which
$V = 0$. 
The estimation error is therefore linear in $V$ 
(i.e., is considered as the ratio of independent pairs of values $(a,b)$
to the $d\times d$ pairs). 
In contrast, the MAE for \textsf{RR-Ind-Joint} does not depend on $V$. 
It estimates the joint probabilities accurately,
irrespective of any attributes correlations.

Figure~\ref{fig-eps} shows that the MAE of \textsf{RR-Ind-Joint} 
decreases as the privacy budget $\epsilon$
increases, which follows in turn the increases in
the probabilities of retaining.
It also shows that the MAE for \textsf{RR-Independent} is constant
because the primary part of the estimation error is caused by
the strength of correlation between attributes and the effect
of the privacy budget is to hide the other errors.  

The MAE for \textsf{RR-Ind-Joint} depends on the number of respondents $n$
and the domain size  $d = |A|$.
There is a reduction in MAE with decreasing $n$ in Fig.~\ref{fig-n}. 
The MAE of \textsf{RR-Ind-Joint} decreases according to $1/n$
when $n > 1000$.
The MAE also tends to increase with increasing $d$ in Fig.~\ref{fig-d}.
We conclude that \textsf{RR-Ind-Joint}
estimation needs a sufficiently large number of respondents and has
a limit of the dimensionality. 

The reduction of MAE with increasing $d$ is consistent
with Theorem~\ref{th.rrind.mse}, which states that the MAE is
linear with respect to $1/\sqrt{d}$.

\subsection{Continuous attribute}

\begin{figure*}[tb]\centering
  \begin{subfigure}[t]{\wdq\linewidth}\centering
    \includegraphics[width=\linewidth]{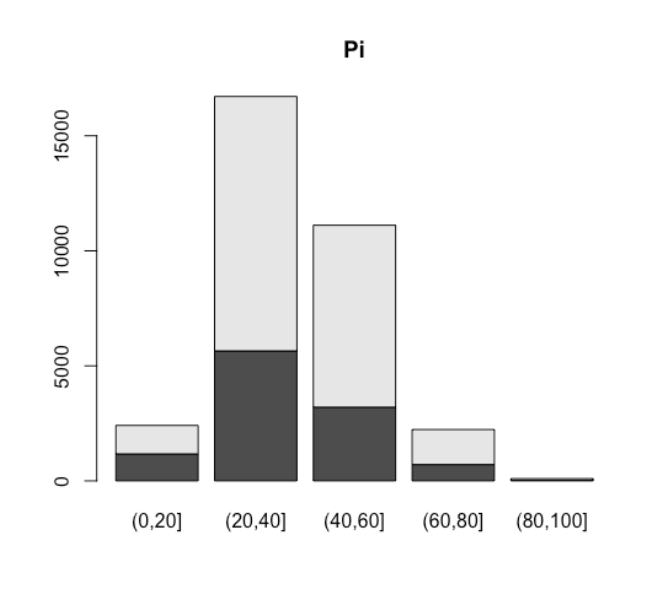}
    \caption{Histogram for Age and Sex}\label{fig-age-pi}
  \end{subfigure}
  \hfill
  \begin{subfigure}[t]{\wdq\linewidth}\centering
    \includegraphics[width=\linewidth]{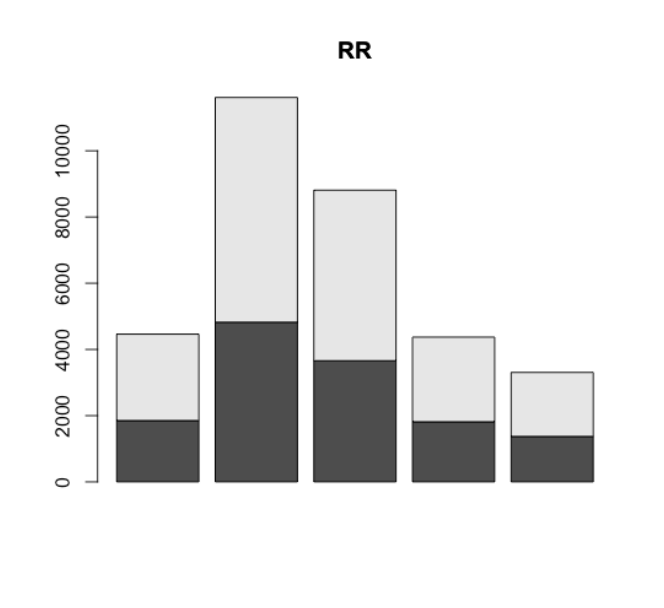}
    \caption{Randomized $Y$}\label{fig-age-rr}
  \end{subfigure}
  \hfill
  \begin{subfigure}[t]{\wdq\linewidth}\centering
    \includegraphics[width=\linewidth]{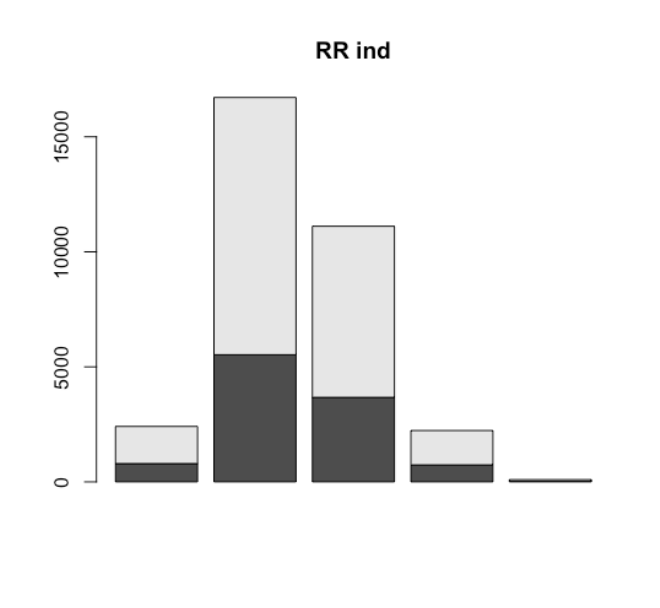}
    \caption{Estimated $\hat\Pi_{\sf RR Ind}^S$}\label{fig-age-ind}
  \end{subfigure}
  \hfill
  \begin{subfigure}[t]{\wdq\linewidth}\centering
    \includegraphics[width=\linewidth]{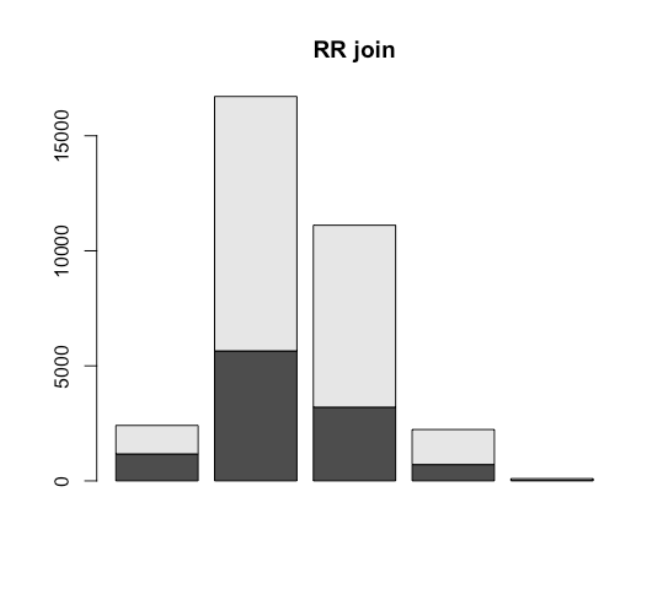}
    \caption{Estimated $\hat\Pi_{\sf RR Ind Joint}^S$} \label{fig-age-joint}
  \end{subfigure}
  \caption{Examples of continuous attribute Age and nominal attribute Sex}\label{fig-age}
\end{figure*}

Continuous data can be quantified into several categories if necessary. 
Fig.~\ref{fig-age} shows the frequency distributions of Male (light) and Female (dark) 
respondents and for
Age (categorized into 20-year bins (Fig.~\ref{fig-age-pi})), 
the frequency distributions performed via \textsf{RR(X)} (Fig.~\ref{fig-age-rr})
, and the estimated distributions via \textsf{RR-Independent} (Fig.~\ref{fig-age-ind})
and via \textsf{RR-Ind-Joint} (Fig.~\ref{fig-age-joint}). 
The estimations for \textsf{RR-Ind-Joint} are close to the original distribution $\Pi$.

\subsection{Code availability}

The source code of {\sf RR-Ind-Joint} in R is available 
at ().

\end{document}